\pdfoutput=1
\RequirePackage{ifpdf}
\ifpdf 
\documentclass[pdftex]{sigma}
\else
\documentclass{sigma}
\fi

\numberwithin{equation}{section}

\newtheorem{Theorem}{Theorem}[section]
\newtheorem*{Theorem*}{Theorem}
\newtheorem{Corollary}[Theorem]{Corollary}
\newtheorem{Lemma}[Theorem]{Lemma}
\newtheorem{Proposition}[Theorem]{Proposition}
\newtheorem{problem}[Theorem]{Problem}

\theoremstyle{definition}
\newtheorem{Definition}[Theorem]{Definition}

\newtheorem{Remark}[Theorem]{Remark}
\usepackage{enumitem}

\usepackage{cancel,mathrsfs}

\usepackage{tikz, pgfplots, todonotes}
\usetikzlibrary{decorations.markings,arrows, calc, shapes, intersections, patterns}

\def \scr{\mathscr}
\def\Abel{\mathfrak u}

\def \DD{\mathbb D}

\def\tr{{\rm tr}}

\def \ds{\displaystyle}
\def\le{\left}
\def \bs{\boldsymbol}

\def\ri{\right}

\def\bea#1\eea{\begin{align}#1\end{align}}
\def \wt{ \widetilde }
\def \d{{\mathrm d}}
\def\res{\mathop{\mathrm {res}}}
\def \wh{\widehat}

\def\nn{\nonumber}
\def\be{\begin{equation}}
\def\ee{\end{equation}}
\def\ben{\begin{displaymath}}
\def\een{\end{displaymath}}

\def\ba{\begin{array}}
\def\ea{\end{array}}
\makeatletter
\@addtoreset{equation}{section}
\makeatother

\def \1{\mathbf 1}
\def \br{\begin{remark}}
\def\er{\end{remark}}

\def\M{{\mathcal M}}
\def\C{{\mathbb C}}
\def\Z{{\mathbb Z}}
\def\R{{\mathbb R}}

\def\s{\sigma}

\def\N{{\mathbb N}}

\def\pa{\partial}

\def \Re{\operatorname{Re} }
\def \Im {\operatorname{Im} }

\def\la{\label}

\begin{document}
\allowdisplaybreaks

\newcommand{\arXivNumber}{2411.08853}

\renewcommand{\PaperNumber}{097}

\FirstPageHeading

\ShortArticleName{Rational Solutions of Painlev\'e~V}

\ArticleName{Rational Solutions of Painlev\'e~V\\ from Hankel Determinants
and the Asymptotics\\ of Their Pole Locations}

\Author{Malik BALOGOUN and Marco BERTOLA}

\AuthorNameForHeading{M.~Balogoun and M.~Bertola}

\Address{Department of Mathematics and Statistics, Concordia University, \\ 1455 de Maisonneuve W., Montr\'eal, Qu\'ebec,
 H3G 1M8 Canada}
\Email{\mail{malik.balogoun@concordia.ca}, \mail{marco.bertola@concordia.ca}}

\ArticleDates{Received November 14, 2024, in final form October 28, 2025; Published online November 14, 2025}

\Abstract{In this paper, we analyze the asymptotic behaviour of the poles of certain ra\-tio\-nal solutions of the fifth Painlev\'e equation. These solutions are constructed by relating the corresponding tau function to a Hankel determinant of a certain sequence of moments. This approach was also used by one of the authors and collaborators in the study of the rational solutions of the second Painlev\'e equation. More specifically, we study the roots of the corresponding polynomial tau function, whose location corresponds to the poles of the associated rational solution. We show that, upon suitable rescaling, the roots asymptotically fill a region bounded by analytic arcs when the degree of the polynomial tau function tends to infinity and the other parameters are kept fixed. Moreover, we provide an approximate location of these roots within the region in terms of suitable quantization conditions.}


\Classification{33E17; 34M55; 33C47}

\tableofcontents

\vspace{-1mm}

\section{Introduction}\vspace{-1mm}

The six Painlev\'e equations were classified by Painlev\'e and his student Gambier \cite{Gambier,Painleve} more than a century ago. They were searching for second-order ODEs in the complex plane whose solutions, roughly speaking, have the property that all movable singularities are isolated poles. This property has now become known, and referred to, as the {\it Painlev\'e property}.

While this might have remained a purely mathematical investigation, it was much later recognized that these equations have significant applications in mathematical physics, with the resurgence in the late 70s with the works connecting with Ising model and conformal field theory~\cite{McCoyTracyWu1, McCoyTracyWu2}. Another momentous resurgence happened in the 90s when Tracy and Widom~\cite{TracyWidom94} used a special second Painlev\'e transcendent (the Hastings--McLeod solution \cite{HasMc}) to describe the fluctuations of the largest eigenvalue of a large random Hermitean matrix.

Amongst special solutions of the Painlev\'e equation, the rational ones attract a natural interest. The literature is extensive and seems to start with \cite{Vorobev} who discussed rational solutions of the second Painlev\'e equation and defined a special sequence of polynomials that are now called {\it Vorob'ev--Yablonskii} after their discoverers (it appears that Yablonskii defined them slightly earlier but the reference is difficult to find \cite{Yablonskii}).
Rational solutions also appear in semiclassical limits of integrable PDEs. In the one-dimensional sine-Gordon equation near a separatrix, for example, one finds that a suitable scaling of the solution is expressible in terms of a rational solution of the second Painlev\'e equation \cite{BuckMill12}.

Rational solutions exist for all but the first Painlev\'e equation. In all five remaining cases~II--VI, there is a classification of the values of parameters for which there are rational solutions, and for particular cases there are constructions of special families of rational solutions (for Painlev\'e~II \cite{KajiwaraOhta,Vorobev, Yablonskii}, for Painlev\'e III, V, VI \cite{Masuda,MR1942391, MR1650509, Okamoto2,Okamoto3, Umemura1}, for Painlev\'e IV \cite{Okamoto1}). See a~comprehensive overview in~\cite{DLMF} (\href{https://dlmf.nist.gov/32.8}{https://dlmf.nist.gov/32.8}). Note that in loc.\ cit.\ the conditions for existence of rational solutions of Painlev\'e VI appears to be only a sufficient condition. However, necessary and sufficient conditions exist in \cite{Marta}.

The literature that investigates the asymptotic behaviour of the rational solutions, and the pole distribution thereof, is more recent, probably due to the interest spurred by numerical investigations and the appearance of well-defined patterns. For the zeros of Okamoto polynomials (which are poles of rational solutions of PIV) see \cite{Ref1_3, BuckMill22, Clakxon_IV,Novok14}, for the zeros of Vorob'ev--Yablonskii polynomials and Painlev\'e II see \cite{BertolaBothner, BuckMill14, BuckMill15,Clarxon,Ref1_4}, for the second Painlev\'e hierarchy see \cite{BalBerBo,Clarxon}. There are also studies of explicit families of rational solutions of various Painlev\'e equations: for Painlev\'e III \cite{Ref1_2,Ref1_1, ClaxonIII}, for Painlev\'e~V \cite{Ar1, Ar2,ClarksonV, ClarksonDunning}, for Painlev\'e~IV \cite{Ref1_3, Clakxon_IV, MasoeroIV, MasoeroIV2} and for higher Painlev\'e equations \cite{Ar3,BalBerBo}.

The approach to asymptotic analysis relies on the formulation of an associated Riemann--Hilbert problem, namely, a boundary value problem for a piecewise analytic matrix-valued function.
Within this framework there are two logical distinct approaches that can be used. We can categorize them under the following banners:
\begin{enumerate}\itemsep=0pt\samepage
\item[$(1)$] the isomonodromic approach,
\item[$(2)$] the orthogonal polynomial approach.
\end{enumerate}

\pagebreak

\noindent
The isomonodromic approach relies on the general fact that any Painlev\'e equation appears as the compatibility between a $2\times 2$ system of ODEs with rational coefficients in the complex plane and an additional PDE in an auxiliary parameter \cite{JMU1}. The different solutions are parametrized by (generalized) monodromy data of the ODE, which is the starting point for the Riemann--Hilbert analysis. Typically, the degree of the rational solution appears explicitly as one of the parameters in the monodromy data, and can be used as large parameter in the asymptotics. This is the philosophy behind the works \cite{BuckMill14, BuckMill15}.

The second approach was used, possibly for the first time, in \cite{BertolaBothner} and then also applied to the generalized Vorob'ev--Yablonskii polynomials in \cite{BalBerBo}. It is also the approach we follow in this paper.
The main connection between orthogonal polynomials and equations of Painlev\'e type was established in \cite{Bertola:Semiiso}, where it was shown that Hankel determinants built out of the moments of ``semiclassical'' moment functionals are always isomonodromic tau functions in the sense of Jimbo--Miwa--Ueno \cite{JMU1}. It was a remark \cite[Remark~5.3]{Bertola:Semiiso} that special choices of semiclassical moment functionals lead automatically to tau functions of Painlev\'e equations (all, except possibly for Painlev\'e I). In general, however, these solutions correspond to transcendental solutions like, for example, the solutions of PII constructed out of determinants of derivatives of Airy functions, see \cite{JoshiKajiwara}.

It is possible to further restrict the setup of orthogonal polynomials in such a way that the moments of the moment functional become {\it polynomials} in a parameter, which then guarantees that the Hankel determinant (automatically an isomonodromic tau function) is a polynomial tau function of an equation of Painlev\'e type. This is what works ``behind the scenes'' in \cite{BertolaBothner}.

The advantage of this reformulation in terms of associated orthogonal polynomials is that there is a solid and well-developed framework for studying their large degree asymptotics, with an extensive literature that starts with the seminal work of Deift et al.~\cite{DKMVZ}.

Before going into any further detail, let us discuss the known literature and results about the rational solutions of the fifth Painlev\'e equation.

{\bf Rational solutions of PV.}
The fifth Painlev\'e equation is the following nonlinear, second-order ODE in the complex domain for the unknown function $y(t)$
\begin{gather}
y'' = \le(\frac 1{2y} + \frac 1{y-1}\ri) \bigl(y'\bigr)^2 - \frac {y'}t + \frac {(y-1)^2}{t^2} \le(\alpha y + \frac \beta y\ri) + \frac {\gamma y}t + \frac {\delta y (y+1)}{y-1},
\label{PVrat}
\end{gather}
where the prime denotes $\d/{\d t}$ and $\alpha, \beta, \gamma, \delta \in \C$ are parameters: we shall refer to \eqref{PVrat} as $P_5(\alpha, \beta, \gamma,\delta)$.
The equation admits certain symmetries that change the value of the parameters. If $y(t)$ is a solution of $P_5(\alpha,\beta,\gamma,\delta)$, then
\begin{enumerate}\itemsep=0pt
\item[(1)] $y(-t)$ is a solution of $P_5(\alpha, \beta, - \gamma, \delta)$,
\item[(2)] $\frac 1{y(t)}$ is a solution of $P_5(-\beta,-\alpha, -\gamma, \delta)$,
\item[(3)] $y(\lambda t)$ is a solution of $P_5\bigl(\alpha, \beta, \lambda\gamma, \lambda^2 \delta\bigr)$, for any $\lambda\in \C\setminus\{0\}$.
\end{enumerate}
Using the above symmetries, the analysis is reduced to only two families: the family where $\delta=0$ (which is called ``degenerate'' and can be reduced to Painlev\'e~III) and the case $\delta\neq 0$ which, by virtue of the last of the above symmetries, is customarily set to $\delta = -\frac 1 2$.

The classification of rational solutions is contained in~\cite{Kitaev94}, where the authors show that rational solutions exist only if the parameters satisfy certain relations. More precisely (paraphrasing and condensing their results), we have the following.

\begin{Theorem}[{\cite[Theorems~1.1 and~1.2]{Kitaev94}}]
\label{thmKitaev}
The equation $P_5\bigl(\alpha,\beta,\gamma,-\frac 1 2\bigr)$ \eqref{PVrat} admits rational solutions if and only if there are integers $k,m\in \Z$ such that
\begin{enumerate}[leftmargin=31pt]\itemsep=0pt
\item[$(I)$] $\alpha = \frac 1 2 (\gamma +k)^2$, $\beta = - \frac {m^2}2$, $k+m$ odd, and $\alpha\neq 0 $ when $|k|<m$,
\item[$(II)$] $\alpha = \frac 1 2 m^2$, $\beta = - \frac {(\gamma +k)^2}2$, $k+m$ odd, and $\beta\neq 0 $ when $|k|<m$,
\item [$(III)$] $\beta = -\frac 1 2(\alpha_1+m)^2$, $\gamma=k$ with $\alpha_1^2=2\alpha$ so that $m\geq 0 $ and $k+m$ even,
\item [$(IV)$] $\alpha =\frac {k^2}4$, $\beta = -\frac {m^2}8$, $\gamma \not\in \mathbb Z$ where $k,m>0$ and $k$, $m$ both odd.
\end{enumerate}
In cases $(I)$ and $(II)$, the solution is unique if $\gamma \not\in\Z$ and there are at most two rational solutions otherwise.
\end{Theorem}
Observe that the cases (I) and (II) of the above result are really the same family up to the application of the symmetry \smash{$y(t)\leftrightarrow \frac 1{y(-t)}$} which transforms solution of $P_5\bigl(\alpha, \beta, \gamma, -\frac 1 2\bigr)$ into solutions of $P_5\bigl(-\beta, -\alpha, \gamma, -\frac 1 2\bigr)$.

In the recent \cite{ClarksonDunning}, the authors construct the rational solutions corresponding to the case (II) of \cite{Kitaev94} above (which is case (i) in \cite[Theorem~4.1]{ClarksonDunning}).
More precisely, they construct the tau functions
\be
\label{tauClarkson}
\tau^{(\mu)}_{m,n} := \det \le[\le(t \frac \d{\d t}\ri)^{j+k} L^{(n+\mu)}_{m+n} (t)\ri]_{j,k=0}^{n-1},
\ee
where $L^{( \alpha)}_n(t)$ are the associated Laguerre polynomials (see \href{https://dlmf.nist.gov/18.5.E12}{DLMF:18.5.12}, \cite{DLMF})
\[
L^{(\alpha)}_{n}\left(t\right)=\sum_{\ell=0}^{n}\frac{{\left(\alpha+\ell+1%
\right)_{n-\ell}}}{(n-\ell)!\;\ell!}(-x)^{\ell}=\frac{{\left(\alpha+1\right)_{%
n}}}{n!}{{}_{1}F_{1}}\left({-n\atop\alpha+1};t\right).
\]
Then their result implies that \eqref{tauClarkson} is the tau function of
\[
P_5\le( \frac {m^2}2,- \frac{(m+2n+1+\mu)^2} 2, \mu, -\frac 1 2\ri),
\]
 which is the case (II) in Theorem~\ref{thmKitaev}.
Many interesting properties are discussed but no description of the asymptotic behaviour for large degree is undertaken.

{\bf Results.}
In the present paper, we will provide a description for the tau function of the case~(I) of Theorem~\ref{thmKitaev}, with the main goal of describing the distribution of its zeros as $n\to \infty$.
More precisely, we construct the tau function corresponding to the parameters
$k = -3n - {K} -1$, $ m = n+ {K}$, $ \gamma = 1+2n + {K} -{\rho}$, $ n\in \N$, $ {K} \in \Z$, $ {\rho} \in \C$,
namely
\[
\alpha = \frac {(n+ { {\rho}})^2}2,\qquad
\beta = -\frac {(n + { {K}})^2} 2,\qquad
\gamma = 1+2n + { {K}} - { {\rho}}.
\]
Thus the tau function we construct is for the equation
\[
P_5\le( \frac {(n+ { {\rho}})^2}2, -\frac {(n + { {K}})^2} 2, 1+2n + { {K}} - { {\rho}}, - \frac 1 2\ri).
\]
\begin{Remark}
The rational solutions we consider here are for case (I) while those in \cite{ClarksonDunning} are for case (II) of Theorem~\ref{thmKitaev}. Of course, the two cases are related: at the level of the transcendents~the solutions are related by the simple transformation $y(t)\mapsto \frac 1{y(t)}$ and $(\alpha,\beta, \gamma)\mapsto (-\beta,-\alpha,-\gamma)$, but at the level of the tau functions the relation is slightly less straightforward and given in Remark~\ref{CDcomp} below.
The comparison of the parameters $\mu$, $n$, $m$ of \cite{ClarksonDunning} (which we now have to rename $\wt \mu$, $\wt n$, $\wt m$ to avoid confusion) is then as follows
$
\wt m= n+ { {K}} $, $ \wt n= n$, $
\wt \mu =- 1-2n- { {K}}+ { {\rho}}$.
\end{Remark}
To construct this $\tau$-function, we consider the following sequence of {\it moments} depending on the parameters ${K} \in \Z$, ${\rho} \in \C$:
\be
\label{muj}
\mu_j(t):= \oint_{|z|=R}z^j z^{ {K}} \le(1 - \frac 1 z \ri)^{{\rho}} {\rm e}^{\frac t z}\d z, \qquad j=0,1,2,\dots.
\ee
As we detail in Remark \ref{CDcomp}, the moments $\mu_j(t)$ are Laguerre polynomials up to a sign.
Here $R>1$ can be chosen arbitrarily, thanks to Cauchy's residue theorem.
We interpret the $\mu_j$'s as the moments of the complex measure \smash{$\d\mu(z) = z^{ {K}} \le(1 - \frac 1 z \ri)^{{\rho}} {\rm e}^{\frac t z}\d z$} on $|z|=R$. Observe that the~$\mu_j(t)$'s are polynomials in~$t$, and can also be obtained from a generating function
\[
\sum_{j=0}^\infty \mu_j(t) \frac{s^j}{j!} = \oint_{|z|=R} z^{ {K}} \le(1 - \frac 1 z \ri)^{{\rho}} {\rm e}^{\frac t z + sz}\d z
\]
 or also (by replacing $z= \frac 1 w$ in the definition of moments \eqref{muj} and then by evaluating them as residues)
 \[
 F(s;t):=-2{\rm i}\pi s^{-{K}-1} \le(1-s\ri)^{{\rho}} {\rm e}^{st} = \sum_{j=0}^{\infty} \mu_j(t) \frac { s^j}{j!}.
 \]
The map that associates to any polynomial $p(z)$ the value \smash{$\mathcal M[p]:= \oint_{|z|=R} p(z) \d \mu(z)$} is an example of {\it semiclassical moment functional} \cite{Marcellan0, Marcellan, Maroni, Shohat} and thus fits naturally in the general theory of \cite{Bertola:Semiiso} which guarantees that the Hankel determinant
\[
\tau_n(t; {K}, {\rho} ):= \det \le[\mu_{a+b-2}(t)\ri]_{a,b=1}^n
\]
is a tau function of an equation of Painlev\'e type. The matching with PV is explained in Section~\ref{SecOPLP}.
More explicitly, from the identification it will follow that it satisfies the $\sigma$-form of the PV equation \cite[Appendix~C]{JMU2} in the following way. Let us define the quantities
\begin{gather}%
H_{_V}:= \frac {\d}{\d t} \ln \tau_n(t; {K}, {\rho} ) + \frac {\rho} 2,
\qquad
 \s(t) := t H_V + \frac t 2 (\theta_0 + \theta_\infty) + \frac { (\theta_0 + \theta_\infty)^2 - \theta_1^2}4,\nonumber
 \\
\theta_0 :=2n + { {K}}, \qquad
\theta_1 := - { {\rho}},\qquad
\theta_\infty:={ {\rho}} - { {K}}.\label{defsigmaHV}
\end{gather}
Then the $\sigma$-form of PV is the following ODE for $\sigma$ defined in \eqref{defsigmaHV}
\begin{align*}
\le(t \frac {\d^2 \s}{\d t^2}\ri)^2 ={}& \le(
 \s - t\frac {\d \s}{\d t} + 2\le(\frac {\d \s}{\d t}\ri)^2 - (\theta_\infty + 2\theta_0) \frac {\d \s}{\d t}
 \ri)^2 \nonumber\\
 &- 4 \le(\frac {\d \s}{\d t}\ri)
 \le(\frac {\d \s}{\d t} - \frac { \theta_0 -\theta_1 + \theta_\infty}2 \ri)
 \le( \frac {\d \s}{\d t} - \theta_0 \ri)
 \le(\frac {\d \s}{\d t} - \frac { \theta_0 + \theta_1 + \theta_\infty}2 \ri).
 \end{align*}

In the second part of the paper, we exploit the connection between $\tau_n(t)$ and Hankel determinants to describe the asymptotic location of its zeros as $n\to\infty$.

\begin{figure}[h]\centering
\begin{minipage}{0.32\textwidth}
\begin{center}
\includegraphics[width=1\textwidth]{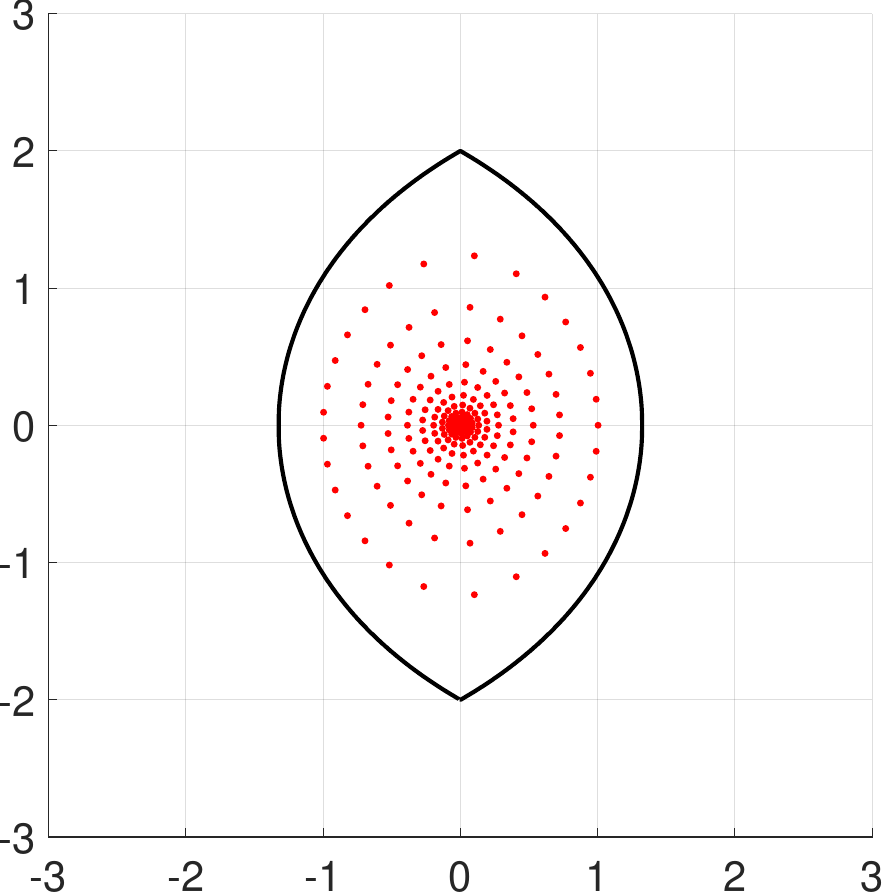}
$n=15, \ {\rho} = \frac 1{100}$
\end{center}
\end{minipage}
\begin{minipage}{0.32\textwidth}
\begin{center}
\includegraphics[width=1\textwidth]{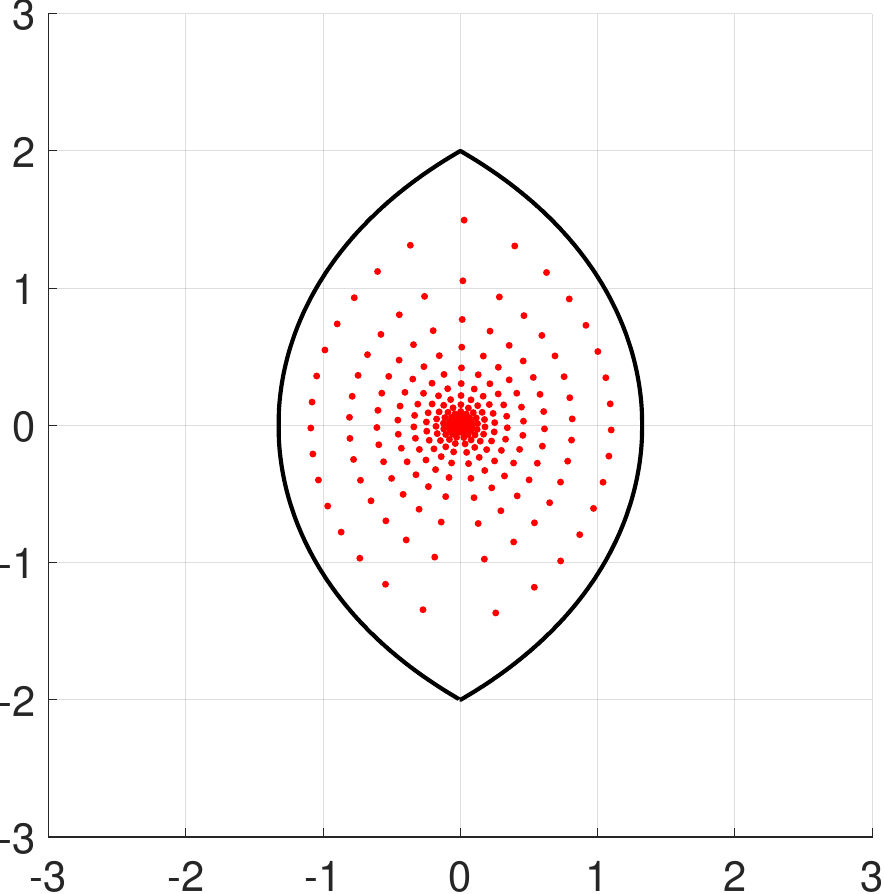}
$n=16, \ {\rho} = \frac 3 {100} + \frac {13{\rm i}}{100}$
\end{center}
\end{minipage}
\begin{minipage}{0.32\textwidth}
\begin{center}
\includegraphics[width=1\textwidth]{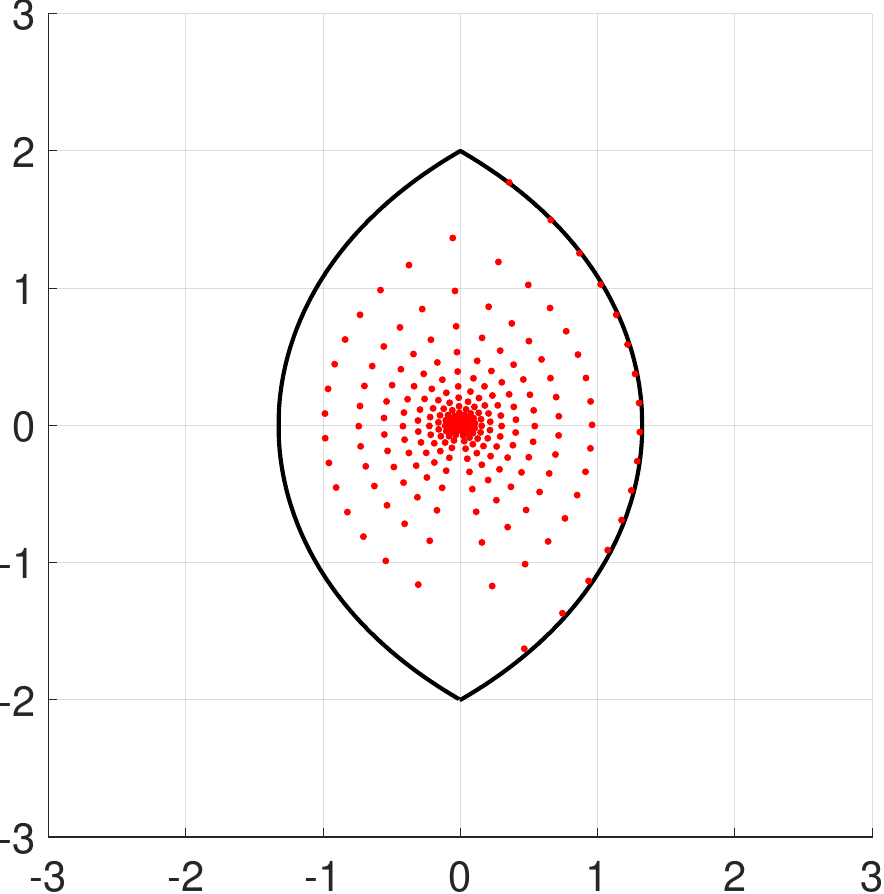}
$n=16, \ {\rho} = \frac {101} {100} + \frac {33{\rm i}}{100}$
\end{center}
\end{minipage}
\begin{minipage}{0.32\textwidth}
\begin{center}
\includegraphics[width=1\textwidth]{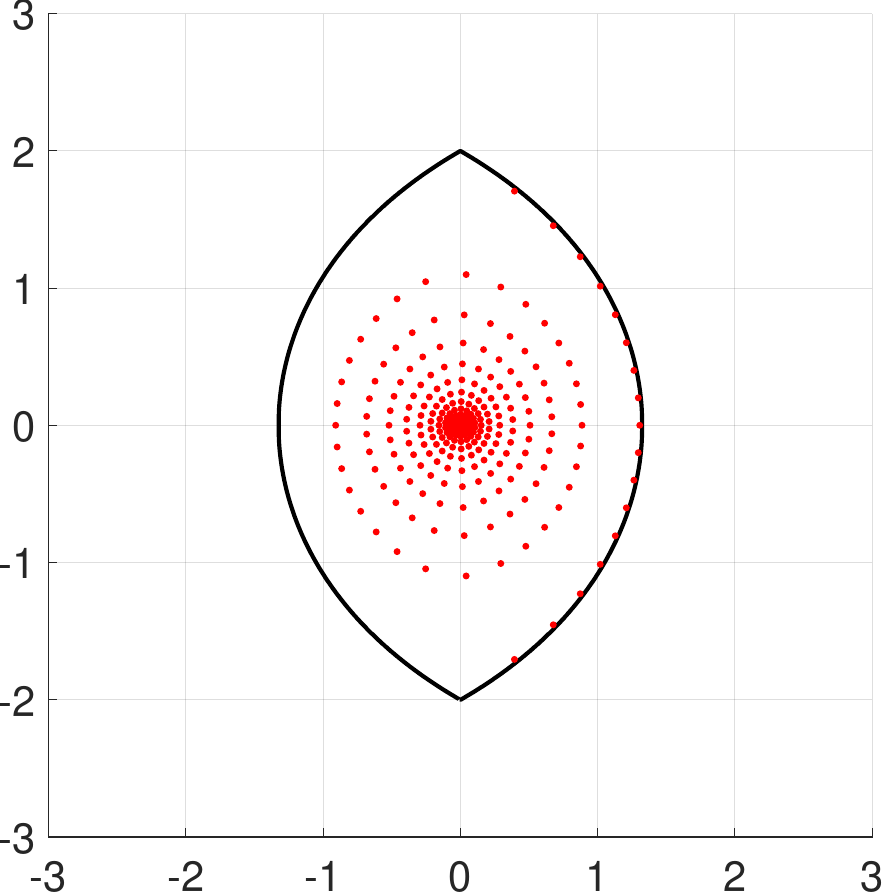}
$n=17, \ {\rho} = \frac {101} {100} $
\end{center}
\end{minipage}
\begin{minipage}{0.32\textwidth}
\begin{center}
\includegraphics[width=1\textwidth]{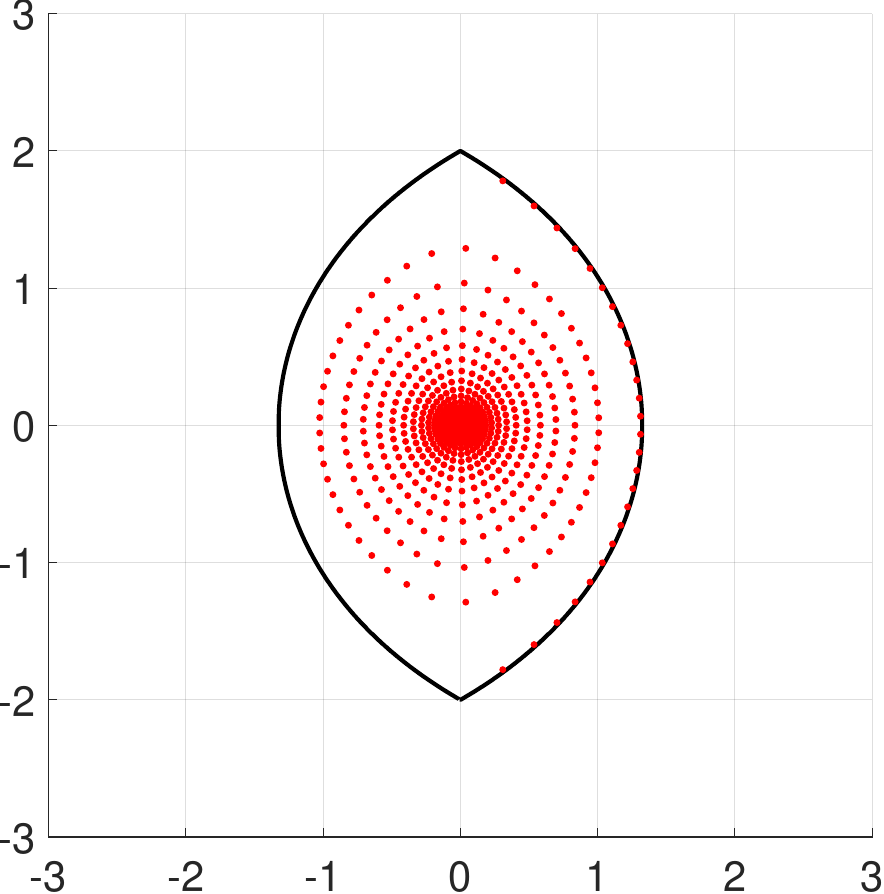}
$n=26, \ {\rho} = \frac {101} {100} $
\end{center}
\end{minipage}
\begin{minipage}{0.32\textwidth}
\begin{center}
\includegraphics[width=1\textwidth]{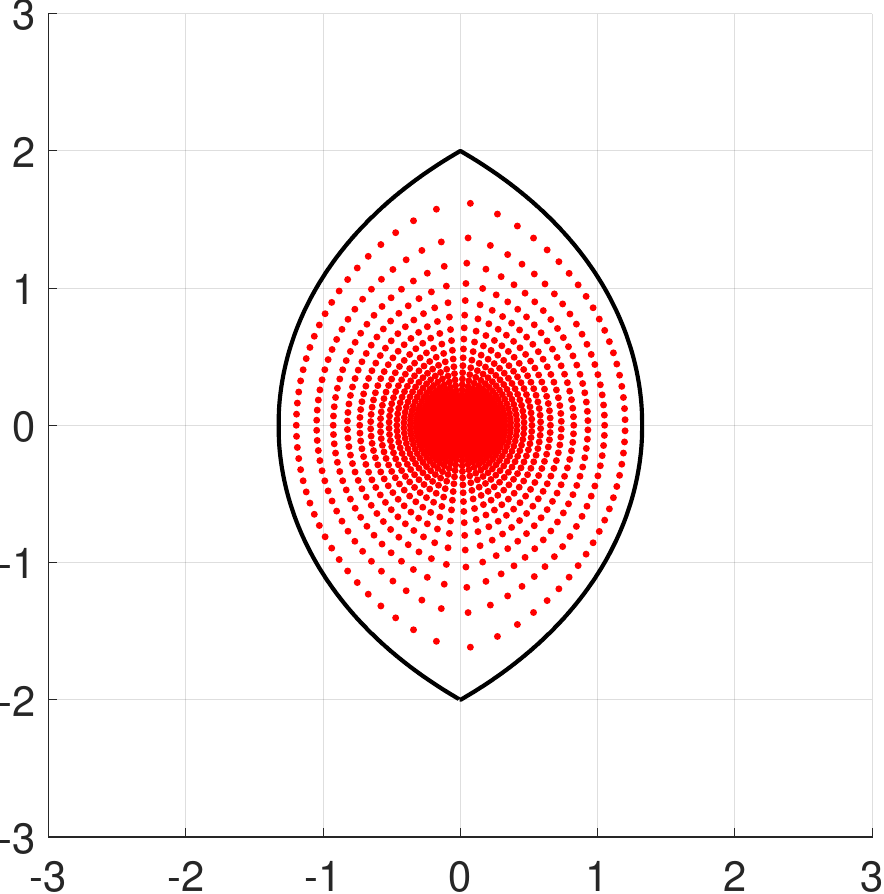}
$n=40, \ {\rho} = \frac {3} {100} $
\end{center}
\end{minipage}
\caption{The zeros of several instances of polynomial tau functions (all for ${K}=0$).}\label{FigEyes}
\end{figure}

More precisely, we describe the asymptotic behaviour of the zeros of the polynomials
\be
\mathcal T_n(s):=
\tau_n(ns; {K}, {\rho} ), \qquad {K}\in \Z,\quad {\rho}\not\in \Z,
\label{Pns}
\ee
as $n$ tends to infinity in the $s$-plane (these zeros are the same as the zeros of $\tau_n$ but homothetically rescaled by a factor $1/n$).
We observe that we do not consider a double-scaling limit where the either one of the parameters ${K}$, ${\rho}$ or both are proportional to $n$ as $n\to\infty$. While the general setup is conducive to such an analysis, the details of the construction of the $g$-function would have to be changed significantly and this is deferred to a future publication. The logic of the analysis is parallel to the one employed in \cite{BalBerBo,BertolaBothner} and
the main goal is to explain the emerging shape which clearly appears, see Figure~\ref{FigEyes}.

In the figure, an almond-shaped region, which we call ``Eye of the Tiger'' (${\rm EoT}$) is visible.
We will show that for fixed values of $K$, $\rho$ and as $n$ tends to infinity, the zeros of $\mathcal T_n(s)$ will eventually vacate any closed set $\mathcal K$
not intersecting the (closure of) ${\rm EoT}$ or, which is the same,
the zeros will eventually lie within $\epsilon$-distance of ${\rm EoT}$.

Namely, for every $K\in \N$, $ \rho\in \C\setminus \Z$, $\epsilon>0$, there is an $N_0= N_0(K, \rho, \epsilon)$ such that all the zeros of $\mathcal T_n(s)$ are inside the $\epsilon$-fattening ${\rm EoT}^\epsilon$. We recall that an $\epsilon$-fattening of a bounded set~${X\subset \C}$ is the (open) set
$
X^\epsilon:= \bigcup_{z\in X} \DD_z(\epsilon )$, $ \DD_z(\epsilon):=\{w\mid |w-z|<\epsilon\}$.
We do not provide an estimate of the threshold value $N_0$.

This phenomenon is well illustrated by Figure~\ref{Figone}, top right pane. Indeed, we can see that certain zeros lie outside the region (for the specific value of $n=17$). If we were to plot the zeros for the same values of the parameters ${K}$, ${\rho}$, but for larger values of $n$, the zeros would gradually drift towards the region ${\rm EoT}$. The method of proof does not allow us to conclusively show that all the zeros will eventually fall in the interior of ${\rm EoT}$. See Theorem~\ref{theoremnozeros}.

 The boundary of the region ${\rm EoT}$ (marked in black) is determined by the implicit equation in the $s$-plane
\be
\label{EoTbdry}
 \ln\le|\frac {2}{s} + \sqrt{\frac {4}{s^2}+1}\ri| - \Re\le(\frac s 2\sqrt{ \frac {4}{s^2} + 1} \ri)=0,
\ee
which follows from \eqref{phi0}, \eqref{EoTcurve} below. Note that in the above expression the square roots need to be understood with branch-cuts extending from $2{\rm i}$ vertically upwards, and from $-2{\rm i}$ vertically downwards. With this choice the function on the left side of \eqref{EoTbdry} is continuous in~${\C\setminus\{0\}}$ and even.

Within the set ${\rm EoT}$, the zeros of $\mathcal T_n(s)$ \eqref{Pns} arrange themselves in a semi-regular lattice. We give a description of this lattice in terms of a ``quantization condition'' described in Section~\ref{summary}, in particular \eqref{quantcond} (for $\Re (s)>0$) or \eqref{quantcond<0} (for $\Re (s)<0$). The two quantization conditions~\eqref{quantcond}, \eqref{quantcond<0} determine an asymptotic grid (shown in Figure~\ref{Figone} in green and blue thin lines) at whose vertices the zeros are approximately located. We do not estimate rigorously the rate of convergence, but the numerical evidence is quite striking even for relatively small values~of~$n$.\looseness=-1

\begin{figure}[t]\centering
\begin{minipage}{0.32\textwidth}
\centering
\includegraphics[width=1\textwidth]{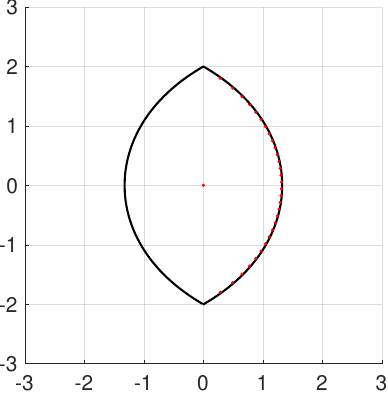}
$n=30$, $ {\rho} = 1 $
\end{minipage}
\begin{minipage}{0.32\textwidth}
\centering
\includegraphics[width=1\textwidth]{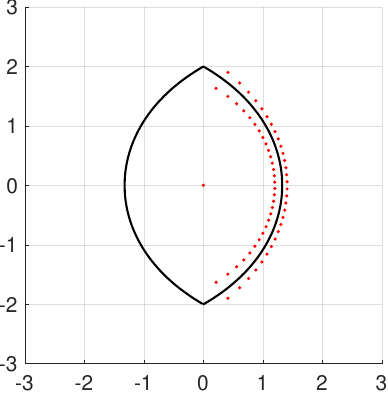}
$n=30$, $ {\rho} = 2 $
\end{minipage}
\begin{minipage}{0.32\textwidth}
\centering
\includegraphics[width=1\textwidth]{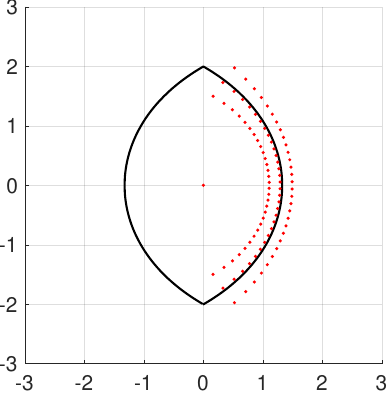}
$n=30$, $ {\rho} = {3}$
\end{minipage}
\caption{Some plot of the zeros of the tau function $\mathcal T_n(s)$ \eqref{Pns} for $(n,\rho) = (30,1), (30,2), (30,3)$ ($K=0$ in all cases). Observe that for $\rho$ integer our description of the distribution of zeros does not apply and in particular the region ${\rm EoT}$ remains largely empty, with an ``eyelash'' effect. There is a high multiplicity zero at the origin. As $n$ increases but $\rho$ remains fixed, the ``eyelashes'' become thinner around the edge of ${\rm EoT}$. A similar phenomenon was observed, for rational solutions of PIII, in \cite{Ref1_1}. We are not going to discuss the case $\rho\in \Z$ in this paper. Clearly the behaviour of the zeros undergoes a substantial change, as they appear to be accumulating along part of the boundary of ${\rm EoT}$.}\label{plotrhoint}
\end{figure}

\begin{Remark}\la{rhoint}
Our analysis assumes that $\rho$ is not integer. This is an essential assumption, not of technical nature. It is sufficient to show a few plots of the zeros of $\mathcal T_n(s)$ \eqref{Pns} to see that the pattern of zeros inside ${\rm EoT}$ has completely collapsed. For example, it is easy to see that if $\rho=0$ the tau function reduces to a monomial. On the technical level, for the asymptotic analysis in the interior of ${\rm EoT}$ we need the density of our measure to have a branch-cut discontinuity from~${z=0}$ to $z=1$. A few plots with integer $\rho$ are indicated in Figure~\ref{plotrhoint} to illustrate the change of behaviour.
\end{Remark}

\section[Semiclassical orthogonal polynomials and tau functions of Painlev\'e type]{Semiclassical orthogonal polynomials and tau functions\\ of Painlev\'e type}

Let us recall the notion of semiclassical orthogonal polynomials \cite{Marcellan0,Marcellan,Maroni,Shohat}.
\begin{Definition}\label{defSOP}
Given a pair of polynomials $(A,B)$, with $B$ monic, a semiclassical moment functional of type $(A,B)$ is a linear map $\M\colon\C[z]\to \C$ on the space of polynomials in the indeterminate variable $z$ such that for all $p\in \C[z]$
\be
\label{defsemi}
\M\big[A(z) p(z)\big] = \M \big[B(z) p'(z)\big].
\ee
A polynomial $p_n(z)$ of degree $n$ is called {\rm orthogonal} for the moment functional $\mathcal M$ if
\[
\M\big[p_n(z) z^j\big]=0, \qquad j=0,1,\dots, n-1.
\]
\end{Definition}
Given a moment functional $\M$ its {\it moments} are the (in general complex) numbers
$
\mu_j:= \mathcal M\big[z^j\big]$.
 It is possible to show that these moment functionals admit an integral representation as follows: define the {\it symbol}\footnote{We adopt the terminology common in the literature on Toeplitz operators. } as the function $\theta$ such that
\[
\theta'(z):= -\frac {B'(z)+A(z)}{B(z)}.
\]
Then we can express a semiclassical moment functional of type $(A,B)$ in terms of an integral of the form
\[
\M\le[p(z)\ri] = \int_\gamma p(z) {\rm e}^{\theta(z)} \d z.
\]
The contour of integration can be selected from several {\it homology classes}, each providing a~linearly independent moment functional of type $(A,B)$. The allowable contours have the defining property that the integration of the right side of \eqref{defsemi} can be performed by parts and with vanishing boundary terms.
It is known that there are $d = \max\{\deg A, \deg B-1\}$ independent such homology classes (generically) and a description of them can be found, for example, in \cite{Bertola:Semiiso}.
It was also shown in loc.\ cit.\ that semiclassical moment functionals are inextricably related with the theory of isomonodromic deformations, and hence {\it in particular} with the theory of Painlev\'e equations, as we briefly recall.

Indeed, it was shown in \cite{Bertola:Semiiso} that any deformation of the coefficients of $A$, $B$ that preserves all the residues of $\theta'$ is an {\it isomonodromic deformation} for a suitably defined differential equation in the complex plane of rank $2$. It was determined by the Japanese school several decades ago~\cite{JMU1,JMU2} that to any such isomonodromic deformation we can associate a tau function, namely a function of the isomonodromic deformation parameters (i.e., of the coefficients of $A$, $B$ in this case). While this is not an appropriate venue to review all applications of tau functions, suffice to say here (see \cite[Remark~5.3]{Bertola:Semiiso}) that by appropriate choices of the symbol and corresponding contours of integration $\gamma$ we can construct tau function for all the Painlev\'e equations II--VI. We will make below one such choice which has the additional property of producing {\it polynomial} moments, and hence polynomial tau functions.

\subsection{Special case of semiclassical functional of PV type}
We specialize the previous description to the case of the following symbol depending on the parameters $t$, ${ {K}}$, ${ {\rho}}$:
\be
\label{deftheta}
{\theta(z)} := \frac t z +{ {\rho}} \ln \le(1- \frac 1 z\ri) + { {K}} \ln z, \qquad {\rm e}^{\theta(z)} = \le(1 - \frac 1 z\ri)^{ {\rho}} z^{{ {K}}} {\rm e}^{\frac t z}.
\ee
In \eqref{deftheta}, all the logarithms are principal and $\theta$ is analytic in $\C\setminus (-\infty, 1]$.
Since ${K}$, $ {\rho}$ are the residues of $\theta'\d z$, the only isomonodromic parameter is $t$ in the above expression.
 Note that if~${ {K}} \in \Z$ (which is going to be our main focus shortly) the weight function ${\rm e}^{\theta(z)}$ is analytic in~${\C\setminus [0,1]}$. Furthermore, we will avail ourselves of the freedom of deforming the segment $[0,1]$ to a smooth arc with the same endpoints when needed.
The corresponding moment functionals, in the terminology of Definition~\ref{defSOP} above, are of type
\be
(A,B) = \bigl({ {K}} z^2 + ({ {\rho}} + { {K}}+t)z + t, z^2(z-1)\bigr).
\la{AB}
\ee
We can choose the contours of integration for defining the specific moment functional as follows:%
\begin{enumerate}\itemsep=0pt
\item[(1)] $ \gamma$ a closed contour originating at $z=0$ along the direction $\arg (-t)$, looping around $z=0$ and terminating at $z=0$ along the same direction and leaving $z=1$ in the inside.
\item[(2)] $\wt \gamma$ the same as above but leaving $z=1$ on the outside.
\end{enumerate}
There are some special instances, depending on the values of the parameters ${ {K}}, { {\rho}}$. If ${ {\rho}}=0,1,2,\dots $, then we can replace $\wt \gamma$ with a contour terminating at $z=1$.

Most important for our consideration is the case ${K} \in \Z$, ${ {\rho}} \not\in \Z$. In these cases, we can replace~$\gamma$ with the homotopy class of a circle of radius $R>1$ in $\C\setminus [0,1]$.

The latter case is relevant for the construction of rational solutions of the fifth Painlev\'e equation. Indeed, if ${K}\in \Z$ the moments are {\it polynomials} in the variable $t$,
\be
\la{defmuj}
\mu_j(t) = \oint_{|z|=R} z^j {\rm e}^{\theta(z)}\d z, \qquad \deg \mu_j (t) = j+{K}+1.
\ee

In our paper, we consider the Hankel determinants
\be
\label{Dn}
\tau_n(t; {K}, {\rho} ):= \det \bigl[\mu_{a+b-2}(t)\bigr]_{a,b=1}^n.
\ee
They are also polynomials of degree
$
\deg \tau_n(t) = n(n+{K})$.
In fact, the determinant of leading terms in $t$ of the matrix of moments gives
\[
\tau_n(t; {K}, {\rho}) = t^{n(n+{K})} \det \le[\frac 1{( {K} + a+b-1)!} \ri]_{a,b=1}^n \le(1 + \mathcal O\bigl(t^{-1}\bigr)\ri).
\]
 It is simple to establish that the leading coefficient does not vanish, so that the degree of $\tau_n$ is exactly $n(n+{K})$. Indeed, one may notice that the determinant is the Wronskian, evaluated at~${x=1}$ and up to reversal of order of rows, of the monomials
\smash{$f_j(x) =\frac{ x^{{K}+ 2n-j}}{ ({K}+ 2n-j)!}$}, $j=1,\dots, n$. This Wronskian is itself a monomial, and hence it does not vanish at $x=1$.

\begin{Remark}[comparison with \cite{ClarksonDunning}]\label{CDcomp}
To connect the moments with known functions, we perform the change of variable $\frac 1 z = \frac s{s-1}$ in the defining integral of the moments $\mu_j(t)$ \eqref{defmuj}. After a simple computation, the expression for the moments becomes
\[
\mu_j(t) = (-1)^{j+{K}} \oint_{|s|= r} \frac { {\rm e}^{t \frac s{s-1}}}{(1-s)^{\nu + 1} s^{n+1}} \frac{\d s}{2{\rm i}\pi}, \qquad \nu = {\rho}-j-{K}-1,\qquad n = j + {K}+1.
\]
We compare with the known integral representation of generalized Laguerre polynomials
\[
L^{(\nu)}_n(t) = \oint_{|s|=r} \frac{{\rm e}^{t \frac s{s-1}}}{(1-s)^{\nu+1} s^{n+1}} \frac{ \d s}{2{\rm i}\pi},
\qquad 0<r<1,
\]
from which we conclude
\smash{$
\mu_j(t) = (-1)^{j+{K}} L^{{\rho}-j-{K}-1}_{j+{K}+1}(t)$}.
Using finally the simple relation $\pa_t^k \mu_j(t) \allowbreak= \mu_{j-k}(t)$, we can rewrite the Hankel determinant~\eqref{Dn} as
\begin{align*}
\tau_n(t; {K}, {\rho}) = {}&(-1)^{ {K} n} \det \le[ \le(\frac{\d}{\d t} \ri)^{a+b}\mu_{2n-2}(t)
\ri]_{a,b=0}^{n-1}\\
={}& (-1)^{ {K} n} \det \le[ \le(\frac{\d}{\d t} \ri)^{a+b} L^{ ({\rho}-{K} - 2n+1)} _{2n-1 + {K}} (t)
\ri]_{a,b=0}^{n-1}.
\end{align*}
With reference to the notation of \cite[Definition 3.1]{ClarksonDunning}, the relationship is then
\[
\tau_n(t; {K}, {\rho})= (-1)^{ {K} n} T^{({\rho}-{K} - 2n)}_{n+ {K}-1, n}(t)= (-1)^{ {K} n} t^{\frac{n(1-n)}2} a_{n+ {K}-1,n}
\tau^{({\rho}-{K} - 2n)}_{n+ {K}-1, n}(t),
\]
where in the last identity we have used Lemma 4.5 of loc.\ cit.\ to connect with the definition~\eqref{tauClarkson}. The proportionality constant $a_{m,n}$ is defined in the same lemma.
\end{Remark}

The results of \cite{Bertola:Semiiso} imply that these Hankel determinants are tau functions of an isomonodromic system, and we are going to identify this with the Lax pair of the fifth Painlev\'e equation in the next section.

\begin{Remark}
We are considering the moment functional of type \eqref{AB} that corresponds to the choice of contour $\gamma$ surrounding the segment $[0,1]$ in the complex plane. The general theory of moment functionals discussed earlier allows for other choices. While these do not lead to {\it polynomial} moments (in the variable $t$), their Hankel determinants would, by the same reasoning in \cite{Bertola:Semiiso}, provide isomonodromic tau functions of Painlev\'e~V. Up to the map $z\mapsto 1/z$ and setting ${\rho}= {K} =0$ and choosing the contour of integration as the segment $[0,1]$ (the so-called ``hard-edge case''), one would recover the moment functional appearing in \cite{Deano}. In that case, the orthogonal polynomials are the so-called ``kissing polynomials''. It would be possible to perform a similar asymptotic analysis of the corresponding tau functions for large~$n$ and determine the zero distribution, despite the tau functions not being polynomials.
\end{Remark}
\subsection{From orthogonal polynomials to Lax pairs}
\label{SecOPLP}
The bridge is provided by the Riemann--Hilbert formulation of orthogonal polynomials \cite{FIK} which, in our case reads as follows.
\begin{problem}
\label{RHPY}
Given $n\in \mathbb N$,
find a $2\times 2$ matrix-valued function $Y(z)= Y_n(z)$ such that
\begin{enumerate}\itemsep=0pt
\item[$(1)$] $Y(z)$ and $Y^{-1}(z)$ are holomorphic and bounded in $\C\setminus \gamma$.
\item[$(2)$] The boundary values along $z\in \gamma$ satisfy
\[
Y(z_+) = Y(z_-) \begin{bmatrix}
1&{\rm e}^{\theta(z)}\\
0 & 1
\end{bmatrix},\qquad \forall z\in \gamma
\]
with $\theta$ as in \eqref{deftheta}.
\item[$(3)$] As $z\to \infty$, the matrix $Y_n(z)$ admits an asymptotic expansion of the form
\be
\label{asympY}
Y_n(z) = \le(\1 + \sum_{\ell=1}^\infty \frac {Y_n^{(\ell)}} {z^\ell} \ri)z^{n\s_3}, \qquad \s_3:= \begin{bmatrix} 1&0\\ 0&-1\end{bmatrix},
\ee
where the coefficient matrices \smash{$Y_n^{(\ell)}$} are independent of $z$ and $\1$ denotes the identity matrix of the appropriate size $($here, $2\times 2)$.
\end{enumerate}
\end{problem}

In the following theorem, we condense the essential results that are at the core of much of the theory of orthogonal polynomials. The theorem is mostly the reformulation of results of \cite{FIK} together with notational adaptations to the current context.
\begin{Theorem}[\cite{FIK}]
\label{thmY}
Consider the Riemann--Hilber Problem~{\rm\ref{RHPY}}.
\begin{enumerate}
\item[$(1)$]
If the solution exists, it is unique and $\det Y_n(z) \equiv 1$.
\item[$(2)$] The solution exists if and only if the Hankel determinant $\tau_n$ in \eqref{Dn} is different from zero.
\item[$(3)$] The solution has the form
\be
\label{Ynsol}
Y_n(z) = \begin{bmatrix}
p_n(z) & \displaystyle \oint_{\gamma} \frac {p_n(w) {\rm e}^{\theta(w)} \d w}{(w-z)2{\rm i}\pi}
\vspace{1mm}\\
\wt p_{n-1}(z) & \displaystyle \oint_{\gamma} \frac {\wt p_{n-1}(w) {\rm e}^{\theta(w)} \d w}{(w-z)2{\rm i}\pi}
\end{bmatrix},
\ee
where $p_n(z)$ is the monic orthogonal polynomial of degree $n$ and $\wt p_{n-1}$ a polynomial of degree~${\deg \wt p_{n-1} \leq n-1}$. They are expressible explicitly in terms of the moments as follows:
\begin{gather}
p_n(z) = \frac 1{\tau_n} \det \begin{bmatrix}
\mu_0 &\mu_1 &\mu_2 &\cdots & \mu_n\\
\mu_1 &\mu_2 & \cdots &&\mu_{n+1}\\
\mu_2 & \cdots &&&\mu_{n+2}\\
\vdots\\
\mu_{n-1} & \mu_{n} & \cdots & &\mu_{2n-1}\\
1 & z & z^2 & \cdots & z^n
\end{bmatrix},
\label{pn}
\\
\wt p_{n-1}(z) = \frac {-2{\rm i}\pi} {\tau_n} \det \begin{bmatrix}
\mu_0 &\mu_1 &\cdots & \mu_{n-1}\\
\mu_1 &\mu_2 & \cdots &\mu_{n}\\
\mu_2 & \cdots &&\mu_{n+1}\\
\vdots\\
\mu_{n-2} & \cdots & &\mu_{2n-3}\\
1 & z & \cdots & z^{n-1}
\end{bmatrix}
\label{wtpn-1},
\end{gather}
\end{enumerate}
with $\tau_n$ the Hankel determinant of moments~\eqref{Dn}.
\end{Theorem}
The matrix $Y_n(z;t)$ (we emphasize its dependence on $t$ as well) satisfies a first-order ODE. Indeed, let us define
\be
\Psi(z;t):= Y_n(z;t) {\rm e}^{\frac 1 2 \theta(z) \s_3} =
 Y_n(z;t) z^{\frac{{ {K}}}2\s_3}
 \le(1- \frac 1 z\ri)^{\frac{{ {\rho}}}2\s_3} {\rm e}^{\frac t{2z} \s_3}.
\label{defPsi}
\ee
We note that near $z=0,1,\infty$, the matrix $\Psi$ has the following expansion, which follows from the properties of $Y_n$ and the definition of $\theta$ \eqref{deftheta}:
\be
\Psi(z;t) =
\begin{cases}
{\mathcal O}^\times(1) z^{\frac {{ {K}}-{ {\rho}}}2\s_3} {\rm e}^{\frac t {2z}\s_3},& z\to 0,\\
{\mathcal O}^\times(1) (z-1)^{\frac {{ {\rho}}}2\s_3}, & z\to 1,\\
\bigl(\1 + \mathcal O\bigl(z^{-1}\bigr) \bigr) z^{ (n+ \frac { {K}} 2)\s_3}, & z\to \infty,
\end{cases}
\label{Psiasy}
\ee
where $\mathcal O^\times(1)$ means a locally analytic and invertible matrix.
\begin{Proposition}
The matrix $\Psi(z;t)$ in \eqref{defPsi} satisfies the pair of first-order PDEs
\be
\label{Laxour}
\pa_z \Psi(z;t) = A(z;t) \Psi(z;t),\qquad
\pa_t \Psi(z;t) = B(z;t) \Psi(z;t),
\ee
where the matrices $A$, $B$ have the form
\begin{gather*}
A(z;t):= -\frac t {z^2} G_0 \s_3 G_0^{-1} + G_0\le(\frac{{ {K}}-{ {\rho}}}z\s_3 - \frac t z \big[G_0^{-1}G_1, \s_3\big]\ri) G_0^{-1} \\
\phantom{A(z;t):=}{}+\frac {{ {K}}}{z-1} H_0 \s_3 H_0^{-1} + \frac{n+{ {K}}}z \s_3,
\\
B(z;t):= \frac 1 {z} G_0 \s_3 G_0^{-1}.
\end{gather*}
Here the matrices $G_0$, $G_1$ are constant with respect to $z$.
\end{Proposition}
\begin{proof}
It follows from the Riemann--Hilber Problem~\ref{RHPY} that $\Psi$ satisfies a boundary value problem
\[
\Psi(z_+;t) = \Psi(z_-;t) \begin{bmatrix}1 & 1 \\ 0& 1\end{bmatrix}.
\]
Since the jump matrices are constant with respect to $z$, both $\Psi$ and $\pa_z \Psi$ satisfy the same jump relation and hence the matrix
$
A(z;t):= \pa_z\Psi(z;t) \Psi^{-1}(z;t)
$
extends analytically to an analytic function with at most isolated singularities at $z=0,1$.
Thus, we have
\[
A(z;t) = Y_n' Y_n^{-1} + Y_n\le(\frac {-t}{z^2} + \frac{{ {\rho}}}{z(z-1)} + \frac {{ {K}}	}z \ri)\s_3 Y_n^{-1}.
\]
This shows that the only singularities of $A(z)$ are a double pole at $z=0$ and a simple pole at~${z=1}$. Near $z=\infty$, using the asymptotic behaviour of $Y_n$ as in \eqref{asympY}, we conclude that
\begin{gather*}
A(z;t) =-\frac t {z^2} G_0 \s_3 G_0^{-1} + G_0\le(\frac{{ {K}}-{ {\rho}}}z\s_3 - \frac t z \Big[G_0^{-1}G_1, \s_3\Big]\ri) G_0^{-1} \\
\phantom{A(z;t) =}{}+\frac {{ {K}}}{z-1} H_0 \s_3 H_0^{-1} + \frac{n+{ {K}}}z \s_3,\\
G_0 := Y_n(0), \qquad G_1:= Y_n'(0), \qquad H_0 := Y_n(1).
\end{gather*}
A similar argument produces the expression for $B(z;t)$.
\end{proof}

In the pair of PDEs \eqref{Laxour}, the $z$-equation is an ODE (considering $t$ as parameter) with an irregular singularity of Poincar\'e\ rank $2$ at $z=0$ and two Fuchsian singularities at $z=1,\infty$, respectively.

{\bf Comparison with the Japanese Lax pair.}
In the work of the Japanese school~\cite{JMU2}, the $z$-component of the Lax pair for Painlev\'e V has also two Fuchsian and one second rank singularities, but with the positions reversed.
More specifically, the Lax pair proposed in~\cite{JMU2} is as follows\footnote{We transcribe the results of \cite{JMU2} but we adapt their notation to our conventions. Note that the paper contains a couple of minor typographical errors: in (C.39) there should be an $x$ in front of the first term of $B$ and the sign of the $(1,2)$ entry of the next term should be the opposite.} (see formulas (C.38)--(C.45) in loc.\ cit.):
 \begin{gather}%
\pa_\zeta \Phi(\zeta;t) = A_{_{\rm JMU}}(\zeta;t) \Phi(\zeta;t),\qquad
\pa_t \Phi(\zeta;t) = B_{_{\rm JMU}}(\zeta;t) \Phi(\zeta;t),
\nn
\\
 A_{_{\rm JMU}}(\zeta;t) = \frac t 2 \s_3 + \frac 1 \zeta \begin{bmatrix}
 \displaystyle Z + \frac {\theta_0}2 & -U(Z+\theta_0)\vspace{1mm}\\
 \displaystyle \frac Z U & \displaystyle -Z - \frac {\theta_0}2
 \end{bmatrix}\nn\\
 \phantom{ A_{_{\rm JMU}}(\zeta;t) = }{}
 +\frac 1{\zeta-1}\begin{bmatrix}
 \displaystyle-Z-\frac {\theta_0 + \theta_\infty}2 &
\displaystyle UY\left(Z+ \frac {\theta_0-\theta_1+ \theta_\infty}2\right) \\
\displaystyle-\frac{Z+\frac {\theta_0 + \theta_1 + \theta_\infty}2}{UY} &
\displaystyle Z+\frac {\theta_0 + \theta_\infty}2
 \end{bmatrix},
\nn
\\
 B_{_{\rm JMU}}(\zeta;t) = \frac \zeta 2 \s_3 \label{LaxJMU}\\
 \quad{}+ \begin{bmatrix}
 0
 &
 \displaystyle -\frac {U}t\left( Z +\theta_0 - Y\left(Z + \frac {\theta_0 - \theta_1 + \theta_\infty}2\right)\right)\\
 \displaystyle \frac 1{t U Y} \left(( Y -1) Z + \frac {\theta_0 + \theta_1 +\theta_\infty}2\right)
 & 0
 \end{bmatrix},\nn
 \end{gather}
 where $Z = Z(t)$, $ Y = Y(t)$, $ U=U(t)$ satisfy a nonlinear first-order system of ODEs in $t$ ((C.40) in loc.\ cit.), which implies the fifth Painlev\'e equation for $Y$
 \begin{gather}
\frac {\d^2 Y}{\d t^2 } = \le(\frac 1{2Y} + \frac 1{Y-1} \ri) \le(\frac {\d Y}{\d t}\ri)^2 - \frac 1 t \frac {\d Y}{\d t} + \frac {(Y-1)^2\bigl(\alpha Y+\frac Y \beta\bigr)}{t^2} + \frac {\gamma Y}t + \frac {\delta Y(Y+1)}{Y-1},\nn
\\
\alpha = \frac 1 2 \le(\frac {\theta_0 - \theta_1 + \theta_\infty }2\ri)^2,\qquad
\beta = -\frac 1 2 \le(\frac {\theta_0 - \theta_1 - \theta_\infty }2\ri)^2,\nn\\
\gamma = 1 -\theta_0 - \theta_1,\qquad \delta = -\frac 1 2. \label{PV}
 \end{gather}
 The solution $\Phi(\zeta;t)$ has the following formal expansions near $\zeta=0, 1, \infty$:
 \begin{gather}
 \Phi(\zeta;t) = \mathcal O(1) \zeta^\frac {\theta_0 \s_3}2, \qquad \zeta \to 0,\qquad
 \Phi(\zeta;t) = \mathcal O(1) (\zeta-1)^\frac {\theta_1 \s_3}2, \qquad \zeta \to 1,\nn\\
 \Phi(\zeta;t) =\le(\1 + \frac {\Phi_1}\zeta + \mathcal O\bigl(\zeta^{-2}\bigr)\ri) \zeta ^{-\frac {\theta_\infty}2 \s_3}{\rm e}^{\frac t 2 \zeta\s_3}, \qquad \zeta \to \infty.
 \label{Phiasy}
\end{gather}
 The Hamiltonian function for the Painlev\'e equation is given by
 \be
 \label{HamJMUPV}
 H_V = -\frac 1 2 \tr( \Phi_1\s_3),
 \ee
 and the equation admits the so-called {\it sigma-form}. Indeed, introducing the new function
 \[
 \s(t) = t H_V - \frac t 2 (\theta_0 + \theta_\infty) + \frac { (\theta_0 + \theta_\infty)^2 - \theta_1^2}4,
 \]
 it can be verified that it satisfies \cite[formula (C.45)]{JMU2}
 \begin{align*}
 \le(t \frac {\d^2 \s}{\d t^2}\ri)^2 ={}& \le(
 \s - t\frac {\d \s}{\d t} + 2\le(\frac {\d \s}{\d t}\ri)^2 - (\theta_\infty + 2\theta_0) \frac {\d \s}{\d t}
 \ri)^2 \\
 &- 4 \le(\frac {\d \s}{\d t}\ri)
 \le(\frac {\d \s}{\d t} - \frac { \theta_0 -\theta_1 + \theta_\infty}2 \ri)
 \le( \frac {\d \s}{\d t} - \theta_0 \ri)
 \le(\frac {\d \s}{\d t} - \frac { \theta_0 + \theta_1 + \theta_\infty}2 \ri).
 \end{align*}
In order to identify our Lax pair \eqref{Laxour} with the Japanese one \eqref{LaxJMU}, it suffices to map $\zeta = \frac 1 z$ and suitably normalize our matrix $\Psi(z;t)$.
\begin{Lemma}
The map $\zeta = \frac 1 z$ and
\[
\Phi(\zeta;t) = Y_n(0;t)^{-1}\Psi\le(\frac 1 \zeta;t\ri){\rm e}^{-{\rm i}\frac{\pi}2 { {\rho}} \s_3}
\]
transforms the Lax pair \eqref{Laxour} into \eqref{LaxJMU} with parameters
$
\theta_0 = - 2n - { {K}}$, $
\theta_1 = { {\rho}}$, $
\theta_\infty= { {K}}-{ {\rho}} $.
This corresponds to the parameters $\alpha$, $\beta$, $\gamma$ in \eqref{PV} as follows:
\[
\alpha = \frac {(n+ { {\rho}})^2}2,\qquad
\beta = -\frac {(n + { {K}})^2} 2,\qquad
\gamma = 1+2n + { {K}} - { {\rho}}.
\]
\end{Lemma}
\begin{proof}
The map $\zeta=\frac 1 z$ maps $z=0$ to $\zeta=\infty$, $z=\infty$ to $\zeta=0$ and $z=1 $ to $\zeta=1$. Thus the exponents of (formal) monodromy $\theta_{\{0,1,\infty\}}$ are the read off by matching the exponents in \eqref{Psiasy} and \eqref{Phiasy}.
\end{proof}

The above lemma allows us to identify the Hamiltonian with the logarithmic derivative of the Hankel determinant and, thus, the $\tau$ function with the Hankel determinant itself. This is, a~priori, a result of \cite{Bertola:Semiiso}, but we can here derive it directly from the formulas already reported.
\begin{Proposition}\label{tauPV}
The Hankel determinant $\tau_n(t)$ in \eqref{Dn} is a polynomial tau function of the fifth Painlev\'e equation. In particular, the Hamiltonian is
\[
H_V = -\frac 1 2 \tr \le(Y_n^{-1}(0) Y'_n(0) \s_3\ri) + \frac { {\rho}} 2 =-
\le(Y_n^{-1}(0) Y'_n(0) \ri)_{11} + \frac { {\rho}} 2 = \frac {\d}{\d t} \ln \tau_n(t) + \frac { {\rho}} 2.
\]
\end{Proposition}
\begin{proof}
We need to identify $\Phi_1$ in the expansion at $\zeta=\infty$ in~\eqref{Phiasy} with the suitable expansion of $\Psi$ at $z=0$. Recalling the definition of $\Psi $ \eqref{defPsi}, we see that
\begin{gather*}
Y_n(0)^{-1}\Psi\le(\frac 1 \zeta;t\ri) {\rm e}^{-\frac {{\rm i}\pi}2 { {\rho}} \s_3}\\
\qquad
=
 Y_n(0)^{-1}\le(Y_n(0) + \frac 1 \zeta Y_n'(0) + \mathcal O\bigl(\zeta^{-2}\bigr)\ri)
(\zeta-1)^{\frac { {\rho}} 2 \s_3} \zeta ^{-\frac { {K}} 2 \s_3} {\rm e}^{\frac t 2 \zeta \s_3}\\
\qquad=\le(\1 + \frac {Y_n(0)^{-1} Y_n'(0)}\zeta + \mathcal O\bigl(\zeta ^{-2}\bigr)\ri)
 \le( \1 - \frac {{ {\rho}}}{2\zeta}\s_3 + \mathcal O\bigl(\zeta ^{-2}\bigr)\ri)
 \zeta^{\frac {{ {\rho}}-{ {K}}}2\s_3}
 {\rm e}^{\frac t 2 \zeta \s_3}\\
\qquad=\le(\1 + \frac { Y_n(0)^{-1} Y_n'(0) - \frac { {\rho}} 2\s_3}\zeta + \mathcal O\bigl(\zeta ^{-2}\bigr)\ri) \zeta^{\frac {{ {\rho}}-{ {K}}}2\s_3} {\rm e}^{\frac t 2 \zeta \s_3}.
\end{gather*}
Thus we conclude that
$
\Phi_1 = Y_n(0)^{-1} Y_n'(0) - \frac { {\rho}} 2\s_3$.
Then, according to \eqref{HamJMUPV}, we must compute
\[
H_{_V}=-\frac 1 2 \tr \le(\Phi_1\s_3\ri) = -\frac 1 2 \tr \le( Y_n(0)^{-1} Y_n'(0) \s_3\ri) + \frac { {\rho}} 2
=-\le( Y_n(0)^{-1} Y_n'(0) \ri) _{11} + \frac { {\rho}} 2,
\]
where we have used that $Y_n(0)Y_n'(0)$ is a traceless matrix (due to the fact that $\det Y_n \equiv 1$).

{\bf Computation of $\boldsymbol{\bigl(Y_n(0)^{-1} Y_n'(0)\bigr)_{11}}$.}
Using that $\det Y_n\equiv 1$ and the formula \eqref{Ynsol}, we deduce
\be
\label{YYY}
\bigl(Y_n(0)^{-1} Y_n'(0)\bigr)_{11}= p'_n(0) \oint_\gamma \frac {\wt p_{n-1}(w) {\rm e}^{\theta(w)} \d w }{w2{\rm i}\pi} - {\wt p} '_{\!n-1}(0) \oint_\gamma \frac {p_n(w) {\rm e}^{\theta(w)} \d w }{w2{\rm i}\pi}.
\ee
The derivatives at $z=0$ of $p_n$, $\wt p_{n-1}$ are obtained from their determinantal representations \eqref{pn}, \eqref{wtpn-1}
\begin{gather*}
 p'_n(0)={(-1)^{n+1}}\dfrac{\begin{vmatrix}
 \mu_0 & \mu_2 & \dots & \mu_{n} \\
 \mu_1 & \mu_3 & \dots & \mu_{n+1} \\
 \vdots & \vdots & \ddots & \vdots \\
 \mu_{n-1} & \mu_{n+1}& \dots & \mu_{2n-1}
\end{vmatrix}}{\tau_n},
\label{p'n}
\\
\wt p'_{n-1}(0)=\dfrac{(-1)^{n+1}2\pi {\rm i}}{\tau_n}\begin{vmatrix}
 \mu_0 & \mu_2 & \dots & \mu_{n-1} \\
 \mu_1 & \mu_3 & \dots & \mu_{n} \\
 \vdots & \vdots & \ddots & \vdots \\
 \mu_{n-2} & \mu_{n} & \dots & \mu_{2n-3}
\end{vmatrix}.
\label{wtp'n-1}
\end{gather*}
The contour integrals in \eqref{YYY} can be expressed as well in terms of determinants of the moments~$\mu_j$ provided we extend their definition to all $j\in \Z$
\begin{gather}
\oint_\gamma \frac {\wt p_{n-1}(w) {\rm e}^{\theta(w)} \d w }{2{\rm i}\pi w }
=- \dfrac{\begin{vmatrix}
 \mu_0 & \mu_1 & \mu_2& \dots & \mu_{n-1} \\
 \mu_1 & \mu_2 & \mu_3 & \dots & \mu_{n}\\
 \vdots & \vdots & \vdots & \ddots & \vdots \\
 \mu_{n-2} & \mu_{n-1} & \mu_{n} & \dots & \mu_{2n-3}\\
 \mu_{-1} & \mu_0 & \mu_1 & \dots & \mu_{n-2}
\end{vmatrix}}{\tau_n}\nonumber\\
\phantom{\oint_\gamma \frac {\wt p_{n-1}(w) {\rm e}^{\theta(w)} \d w }{2{\rm i}\pi w }
}{}
=(-1)^{n} \dfrac{\begin{vmatrix}
\mu_{-1} & \mu_0 & \mu_1 & \dots & \mu_{n-2}\\
 \mu_0 & \mu_1 & \mu_2& \dots & \mu_{n-1} \\
 \mu_1 & \mu_2 & \mu_3 & \dots & \mu_{n}\\
 \vdots & \vdots & \vdots & \ddots & \vdots \\
 \mu_{n-2} & \mu_{n-1} & \mu_{n} & \dots & \mu_{2n-3}
\end{vmatrix}}{\tau_n},\label{intwtp}
\\
\oint_\gamma \frac {p_n(w) {\rm e}^{\theta(w)} \d w }{2{\rm i}\pi w}
= \dfrac{1}{2{\rm i}\pi }\dfrac{\begin{vmatrix}
 \mu_0 & \mu_1 & \mu_2& \dots & \mu_{n} \\
 \mu_1 & \mu_2 & \mu_3 & \dots & \mu_{n+1} \\
 \vdots & \vdots & \vdots & \ddots & \vdots \\
 \mu_{n-1} & \mu_{n} & \mu_{n+1} & \dots & \mu_{2n-1}\\
 \mu_{-1} & \mu_0 & \mu_1 & \dots & \mu_{n-1}
\end{vmatrix}}{\tau_n}\nonumber\\
\phantom{\oint_\gamma \frac {p_n(w) {\rm e}^{\theta(w)} \d w }{2{\rm i}\pi w}
}{}=
 \dfrac{(-1)^n }{2{\rm i}\pi }\dfrac{\begin{vmatrix}
 \mu_{-1} & \mu_0 & \mu_1 & \dots & \mu_{n-1}\\
 \mu_0 & \mu_1 & \mu_2& \dots & \mu_{n} \\
 \mu_1 & \mu_2 & \mu_3 & \dots & \mu_{n+1} \\
 \vdots & \vdots & \vdots & \ddots & \vdots \\
 \mu_{n-1} & \mu_{n} & \mu_{n+1} & \dots & \mu_{2n-1}
\end{vmatrix}}{\tau_n}.\label{intp}
\end{gather}
Inserting \eqref{p'n}, \eqref{wtp'n-1}, \eqref{intwtp}, \eqref{intp} into \eqref{YYY}, simplifying and rearranging the rows of the resulting determinants, we obtain
\begin{align*}
\bigl(Y_n(0)^{-1} Y_n'(0)\bigr)_{11}={}&
-
\dfrac{\begin{vmatrix}
 \mu_0 & \mu_2 & \dots & \mu_{n} \\
 \mu_1 & \mu_3 & \dots & \mu_{n+1} \\
 \vdots & \vdots & \ddots & \vdots \\
 \mu_{n-1} & \mu_{n+1}& \dots & \mu_{2n-1}
\end{vmatrix}}{\tau_n}
\dfrac{\begin{vmatrix}
 \mu_{-1} & \mu_0 & \mu_1 & \dots & \mu_{n-2}\\
 \mu_0 & \mu_1 & \mu_2& \dots & \mu_{n-1} \\
 \mu_1 & \mu_2 & \mu_3 & \dots & \mu_{n}\\
 \vdots & \vdots & \vdots & \ddots & \vdots \\
 \mu_{n-2} & \mu_{n-1} & \mu_{n} & \dots & \mu_{2n-3}
\end{vmatrix}}{\tau_n}
\\
&+
\dfrac{\begin{vmatrix}
 \mu_0 & \mu_2 & \dots & \mu_{n-1} \\
 \mu_1 & \mu_3 & \dots & \mu_{n} \\
 \vdots & \vdots & \ddots & \vdots \\
 \mu_{n-2} & \mu_{n} & \dots & \mu_{2n-3}
\end{vmatrix}}{\tau_n}
\dfrac{\begin{vmatrix}
 \mu_{-1} & \mu_0 & \mu_1 & \dots & \mu_{n-1}\\
 \mu_0 & \mu_1 & \mu_2& \dots & \mu_{n} \\
 \mu_1 & \mu_2 & \mu_3 & \dots & \mu_{n+1} \\
 \vdots & \vdots & \vdots & \ddots & \vdots \\
 \mu_{n-1} & \mu_{n} & \mu_{n+1} & \dots & \mu_{2n-1}
 \end{vmatrix}}{\tau_n}.
\end{align*}
We now use the Desnanot--Jacobi identity, which can be stated as follows. Given a square matrix~$M$ denote by $M^{[a_1,\dots, a_r][b_1,\dots, b_r]}$ the matrix obtained by deleting the rows $a_1,\dots, a_r$ and columns~${b_1,\dots, b_r}$. Then
\be
\label{desnanot}
\det M^{[a,b][c,d]} \det M = \det M^{[a][c]} \det M^{[b][d]} - \det M^{[a][d]} \det M^{[b][c]}.
\ee
We apply the identity to the matrix
\[
M := \begin{vmatrix}
 \mu_{-1} & \mu_0 & \mu_1 & \dots & \mu_{n-1}\\
 \mu_0 & \mu_1 & \mu_2 & \dots & \mu_{n} \\
 \mu_1 & \mu_2 & \mu_3 & \dots & \mu_{n+1} \\
 \vdots & \vdots & \vdots & \ddots & \vdots \\
 \mu_{n-1}& \mu_{n}& \mu_{n+1} & \dots & \mu_{2n-1}
\end{vmatrix}\in \mathfrak{gl}(n+1).
\]
Then, using \eqref{desnanot}, we have
\begin{align*}
\bigl(Y_n(0)^{-1} Y_n'(0)\bigr)_{11}&= \frac {-\det M^{[1][2]} \det M^{[n+1][n+1]} + \det M^{[1,n+1][2,n+1]} \det M}{\tau_n^2}
\nn\\
&\mathop{=}^{\eqref{desnanot}}{}
-\frac {\det M^{[1][n+1]} \det M^{[n+1][2]}}{\tau_n^2} =- \frac { \det M^{[n+1][2]}}{\tau_n},
\end{align*}
where the last equality holds on account of the fact that $\tau_n = \det M^{[1][n+1]}$.
The final part of the verification is based on the observation that
\[
\pa_t \mu_j(t) = \oint_{\gamma} z^{j-1} \le(1-\frac 1 z\ri)^{ {\rho}} z^{{K}} {\rm e}^{\frac t z}\d z= \mu_{j-1}(t), \qquad j\in \Z.
\]
This allows us to show that
$
\pa_t \det \tau_n =\det M^{[n+1][2]}
$
and hence
$
\bigl(Y_n(0)^{-1} Y_n'(0)\bigr)_{11}= -\pa_t\ln \tau_n(t)$.
This concludes the proof.
\end{proof}

 \section{Asymptotic analysis}
 The goal of this second part of the paper is to study the behaviour of the poles of the rational solution or, which is the same, the zeros of the Hankel determinant $\tau_n(t)$ as $n\to \infty$.
 More precisely, and in line with similar investigations done for rational solution of PII \cite{BertolaBothner,BuckMill14, BuckMill15},
 we are going to re-scale the zeros concurrently with $n$ and study the behaviour of the zeros in the~$s =\frac t n $-plane. Namely, we study the sequence of functions (polynomials)
 \be
 \la{taun}
 \mathcal T_n(s):= \tau_n(ns; {K}, { {\rho}}),
 \ee
 where
 \[
 \tau_n(t;{K}, { {\rho}}) := \det\le[
 \oint_\gamma z^{a+b+{K}} \le(1-\frac 1 z\ri)^{ {\rho}} {\rm e}^{\frac t z}\frac{\d z}{2{\rm i}\pi}
 \ri]_{a,b=0}^{n-1}.
 \]
 We emphasize that we are not scaling the parameters ${ {\rho}}\in \C\setminus \Z$ nor $K\in \Z$ as $n\to\infty$. A~different asymptotic analysis would be needed if we were to set ${ {\rho}} \propto n$ and/or $K\propto n$. We will briefly point out the main differences when appropriate.

Theorem~\ref{thmY} can be re-formulated for convenience in the following form.
\begin{Theorem} \label{thmYres}
 The function $\mathcal T_n(s):= \tau_n(ns; K, { {\rho}})$ \eqref{taun} is zero if and only if the Riemann--Hilbert Problem~{\rm\ref{RHPY}} has no solution $($with $t= ns)$.
 \end{Theorem}
 The asymptotic analysis of Riemann--Hilbert Problem~\ref{RHPY} with $t=ns$ falls within the purview of the Deift--Zhou asymptotic method that was brought to fruition in \cite{DKMVZ} for orthogonal polynomials, see also~\cite{BleherIts, Deiftbook}.

It consists of several, by now more or less standardized, steps that we can summarize as follows before delving into the details:
\begin{enumerate}\itemsep=0pt
\item[(1)] Construct a ``$g$-function''. This is a function $g(z;s)$ whose domain of analyticity $\scr D_s$ depends (continuously) on the parameter $s$. It can be expressed as an Abelian integral on a hyperelliptic Riemann surface $\mathcal R$ whose genus depends on $s$. The $g$-function is uniquely characterized by a free-boundary value problem and certain inequalities which will be described in Section~\ref{secgfunction}.
\item[(2)] Use the constructed $g$-function to {\it normalize} Riemann--Hilbert Problem~\ref{RHPY} and express a~new matrix-valued function $W(z)$ which satisfies a new RHP.
\item[(3)] The RHP for $W$ then undergoes a sequence of {\it transformations} (or reformulations) into equivalent RHPs.
\item[(4)] The final reformulation is then amenable to a~nonlinear steepest descent analysis using a~standard {\it small norm theorem} for Riemann--Hilbert problems.
\end{enumerate}
The key concept is that the solvability of the initial Riemann--Hilbert Problem~\ref{RHPY} is equivalent to the solvability of its final reformulation hinted at in the list above.
Thus, if we can guarantee that the final reformulation is solvable for $s$ in a suitable domain, we can conclude that there are {\it no zeros} of~$\mathcal T_n(s)$ in that domain (at least for $n$ sufficiently large).

Let us give a visual overview of the result. In Figure~\ref{FigEyes}, we can see several instances of plots of the zeros of $\mathcal T_n(s)$ for various values of ${ {\rho}}$. The common feature, which will be proved, is that all the zeros lie, asymptotically for large $n$, within the ``Eye of the Tiger'' region (${\rm EoT}$), marked by the black arcs. The equation of these black arcs is given implicitly in~\eqref{EoTcurve}.

Also indicated is a grid of lines within the set ${\rm EoT}$. Their intersection is the {\it approximate} location of the zeros as it follows from the asymptotic analysis, and they represent the vanishing of a theta function for an elliptic curve, see Section~\ref{insideEoT}.

Consequently we are going to split the analysis in two cases:
\begin{itemize}\itemsep=0pt
\item the outside of ${\rm EoT}$, Sections \ref{gzero} and \ref{outsideEoT},
\item the inside of ${\rm EoT}$, Sections \ref{gfuncg1} and \ref{insideEoT}.
\end{itemize}

\section[Construction of the $g$-function]{Construction of the $\boldsymbol{g}$-function} \label{secgfunction}

 The method of the steepest descent analysis requires the construction of a scalar function with certain properties that we list below. This function is universally known in the literature as the $g$-function.
Here we recall that the symbol of the moment functional is
 \[
 \theta(z;s) = \frac {ns}z + { {\rho}} \ln \le(1-\frac 1 z \ri) + { {K}} \ln z.
 \]
 Since ${ {K}}$, ${ {\rho}}$ will not be scaling in our setup, we introduce $\theta_0(z;s) = \frac s z$ (the ``scaling'' part) so that
 \begin{gather}
 \label{42}
\theta(z;s) = n\theta_0(z;s)+ { {\rho}} \ln \le(1-\frac 1 z \ri) + { {K}} \ln z.
 \end{gather}
 If ${ {K}} = n{ {K}}_0$ and/or ${ {\rho}} = n{ {\rho}}_0$, namely, if we were to scale ${ {K}}$, ${ {\rho}}$, we would accordingly re-define~$\theta_0$ to contain all terms proportional to $n$.

\begin{Definition}[the $g$-function and its properties]
\label{gdef}
The $g$-function is a locally bounded analytic function on $\C\setminus \Gamma$ where $\Gamma$ is a union of oriented contours (to be determined) extending to infinity satisfying the properties listed hereafter.
\begin{enumerate}\itemsep=0pt
\item[(1)] The contour $\Gamma$ can be written as $\Gamma = \Gamma_{m} \cup \Gamma_{c} \cup \Gamma_{\infty}$ (with $\Gamma_{m}$ denoting the ``main arc(s)'', and~$\Gamma_{c}$ the ``complementary arc(s)'') where each of the components have pairwise disjoint relative interiors and both $\Gamma_{m}$, $\Gamma_{c}$ consist of a finite union of compact arcs: \smash{$\Gamma_{\{m,c\}}= \bigsqcup \Gamma_{\{m,c\}}^{(j)}$}. Finally, $\Gamma_\infty$ is a simple contour extending to infinity from a finite point, traversing eventually the negative real axis and oriented from infinity.
\item[(2)] The contour $\gamma=\{|z|=R, \, R>1\}$ can be homotopically retracted to $\Gamma_m\cup \Gamma_c$ in $\C \setminus [0,1]$, where $[0,1]$ here denotes a smooth simple arc connecting $z=0,1$ (not necessarily the straight segment).
\item[(3)] For each $z\in \Gamma_m\cup \Gamma_c$, we have
\begin{gather*}
g(z_+) + g(z_-) =- \theta_0(z) -\ell + {\rm i}{\varpi}_j, \qquad {\varpi}_j\in\R, \qquad z\in \Gamma_m^{(j)},\\
g(z_+)-g(z_-) = {\rm i} {\wh \varpi}_j, \qquad {\wh \varpi}_j\in\R, \qquad z\in \Gamma_c^{(j)},
\end{gather*}
for some constants $\ell$, $\varpi_j$, $\wh \varpi_j$ (different on each of the connected components\footnote{We will use different notation for these constants in the specific cases we discuss below.} of $\Gamma_m$, $ \Gamma_c$),
while
$
g(z_+)-g(z_-) = 2{\rm i}\pi$, $ z\in \Gamma_\infty$.
\item[(4)] As $z\to\infty$ in $\C\setminus \Gamma $, we have
$
g(z) = \ln(z) + \mathcal O\bigl(z^{-1}\bigr)$.
\item[(5)] The real part of the $g$-function is continuous on $\C$ (including $\Gamma$) and harmonic on $\C \setminus \Gamma_m$ and
satisfies the following inequalities:
\begin{enumerate}\itemsep=0pt
\item[(i)] For all $z\in \Gamma_c$, we have
\[
\Re\le(g(z_+)+g(z_-)+ \theta_0(z) + \ell \ri) = \Re\le(2g(z)+ \theta_0(z) + \ell \ri) \leq 0,
\]
 with the equality holding only at the endpoints of each component of $\Gamma_c$ and possibly at isolated points within the relative interior of $\Gamma_c$.
\item[(ii)] For $z\in \Gamma_m$, we have
\[
\Re \le(g(z_+)+g(z_-)+ \theta_0(z) + \ell \ri)= \Re \le(2g(z)+ \theta_0(z) + \ell \ri) \equiv 0.
\]

\item [(iii)]
The inequality
$
\Re\le(2g(z)+ \theta_0(z) + \ell \ri) \geq 0
$
holds
in an open neighbourhood $U$ of $\dot \Gamma_m$ (the $\dot{}$ indicating the interior set in the relative topology of the collection of arcs $\Gamma_m$, i.e., the arcs minus the end-points) with the equality holding only on $\Gamma_m$ itself.
\end{enumerate}
\end{enumerate}
\end{Definition}

 It is convenient to reformulate Definition~\ref{gdef} in terms of the so-called {\it effective potential}
 \be
\label{defphi}
\varphi(z;s) = 2g(z;s) + \theta_0(z;s) + \ell = 2g(z;s) + \frac s z +\ell.
\ee
\begin{Corollary}
\label{corphicond}
There exist constants $\varpi_j$, $\varsigma_j$ such that the effective potential $\varphi$ satisfies:
\begin{enumerate}\itemsep=0pt
\item[$(1)$]
$\Re(\varphi(z))$ is continuous on $\C\setminus \{0\}$, harmonic on $\C\setminus \Gamma_m\cup \{0\}$.
\item[$(2)$] $\Re(\varphi- \theta_0)$ is harmonic near $z=0$ and $\Re(\varphi -2\ln(z) - \theta_0(z))$ is harmonic at infinity.
\item[$(3)$] The following equalities and inequalities hold:
\begin{gather*}
\varphi(z_+) + \varphi(z_-) = 2{\rm i}{\varpi}_j,\qquad z\in \Gamma_{m}^{(j)},\qquad
\varphi(z_+)-\varphi(z_-) =2 {\rm i}{\varsigma}_j, \qquad z\in \Gamma_{c}^{(j)},\\
\varphi(z_+)-\varphi(z_-) = 4{\rm i}\pi,\qquad z\in \Gamma_\infty,\qquad
\Re\varphi(z)\equiv 0, \qquad z\in \Gamma_m,\\
\Re\varphi(z)\leq 0,\qquad z\in \Gamma_c,\qquad
\Re \varphi(z)\geq 0,\qquad z\in U,
\end{gather*}
with $U$ as in Definition~{\rm\ref{gdef}}\,$(5)\,(iii)$.
\end{enumerate}
\end{Corollary}
In the next two sections, we are going to construct $\varphi$ directly and verify those properties.

 \subsection[Outside EoT: Genus 0 $g$-function and effective potential]{Outside $\boldsymbol{{\rm EoT}}$: Genus $\boldsymbol{0}$ $\boldsymbol{g}$-function and effective potential}
 \label{gzero}
\begin{figure}[t]\centering
\includegraphics[width=0.3\textwidth]{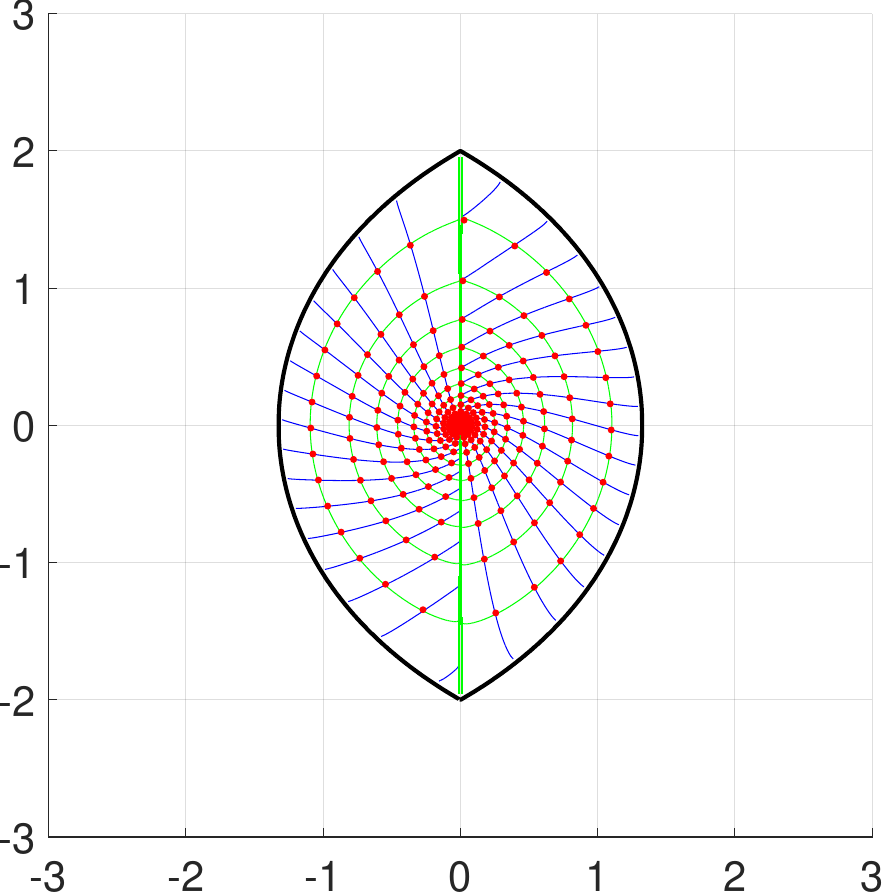}\hfill
\includegraphics[width=0.3\textwidth]{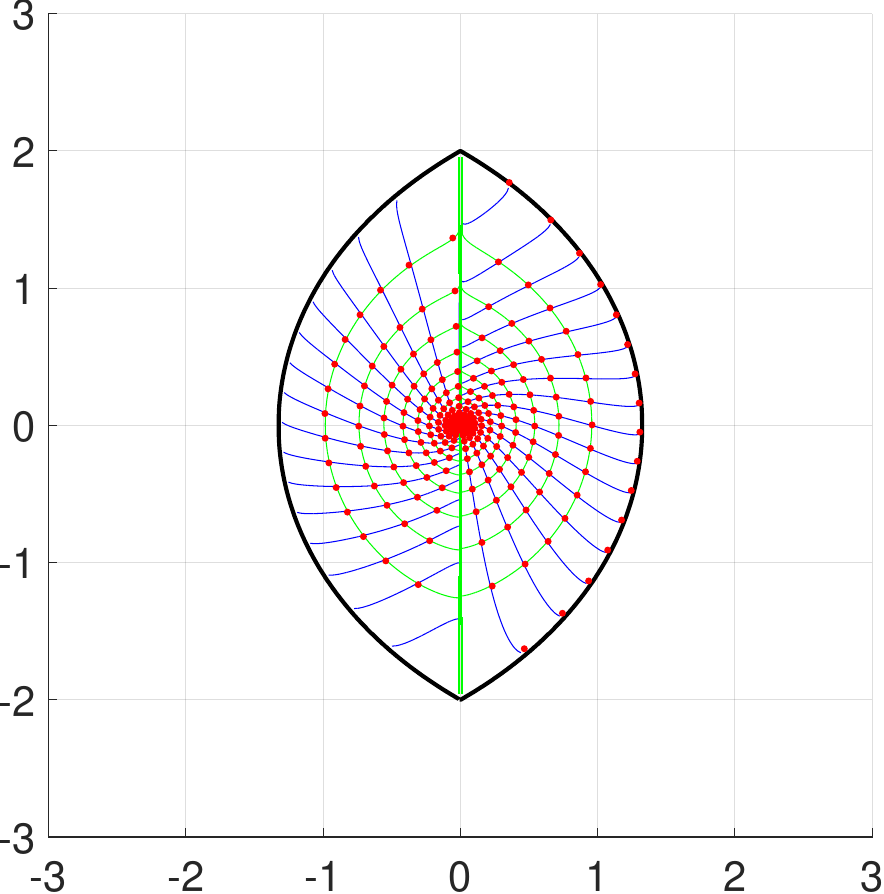}\hfill
\includegraphics[width=0.3\textwidth]{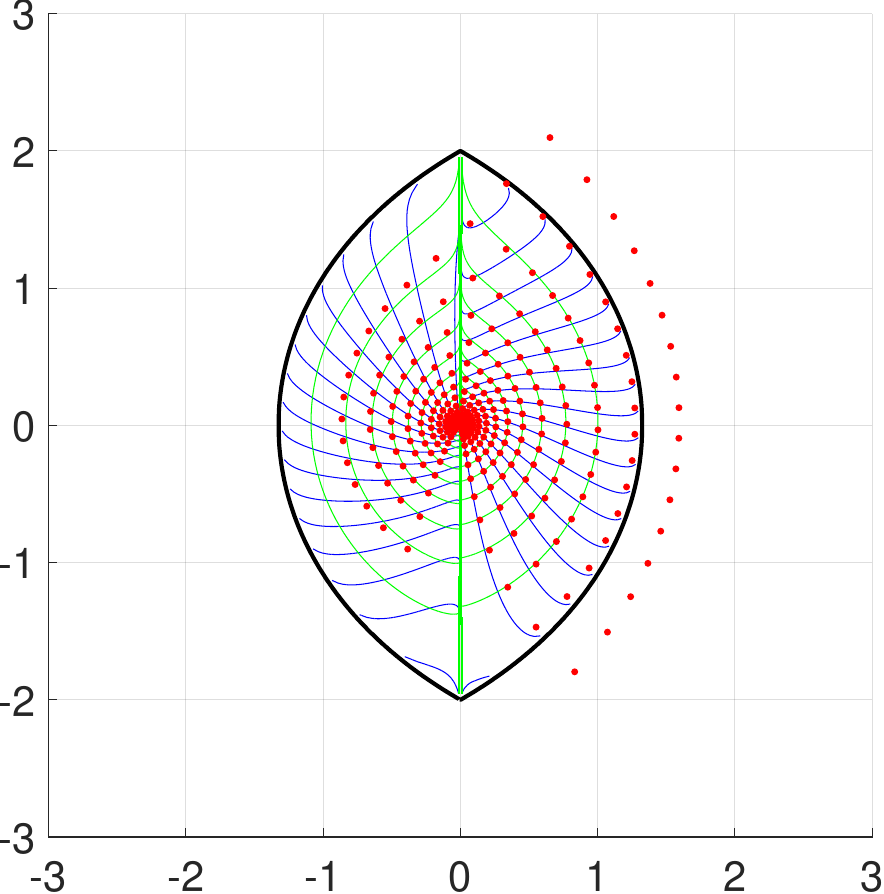}\\
\includegraphics[width=0.3\textwidth]{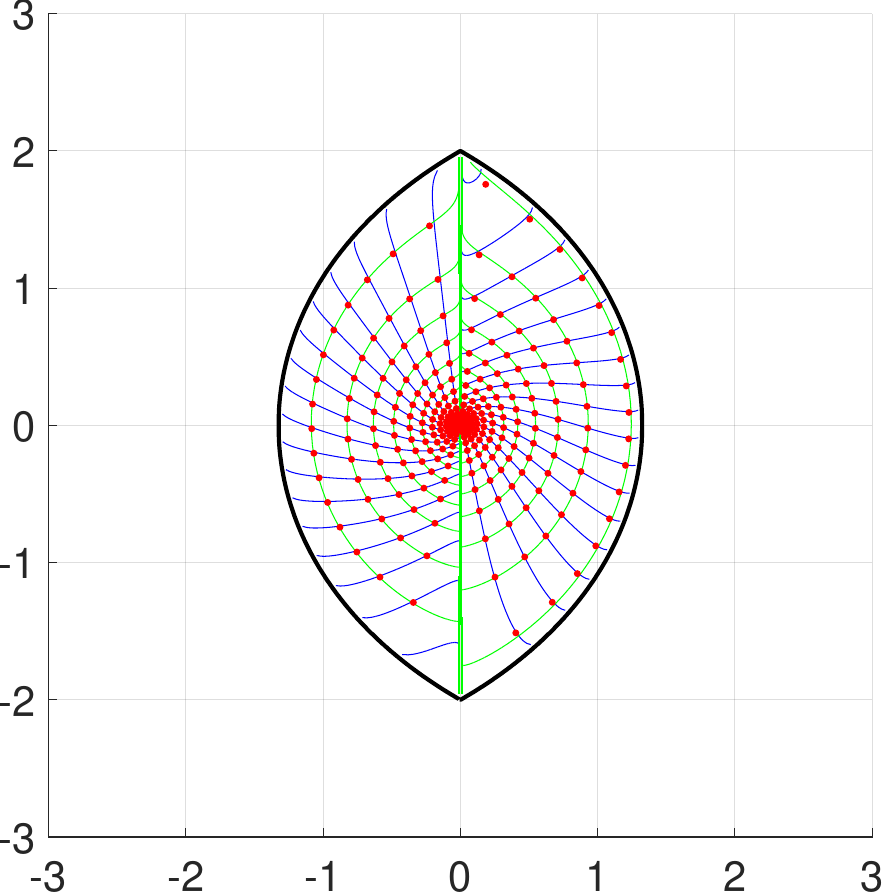}\hfill
\includegraphics[width=0.3\textwidth]{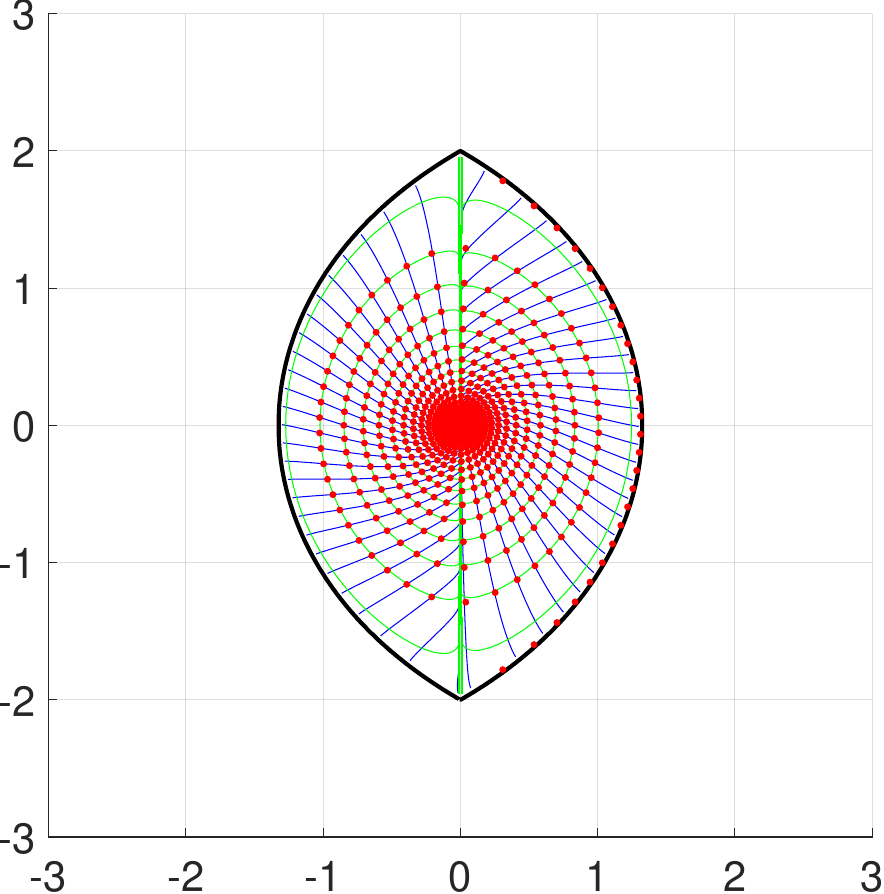}\hfill
\includegraphics[width=0.3\textwidth]{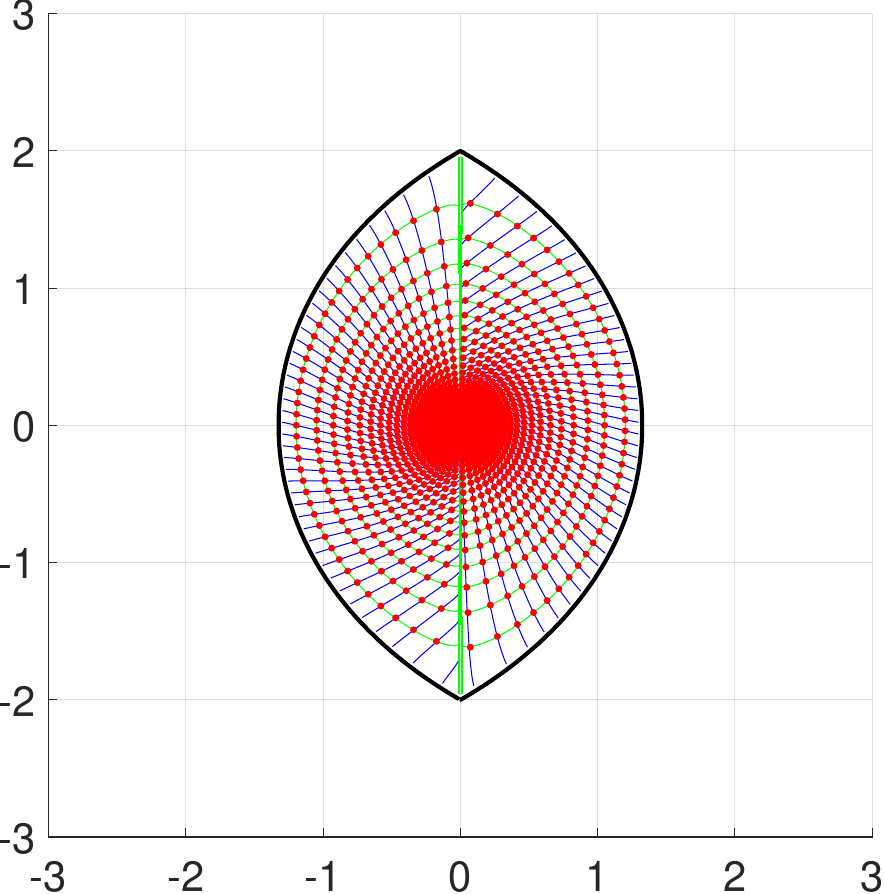}
\caption{From top to bottom, left to right: $(n,{ {\rho}}) = \big(16,\frac{ 3}{100} + \frac{13{\rm i}}{100}\big), \big(16,\frac{ 101}{100} + \frac{13{\rm i}}{100}\big),
(17,3+1{\rm i}\big), \allowbreak \big(17, {\frac 12 +\frac {\rm i}2}\big), \big(26, \frac{101}{100}\big), \big(40, \frac 3{100}\big)$.
Observe that the accuracy of the localization of the zeros by the grid depends on the ratio $|{ {\rho}}|/n$, the smaller the better the approximation. For example, in the case in the top right $(n,{ {\rho}}) = (17,3+ 1{\rm i})$ the ratio is significant (approximately $|{ {\rho}}|/n \simeq 0.19$) and an analysis where ${ {\rho}}$ is considered as scaling would undoubtedly be more appropriate. In all cases, we have ${K}=0$. See the explanation in Remark~\ref{remexp2}. }
\label{Figone}
\end{figure}

 For $|s|$ sufficiently large, we are going to postulate first, and then verify, the form of the effective potential. We will see that the conditions in Corollary~\ref{corphicond} are fulfilled for $s$ ranging in an unbounded region and up to the boundary of ${\rm EoT}$. This is the region bounded by the black arcs in Figure~\ref{Figone}.

We start by postulating the following expression:
 \bea
 \label{phig0}
\varphi(z;s) := 2 \int_{-\frac {{\rm i}s}2}^z \frac{ \sqrt{w^2 + \frac {s^2}4}}{w^2}\d w \ \Rightarrow\ \
\varphi'(z;s) =2\frac{ \sqrt{z^2 + \frac {s^2}4}}{z^2}.
 \eea
 The function $\varphi(z;s)$ can be written explicitly
 \be
 \label{phig0int}
 \varphi(z;s) = 2 \ln\le(\frac {2{\rm i}z}{s} +{\rm i} \sqrt{\frac {4z^2}{s^2}+1}\ri) - \frac s z \sqrt{\frac {4z^2}{s^2} + 1}.
 \ee
 We need to describe the domain of analyticity. We start from the domain of analyticity of $\varphi'$.
A~great simplification is achieved by observing that $\varphi$ is really a function only of $\frac z s$, namely
\[
\varphi(z;s) = \varphi\le(\frac z s;1\ri),
\]
and hence it suffices to describe the domain and properties of $\varphi_0(z):= \varphi\le(\frac z s;1\ri)$ which is given~by
\begin{gather}
\label{phi0}
 \varphi_0(z) = 2 \ln\bigl( {2{\rm i}z} + {\rm i}\sqrt{{4z^2}+1}\bigr) - \frac {\sqrt{4z^2 + 1}}z,\qquad
\varphi_0'(z) = \frac {\sqrt{4z^2+1}}{z^2}.
\end{gather}
The determination of the root is chosen such that $\varphi'_0(z) \simeq \frac 2 z$ at $z=\infty$, with a branch cut connecting the branch-points $\pm \frac {\rm i}2$ to be determined below.

\begin{figure}[t]
\centering
\includegraphics[width=0.46\textwidth, height=0.38\textwidth]{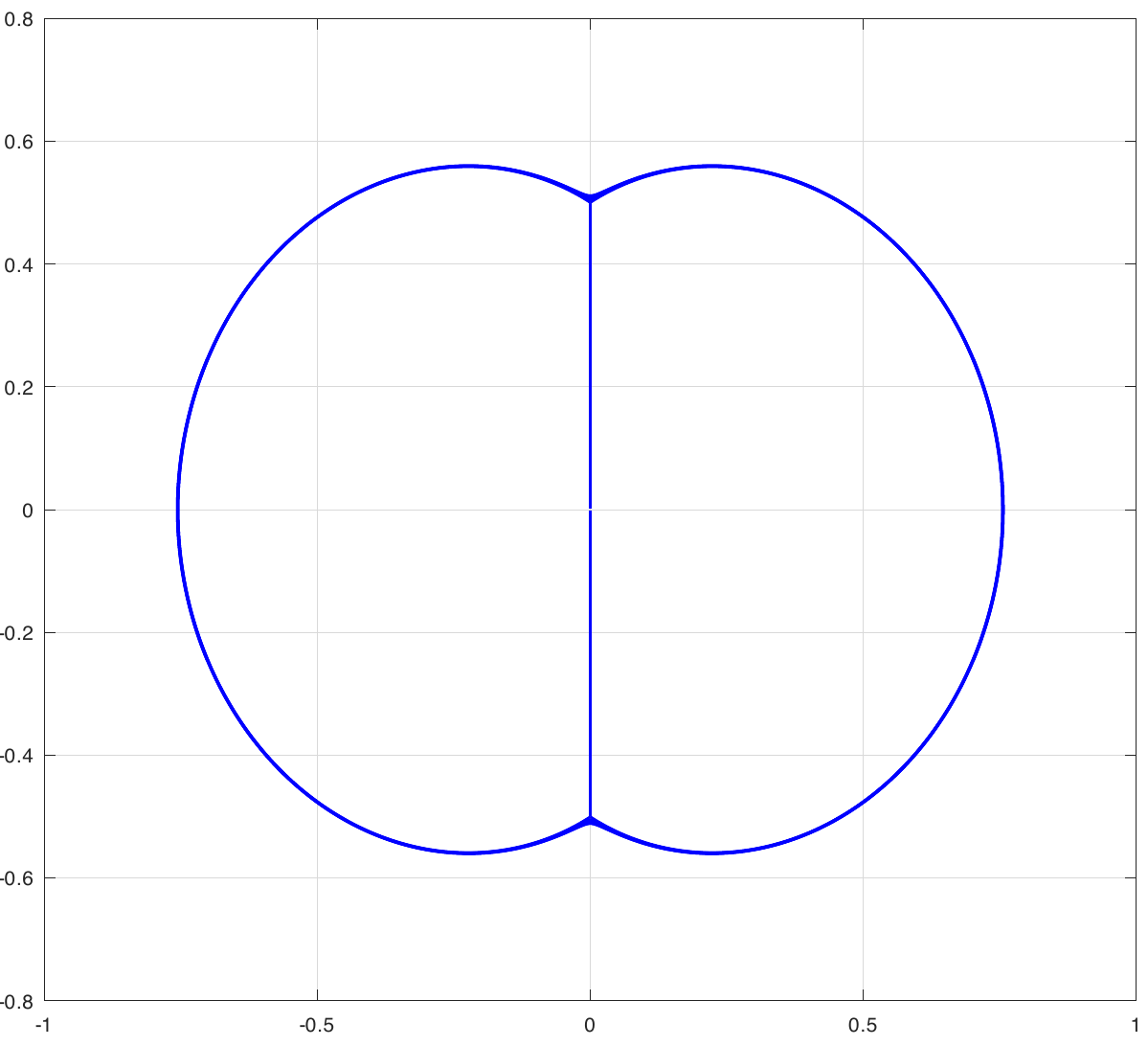}
 \includegraphics[width=0.39\textwidth, height=0.38\textwidth]{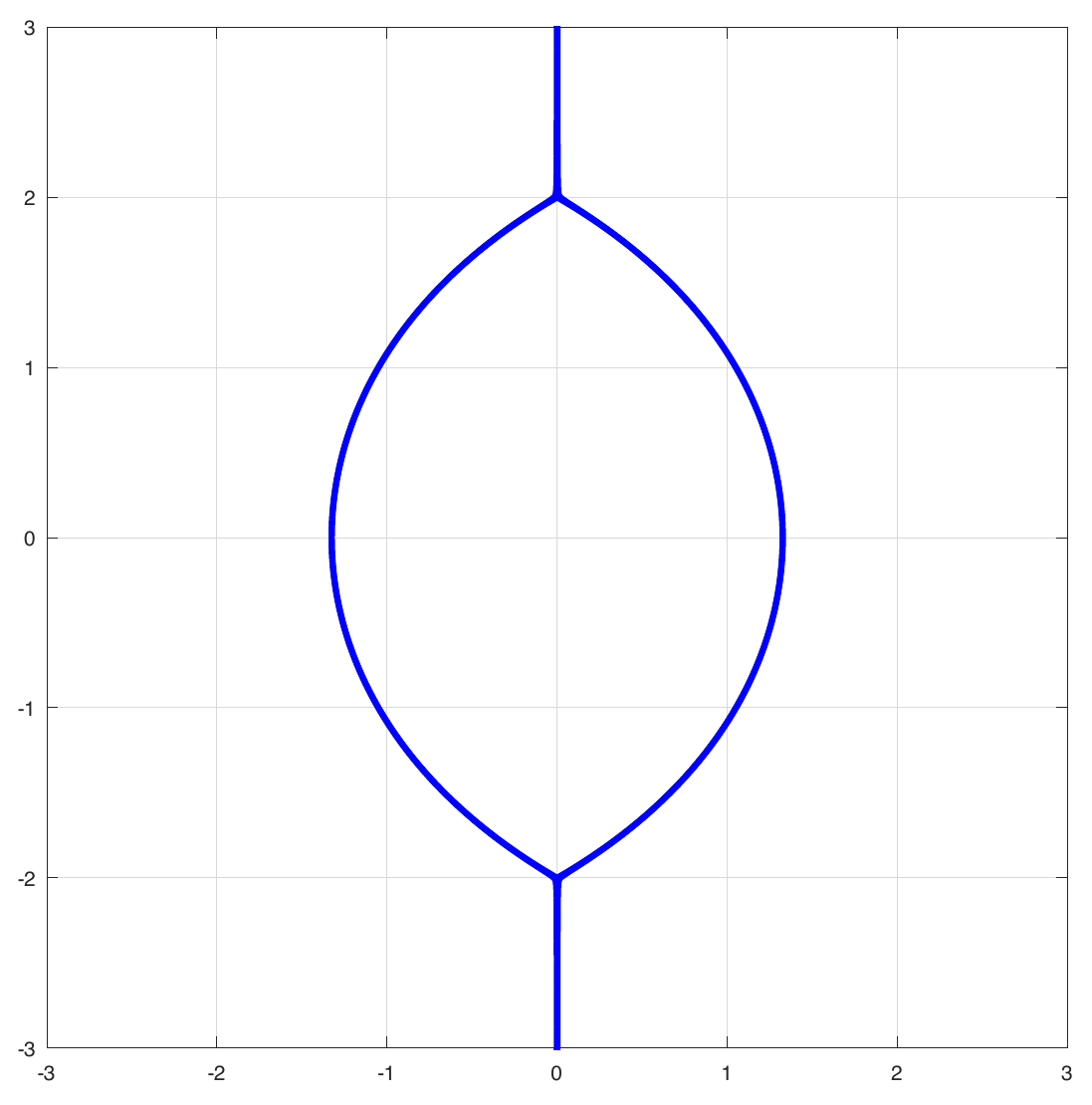}

 \caption{Left: the level sets of $\Re \varphi_0(z)=0$. Right: the boundary of the region of validity of the genus-zero assumption (ignore the vertical rays issuing from $\pm 2{\rm i}$). It is the locus of $\Re \varphi_0(1/s)=0$. The inside of this region we refer to as the ``Eye of the Tiger'' (${\rm EoT}$).}
 \label{apricot}
 \end{figure}

The language of {\it vertical trajectories of quadratic differentials} of \cite{Strebel} is useful in this discussion: by definition these are the arcs of curves where $\Re \varphi_0$ is constant, which, in the plane of the variable $\xi(z):= \varphi_0(z) = \int^z \varphi_0'(w)\d w$ are (by definition) vertical segments, whence the terminology.
We start by observing that $\res_{z=\infty} \varphi'_0(z)\d z=-2{\rm i}\pi$ and $\res_{z=0}\varphi'_0(z)\d z = \pm 2{\rm i}\pi $ (with the sign depending on whether the branch cut leaves $z=0$ to the left or to the right) and hence, no matter how we choose the branch cut $\Gamma_m$ (connecting the branch-points) we have that
the function $\Re \varphi_0(z)$ is single-valued, harmonic in $\C\setminus \Gamma\cup\{0\}$ and continuous in $\C \setminus \{0\}$,\looseness=1
\begin{gather}
\label{obs2}
\text{for $|z|$ sufficiently large $\Re \varphi_0(z) = \ln |z| + $ harmonic and bounded.}
\end{gather}
 The observation \eqref{obs2} implies that the level curves of $\Re \varphi_0$ are deformed circles for $|z|$ sufficiently large.
 One can verify that changing the determination of both radicals in \eqref{phi0} has the effect of flipping the sign of $\Re\varphi_0$ and hence that $\Re \varphi_0$ is a well-defined harmonic function on the Riemann surface of the radical
 $
 w^2= {1 + 4z^2}$.
 Furthermore, $\Re \varphi_0$ is an odd function under the holomorphic involution that maps $(w,z)$ to $(-w,z)$.
 This means that the level sets $\Re \varphi_0=0$ are well defined on the $z$-plane. They consist of ``vertical trajectories'' (in the terminology of \cite{Strebel}) issuing from the points $\pm \frac {\rm i}2$ and forming the pattern illustrated in Figure~\ref{apricot}.

We choose the branch cut of the radical as the arc of Figure~\ref{apricot} joining $\pm \frac {\rm i}2 $ in the right half-plane.
With this choice, we have that
$\varphi_0'(z) \simeq - \frac 1 {z^2} + \mathcal O(1)$, $ z\to 0$,
and in general $\varphi'(z;s) = \frac 1 s \varphi_0'\le(\frac z s\ri)$ satisfies
$\varphi'(z;s) = -\frac s {z^2} + \mathcal O(1)$.

{\bf Verification of the properties of $\boldsymbol{\varphi}$ and range of validity.}
It suffices to verify the properties for $s=1$ since changing $s\in \C$ just amounts to a complex homothety $z\mapsto sz$.
We choose $\Gamma_m$ as the arc joining $\pm \frac {\rm i} 2$ in the right plane, $\Gamma_c$ as an arc joining the two points $\pm \frac {\rm i} 2$ in the left plane, and inside the region bounded by the imaginary axis and the contour $-\Gamma_m$ (see Figure~\ref{apricot}). Finally, we choose $\Gamma_\infty$ as the ray $\bigl(-{\rm i}\infty, -\frac {\rm i}2\bigr]$.
We then proceed with the verification of the properties in Corollary~\ref{corphicond}:
\begin{enumerate}\itemsep=0pt
\item[(1)] On the sole connected component $\Gamma_m$, we have $\varphi_0(z_+)+ \varphi_0(z_-)=0$ since the two boundary values differ by a vanishing period of $\varphi_0'$.
\item[(2)] On $\Gamma_c$, we similarly have $\varphi(z_+)=\varphi(z_-)$.
\item[(3)] On $\Gamma_\infty$, we have $\varphi_0(z_+) = \varphi_0(z_-) - \res_{w=\infty} \varphi_0'(w)\d w =\varphi_0(z_-)+4{\rm i}\pi $.
\item[(4)] Since $\Gamma_m$ is defined as the zero level set of $\Re \varphi_0$, we have $\Re \varphi_0\equiv 0$ on $\Gamma_m$ by definition.
\item[(5)] In the unbounded doubly connected region outside of the ``apricot'' in Figure~\ref{apricot}, we have $\Re \varphi_0 = \ln |z| + \mathcal O(1)$ near $z\to \infty$. Thus inevitably $\Re \varphi_0>0$ in the whole region (which, we remind, is bounded by the zero level sets of $\Re \varphi_0$).
\item[(5)] In the right hemi-apricot, the sign must be also positive because $\Re \varphi_0(z) = \Re\le(\frac 1 z+ \mathcal O(1)\ri)$.
\item[(6)] By the same token, the sign is negative in the left hemi-apricot.
\end{enumerate}
 Thus all conditions except possibly the condition~(2) in Definition~\ref{gdef} are verified, namely, we still need to verify that the union $\Gamma_m\cup \Gamma_c$ is homotopic to a circle $|z|=R$, $R>1$ in the cut plane $\C\setminus [0,1]$.

\begin{figure}[t]
\centering
\includegraphics[width = 0.45\textwidth] {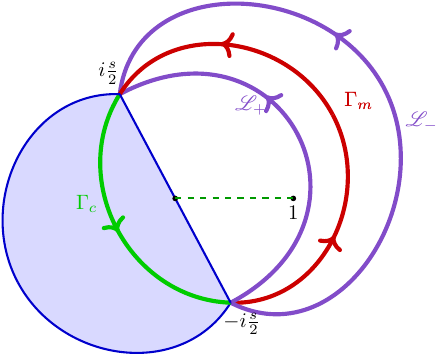}

\caption{The contours $\Gamma_m$, $\Gamma_c$ and the regions where $\Re \varphi(z;s)<0$ (shaded). Indicated also the boundaries, $\scr L_\pm$, of the lens regions $\Lambda_\pm$.}\label{figcontours}
\end{figure}

 Since the level sets in Figure~\ref{apricot} are scaled by $s$, this latter condition is fulfilled as long as the point $z=1$ lies inside the re-scaled apricot. This holds clearly for $|s|$ sufficiently large, and it fails precisely when the point $z=1$ lies on either $\Gamma_m$ or $\Gamma_c$, namely when
 \be
 \label{EoTcurve}
 \Re \varphi(1;s) = 0 = \Re \varphi_0\le(\frac 1 s\ri).
 \ee
 Observing \eqref{EoTcurve}, we conclude that the shape of the locus \eqref{EoTcurve} is simply the image of the apricot in Figure~\ref{apricot} (left pane) under the inversion $z\mapsto \frac 1 z$ (see right pane in Figure~\ref{apricot}).
In Figure~\ref{figcontours},
 we illustrate the contour for a generic value of $s$ outside the set ${\rm EoT}$.
 Note that, as a consequence of the branch-cuts in the $z$-plane, the branch-cuts of $\varphi_0\le(\frac 1 s\ri)$ in the $s$-plane now extend towards infinity (vertically) from the points $s = \pm 2{\rm i}$. However, only the imaginary part is discontinuous.

The discussion of this section can be summarized by the following theorem.
\begin{Theorem}
For $s$ outside the region ${\rm EoT}$, the effective potential is given by \eqref{phig0int} $($and \eqref{phig0}$)$. The contour $\gamma$ can be chosen as the contour $\Gamma_m\cup \Gamma_c$ consisting of the arc $\Gamma_{m}$ from $-\frac {{\rm i}s} 2$ to $\frac {{\rm i}s} 2$ lying on the left of the straight segment, passing through the origin, from $\frac {{\rm i}s}2$ to $-\frac {{\rm i}s}2$ {\rm(}in this orientation$)$. The arc $\Gamma_{c}$ is an arc in the right ``lobe'' keeping $z=0$ to its left. See Figure~{\rm\ref{figcontours}}.

The Robin constant $\ell$ appearing in \eqref{defphi} is given by
\be
\label{Robin0}
\ell = 2 \ln \le(-\frac 4 s\ri).
\ee
\end{Theorem}
The only statement that has not been proven yet is the expression \eqref{Robin0}. The expression for~$g(z;s)$ is derived from that of the effective potential \eqref{phig0int} and the relation \eqref{defphi} expressing the effective potential in terms of the $g$-function.
Since $g(z) = \ln (z) + \mathcal O\bigl(z^{-1}\bigr)$ (note the absence of a constant term in the asymptotic expansion),
we can deduce $\ell$ by the expansion at $z=\infty$ of~${g(z;s)- \ln(z)}$.

From \eqref{phig0int}, \eqref{defphi}, we have
\[
g(z;s)-\ln z = \ln \le(\frac 2 s\ri) + \ln \le(-1 - \sqrt{1 + \frac{s^2}{4z^2}}\ri) - \frac \ell 2 + \mathcal O\bigl(z^{-1}\bigr) = \ln\le(-\frac 4 s\ri) - \frac \ell 2 + \mathcal O\bigl(z^{-1}\bigr).
\]
Setting to zero the constant term in the expansion yields the statement \eqref{Robin0}.

 \subsection[Inside EoT: Genus one $g$-function and effective potential]{Inside $\boldsymbol{{\rm EoT}}$: Genus one $\boldsymbol{g}$-function and effective potential}
 \label{gfuncg1}

The boundary of ${\rm EoT}$ is precisely the condition that the point $z=1$ belongs to the two sub-arcs of the zero-level set of $\Re \varphi(z;s)$ forming the ``rind'' of the apricot. As we move~$s$ inside~${\rm EoT}$, we cannot use the same effective potential described in the previous section because the second condition in Definition~\ref{gdef} ceases to be verified, namely the contour of integration $\gamma$ cannot be homotopically retracted to $\Gamma_m \cup \Gamma_c$ within $\C\setminus[0,1]$ since either $\Gamma_c$ or $\Gamma_m$ intersect the segment~$[0,1]$.

The idea to resolve the impasse is to treat $z=1$ as a ``hard-edge'', using the terminology that has come to pass in the random matrix theory literature \cite{Bertola:wj, Borodin}.
We thus postulate the following form for $\varphi'(z;s)$:
\[
\varphi'(z;s) = \frac 2 {z^2} \sqrt{z^2 + \frac {s^2}4 + \frac {A z^2}{z-1}}
=\frac {2\sqrt{z^2(z-1+A) + \frac {s^2}4 (z-1) }}{ z^2 \sqrt{z-1}}.
\]
The parameter $A=A(s)$ is chosen by the condition that all periods of $\varphi'(z;s)\d z$ on the Riemann surface of the radical are purely imaginary (this is called {\it Boutroux condition}), which is the necessary condition so that $\Re \varphi$ is continuous across the cuts. The Riemann surface of $\varphi'(z;s)$ is an elliptic curve branched at $z=1$ and the other three roots of the radical in the numerator:
\[
\mu^2 =(z-1) \le(z^2(z-1+A) + \frac {s^2}4 (z-1)\ri).
\]
We denote these roots as $b$, $a_+$, $a_-$ with $b$ the closest root to $z=1$. An expression in terms of Cardano's formul\ae\ is possible but not necessary.

Now the complex parameter $A(s)$ is determined implicitly by the two real equations
\be
\label{Boutroux_cond}
\Re \int_b^1\varphi'(z;s)\d z =0,\qquad
\Re \int_b^{a_+} \varphi'(z;s)\d z =0.
\ee
Under these conditions, it then follows that the real part of
\[
\varphi(z;s) = \int_{a_-}^z \varphi'(w;s)\d w
\]
is a well-defined (single valued) harmonic function on the Riemann surface minus the preimages of the points $z=0$ on the two sheets.

\begin{figure}[th!]\centering
\includegraphics[width=0.74\textwidth]{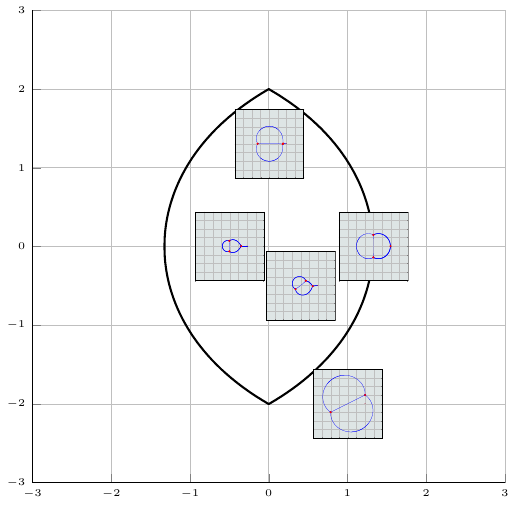}

 \caption{Illustration of the zero level sets of $\Re \varphi(z;s)$ for various values of $s$; the inset vignettes have their centre at the value of $s$ to which they correspond.}
 \label{Crittrajs}
 \end{figure}

{\bf Determination of $\boldsymbol{\Gamma_m}$ and $\boldsymbol{\Gamma_c}$.}
By the same argument already used in the genus-zero case, the zero level sets $\Re \varphi(z;s)=0$ are well defined. They consist of the critical vertical trajectories of the quadratic differential $Q = \varphi'(z;s)^2 \d z^2$ \cite{Strebel}
\[
Q = 4\frac {{z^2(z-1+A) + \frac {s^2}4 (z-1) }}{ z^4 (z-1)}.
\]
The following discussion is best followed by referring to Figure~\ref{Crittrajs}, in particular the smaller inset vignettes.
The main arcs $\Gamma_m$ are sub-arcs of the zero level set of $\Re \varphi$ and we need to discuss their qualitative topology before proceeding.

The critical points are the three (generically) simple zeros and the simple pole $z=1$. There are three vertical trajectories that issue from each simple zero, while from the simple pole there is only one.
The union of the trajectories is a connected planar graph and the unbounded region is put in conformal equivalence with the punctured unit disk by the map
\smash{$\zeta = {\rm e}^{-\frac{ \varphi(z;s)}{4{\rm i}\pi}}$},
which maps the exterior region into the disk $|\zeta|<1$, with $z=\infty$ mapped to $\zeta=0$.
 Some observations are in order:
 \begin{enumerate}\itemsep=0pt
 \item[(1)]
 The level sets of $\Re \varphi$ depend only on $s^2$ and they are conjugated if we conjugate $s$.
 \item[(2)] One of the zeros of $\varphi'$ is connected by a vertical trajectory to $z=1$. We denote this zero by $z=b$. The other two zeros are one in the upper and one in the lower half plane. We denote them by $a_\pm$, respectively.
 \end{enumerate}
While the level sets of $\Re \varphi$ depend on $s^2$ alone, we must choose the branch cuts $\Gamma_m$ differently according to the cases $\Re s>0$ or $\Re s<0$. The reason is that the sign distribution of $\Re \varphi$ differs in the two cases.
This is seen by the following reasoning:
\begin{itemize}\itemsep=0pt
\item In the outside region $\Re \varphi = 2 \ln|z| + \mathcal O(1)$ and hence $\Re \varphi>0$.
\item Near the origin we must have $\varphi(z;s) = -\frac s z + \mathcal O(1)$ and hence for $\Re s>0$ the right ``lobe'' is where $\Re \varphi<0$. Vice versa it is the left one if $\Re s<0$.
\end{itemize}
 The branch cut $\Gamma_m$ is then singled out by the fact that across it the function $\Re \varphi$ is continuous but not differentiable, namely, $\Re \varphi$ has the same sign (positive) on both sides.

 Collecting these observations, we thus have determined that  (see Figure~\ref{Crittrajs})
 \begin{enumerate}\itemsep=0pt
 \item[(1)] For $\Re s>0$, the branch cut $\Gamma_m$ consists of the three arcs of the vertical trajectories connecting $z=b,1$ and $z=b,a_+$ and $z=b,a_-$.
 \item[(2)] For $\Re s<0$, the branch cut $\Gamma_m$ consists of the two arcs of the vertical trajectories connecting $z=b,1$ and $z=a_+,a_-$ (passing to the left of the origin).
 \end{enumerate}

The specification of the constants that appear for the boundary values of $\varphi$ in Corollary~\ref{corphicond} will be given in Sections~\ref{secmodels>0} and~\ref{secmodels<0} for $\Re(s)>0$ and $\Re(s)<0$, respectively.

\section{Deift--Zhou steepest descent analysis}
 We split the asymptotic analysis in three subsections, according to the following cases:
 \begin{enumerate}\itemsep=0pt
 \item[(1)] $s\in \mathcal K_{\rm out}$, where ${\mathcal K_{\rm out}}$ is a closed subset contained outside of ${\rm EoT}$,
 \item[(2)] $s\in \mathcal K_{{\rm in},+}$, where ${\mathcal K_{{\rm in},+}}$ is a compact subset of $\{\Re (s)>0\}\cap {\rm EoT}$,
 \item[(3)] $s\in \mathcal K_{{\rm in},-}$, where ${\mathcal K_{{\rm in},-}}$ is a compact subset of $\{\Re (s)<0\}\cap {\rm EoT}$.
 \end{enumerate}
 Note that for $s\in {\rm i}\R\cap {\rm EoT}$ the elliptic curve is degenerate, namely, two branch-points coincide and the curve becomes of genus zero. Here a different analysis would be needed, analogous to the one needed when $s$ belongs to the boundary of ${\rm EoT}$. Near the corners $s=\pm \frac {\rm i}2$ a yet different analysis would be needed, which involves the construction of special local parametrices. We do not discuss these transitional regions.

The main take-away of the analysis is the following theorem.
\begin{Theorem}
For any closed subset $\mathcal K$ on the outside of ${\rm EoT}$, the Riemann--Hilbert Problem~{\rm\ref{RHPY}} $($with $t=ns)$ is solvable for $n$ sufficiently large, and hence the poles of the rational solutions must be inside a~neighbourhood of~${\rm EoT}$.
\end{Theorem}

The Deift--Zhou steepest descent analysis, formalized in \cite{DZ} and in many articles thereafter requires a number of transformations of the problem into equivalent ones. We describe briefly below these problems.
We recall that our starting point is the following.

\begin{problem}
\label{RHPY1} Find a $2\times 2$ matrix-valued function $Y(z) = Y_n(z;s)$ analytic in $\C \setminus \gamma$, with analytic bounded inverse and such that
\begin{gather*}
Y(z_+) = Y(z_-) \begin{bmatrix}
1 & z^{{ {K}}}\le(1-\frac 1 z\ri)^{ {\rho}} {\rm e}^{\frac {ns}z}\\
0 & 1
\end{bmatrix}, \qquad z\in \gamma,
\\
Y(z) = \le(\1 + \mathcal O\bigl(z^{-1}\bigr)\ri) z^{n\s_3}, \qquad z\to \infty.
\end{gather*}
\end{problem}
We remind the reader that ${ {K}} \in \Z$ while we assume ${ {\rho}} \not\in \Z$.
We will use the notation
\be
\label{defQ}
Q(z) = z^{ {K}}\le(1 - \frac 1 z \ri)^{ {\rho}},
\ee
where the domain is $\C\setminus \Gamma_0^1$, where $\Gamma_0^1$ denotes an arc homotopic to $[0,1]$ at fixed endpoints. For~$s$ outside ${\rm EoT}$, we shall choose $\Gamma_0^1=[0,1]$ (the segment).

It will be necessary for the analysis inside ${\rm EoT}$ to partially homotopically retract the integration contour $\gamma$ along a subarc of $\Gamma_0^1$ (from the left and right of it). In that case, the Riemann--Hilbert Problem~\ref{RHPY1} will take a slightly different form due to the fact that the function $Q$ \eqref{defQ} has a jump discontinuity such that $Q(z_+) = Q(z_-) {\rm e}^{2{\rm i}\pi \rho}$. In this case, the jump matrix of the Riemann--Hilbert Problem~\ref{RHPY1} along such a~subarc of $\Gamma_0^1$ needs to be replaced by
\[
 \begin{bmatrix}
1 & Q(z_+) \le({\rm e}^{-2{\rm i}\pi \rho} - 1\ri) {\rm e}^{\frac {ns}z}\\
0 & 1
\end{bmatrix}.
\]

\subsection[Asymptotic analysis for $s$ outside EoT]{Asymptotic analysis for $\boldsymbol{s}$ outside $\boldsymbol{{\rm EoT}}$}
\label{outsideEoT}

We recall that $\Gamma_m$, $\Gamma_c$ have been defined in Section~\ref{gzero}.

{\bf Normalization: $\boldsymbol{Y\to W}$.}
We define
\be
\label{defW}
W(z)= {\rm e}^{n\frac \ell 2 \s_3} Y(z) {\rm e}^{ -n\le(g(z) + \frac \ell 2\ri)\s_3}.
\ee
A direct verification shows that $W$ solves the following.

\begin{problem}
\label{RHPW}The matrix $W(z)$ is analytic in $\C \setminus \gamma = \C\setminus (\Gamma_m\cup \Gamma_c)$ and satisfies
\[
W(z_+)= W(z_-) J_W(z), \qquad z\in \gamma,\qquad
W(z) = \1 + \mathcal O\le(\frac 1 z \ri), \qquad z\to \infty,
\]
where
\begin{align*}
J_W(z) &= \begin{bmatrix}
{\rm e}^{-n(g(z_+)-g(z_-))} & Q(z) {\rm e}^{ n(\frac s z + g(z_+)+ g(z_-) + \ell)}\\
0 & {\rm e}^{n(g(z_+)-g(z_-))}
\end{bmatrix}\\
&=\begin{bmatrix}
{\rm e}^{-\frac n2( \varphi(z_+)-\varphi(z_-)) } & Q(z) {\rm e}^{ \frac n2 (\varphi(z_+)+\varphi(z_-))}\\
0 & {\rm e}^{\frac n2 (\varphi(z_+)-\varphi(z_-) )}
\end{bmatrix}.
\end{align*}
\end{problem}
{\bf Lens Opening: $\boldsymbol{W\to T}$.}
Refer to Figure~\ref{figcontours}. The {\it lens regions} $\Lambda_\pm$ are the two regions bounded by $\Gamma_m$ and the arcs $\scr L_\pm$ chosen arbitrarily in the regions where $\Re \varphi>0$.
The process of ``opening the lenses'' consists in re-defining the matrix $W$ within those regions.
We thus define
\begin{gather}
\label{defT}
T_0(z):=
\begin{cases}
W(z), & z\not \in \Lambda_\pm,\\
\ds W(z)\begin{bmatrix}
1 & 0 \\
\displaystyle \frac{{\rm e}^{-n\varphi(z)}}{Q(z)} & 1
\end{bmatrix}, & z\in \Lambda_-,
\vspace{1mm}\\
W(z) \begin{bmatrix}
1 & 0 \\
\displaystyle-\frac{ {\rm e}^{-n\varphi(z)}}{Q(z)} & 1
\end{bmatrix}, & z\in \Lambda_+.
\end{cases}
\end{gather}
A direct computation shows that the matrix $T_0(z)$ satisfies the following Riemann--Hilbert problem.

\begin{problem}
\label{RHPT0}
The matrix $T_0(z)$ satisfies the conditions
\[
T_0(z_+)= T_0(z_-) J_{T_0}(z), \qquad z\in \gamma,\qquad
T_0(z) = \1 + \mathcal O\le(\frac 1 z \ri), \qquad z\to \infty,
\]
where
\[
J_{T_0}(z) =
\begin{cases}
J_{_W}(z), & z\in \Gamma_c,\\
 \begin{bmatrix}
1 & 0 \\
\displaystyle\frac{ {\rm e}^{-n\varphi(z)}}{Q(z)} & 1
\end{bmatrix},
& z\in \scr L_\pm,
\vspace{1mm}\\
\begin{bmatrix}
 0 & Q(z) \\
\displaystyle -\frac 1 {Q(z)} & 0
\end{bmatrix},
 & z\in \Gamma_m.
\end{cases}
\]
\end{problem}
At this point, we have obtained a RHP, where the jump matrices on $\Gamma_{c} \cup \scr L_\pm$ converge pointwise in the relative interior to the identity matrix, but not uniformly.

To further normalize the problem, we need to construct the {\it outer parametrix}, namely the (explicit) solution of an auxiliary RHP where we simply drop the jump conditions on $\Gamma_c\cup \scr L_\pm$.

{\bf Outer parametrix.}
We seek the solution of the following ``model problem''.
\begin{problem}
\label{RHPM0}
The matrix $M_{_Q}(z)$ is analytic and analytically invertible in $\C\setminus \Gamma_m$ and satisfies
\begin{gather*}
M_{_Q}(z_+) = M_{_Q}(z_-) \begin{bmatrix}
0 & Q(z)\\ -\frac 1 {Q(z)} & 0
\end{bmatrix},\qquad z\in \Gamma_m
,\qquad
M_{_Q}(z) = \1 + \mathcal O\bigl(z^{-1}\bigr),\qquad |z|\to \infty,\\
M_{_Q}(z) = \mathcal O\le(\frac 1{ \le(z\mp \frac {{\rm i}s}2\ri)^{\frac 1 4} }\ri),\qquad z\to \pm \frac {{\rm i}s}2.
\end{gather*}
\end{problem}
We now solve Riemann--Hilbert Problem~\ref{RHPM0}.
To this end, we need to construct a special scalar function $S(z)$, often called the {\it Szeg\H{o}} function, satisfying the following scalar boundary value problem
$S(z_+) + S(z_-) = \ln Q(z)$, $ z\in \Gamma_m$, $\sup_{z\in \C\setminus \Gamma_m} | S(z)| <+\infty$.
The solution is given by the following expression:
\be
\label{defSg0}
S(z) = R(z) \int_{\Gamma_m} \frac {\ln Q(w) \d w} {R(w_+) (w-z) 2{\rm i}\pi},\qquad R(z):= \sqrt{z^2 + \frac {s^2}4},
\ee
where the branch cut of the radical $R(z)$ is chosen to run along $\Gamma_m$.
We leave to the reader the verification that the proposed expression \eqref{defSg0} fulfills all the required conditions.

We can actually simplify the Szeg\H{o} function by a contour deformation recalling that $\ln Q(w_+)\allowbreak= \ln Q(w_-) + 2{\rm i}\pi { {\rho}}$ for $z$ on the segment $[0,1]$. Indeed, by a contour deformation, we can rewrite~$S(z)$ as follows:
\[
S(z) = R(z) \int_{\Gamma_m} \frac {\ln Q(w) \d w} {R(w_+) (w-z) 2{\rm i}\pi}
= - \frac{R(z)}2 \oint_{\odot \Gamma_m} \frac {\ln Q(w) \d w} {R(w) (w-z) 2{\rm i}\pi},
\]
where the symbol $\odot \Gamma_m$ stands for a counterclockwise loop leaving $\Gamma_m$ in its interior region and the segment $[0,1]$ on the exterior. Using then Cauchy's residue theorem, we get
\begin{align}
S(z) ={}& \frac 1 2 \ln Q(z) - \frac {R(z)}2 \int_0^1 \frac {\le(\ln Q(w_+)- \ln Q(w_-)\ri)\d w}{(w-z)R(w) 2{\rm i}\pi}\nonumber
\\
={}& \frac {\ln Q(z)}2 - \frac{R(z)}2\int_0^1 \frac {{ {\rho}} \d w}{R(w)(w-z)}
- \frac{R(z)}2\int_{-\infty}^0 \frac {{ {K}} \d w}{R(w)(w-z)}
\nonumber
\\={}&
\frac{ {K}}2
\ln\le(\frac {\le (R(z)-z\ri)\le(R(0)+ R(z) \ri)}{s/2}\ri)
+
\frac{ {\rho}}2 \ln\le(
 {\frac { z+ \frac {s^2}4 + R(z) R(1) }{ \frac s2 (R(z) + R(0)
 ) }}
 \ri).
 \label{S(z)}
\end{align}
The simplest way to verify this latest formula is to verify $S(z_+)+S(z_-)=\ln Q(z)$ using that $R(z_+)=-R(z_-)$. Also one needs to verify that the expression has no singularities except the discontinuity across the branch cut of $R(z)$, and that it is bounded at infinity. We leave the instructive but tedious verification to the reader.
\begin{Remark}
The function $S(z)$ has a finite value at $z=\infty$ given by
\[
S(\infty) = \frac { {K}} 2 \ln \le(\frac s4 \ri) + \frac { {\rho}} 2 \ln \le(\frac 2 s + \sqrt{1 + \frac 4{s^2}} \ri).
\]
\end{Remark}
With these preparations, we state the following.

\begin{Proposition}[solution of Riemann--Hilbert Problem~\ref{RHPM0}]
\label{modelMg0}
The solution of Riemann--Hilbert Problem~{\rm\ref{RHPM0}} is given by
\begin{gather}
M_{_Q}(z) := {\rm e}^{-S(\infty)\s_3} M(z) {\rm e}^{S(z) \s_3},\label{defMQ}\\
M(z):= \le(\frac {z-\frac {{\rm i}s}2}{z+\frac {{\rm i}s}2}\ri)^{\frac {\s_2}4}
= \frac 1 2\begin{bmatrix}
{F} + \frac 1 {F} & -{\rm i}\le({F}-\frac 1 {F}\ri)\\
{\rm i}\le({F} - \frac 1 {F}\ri) & {F} + \frac 1 {F}
\end{bmatrix},
\qquad
{F} := \le(\frac {z-\frac {{\rm i}s}2}{z+ \frac {{\rm i}s}2} \ri)^\frac 14,\nonumber
\end{gather}
where the branch cut of ${F}$ runs along $\Gamma_m$ and the determination is chosen so that ${F}(z) \to 1 $ as~${|z|\to \infty}$.
\end{Proposition}
The proof is left to the reader.

\subsubsection{Conclusion of the steepest descent analysis}
\label{Errorg0}
The final steps of the analysis requires the construction of {\it local parametrices} near the endpoints~$\pm \frac {{\rm i}s}2$. Namely, one fixes two disks $\mathbb D_\pm$ centered at each of the two points. This part is a~quite standard construction and therefore we only sketch the main points.
Let $\DD_\pm$ be two disjoint disks centered at $\pm \frac {{\rm i}s}2$ (respectively) and of radii, say, $r=\frac{|s|}4$ (the radius is not important for the discussion as long as they are small enough so as not to contain the origin),
\[
\DD_\pm:= \left \{\le|z\mp \frac{{\rm i}s}2\ri|<\frac {|s|}4 \ri \}.
\]
In a neighbourhood of $a_\pm=\frac {\pm {\rm i}s}2$, the effective potential $\varphi$ \eqref{phig0int} has the following behaviour:
\[
\varphi(z;s) = \frac {2^\frac 52}3 \le(\frac {z}{a_\pm } -1\ri)^\frac 32\le(1 + \mathcal O(z-a_\pm)\ri) \mod 2{\rm i}\pi,\qquad z\sim a_\pm.
\]
In fact, the additive constant is $0$ in the case of $a_-$ and $-2{\rm i}\pi$ for $a_+$, but since $\varphi$ appears always in the exponent, this is irrelevant.

We define the \emph{local coordinates} $\zeta_\pm $ by the formulas
\begin{gather*}
-\frac 4 3\zeta_-^{\frac 32} = n\varphi(z) = n\frac {2^\frac 52}3 \le(\frac {z}{a_- } -1\ri)^\frac 32\le(1 + \mathcal O(z-a_-)\ri),
\\
-\frac 43\zeta_+^{\frac 32} = n\le(\varphi(z) +2{\rm i}\pi \ri)=n \frac {2^\frac 52}3 \le(\frac {z}{a_+ } -1\ri)^\frac 32\le(1 + \mathcal O(z-a_+)\ri).
\end{gather*}
From the above formulas, it appears that both $\zeta_\pm$ define a conformal map from the two disks~$\DD_\pm $ (respectively) to a neighbourhood $\scr D_n$ of the origin which is homothetically expanding with~\smash{$n^\frac 32$}. The determination of the fractional root can be chosen so that $\zeta_\pm$ map the main arc $\Gamma_m$ (where~${\Re \varphi=0}$ and $\Im \varphi(z_+)$ is decreasing from $0$ at $a_-$ to $-2{\rm i}\pi$ at $a_+$) to the negative real axis in the respective $\zeta_\pm$ planes, with the points $a_\pm$ being mapped to $\zeta_\pm =0$. The arc $\Gamma_c$ can be chosen so that it is mapped to the positive $\zeta_\pm$-axis while the arcs of $\scr L_\pm\cap \DD_\pm$ are mapped to two straight segments with slopes $\pm 3\pi/2$.

Then, the jump matrices $J_{T_0}$ in the Riemann--Hilbert Problem~\ref{RHPT0} restricted to $\DD_+$ can be rewritten as (we focus on the case of $\DD_+$ for definiteness)
\begin{gather}
\label{JT0zeta}
J_{T_0}(z) =
\begin{cases}
\begin{bmatrix}
1 & Q(z) {\rm e}^{-\frac 43\zeta_+^\frac 32}\\
0& 1
\end{bmatrix},
 & z\in \Gamma_c\cap \DD_+,\vspace{1mm}\\
 \begin{bmatrix}
1 & 0 \\
\displaystyle \frac{\ds {\rm e}^{\frac 43{\zeta_+}^\frac 32} }{Q(z)} & 1
\end{bmatrix},
& z\in \scr L_\pm \cap \DD_+,
\vspace{1mm}\\
\begin{bmatrix}
 0 & Q(z) \\
\displaystyle -\frac 1 {Q(z)} & 0
\end{bmatrix},
 & z\in \Gamma_m\cap \DD_+.
\end{cases}
\end{gather}
Observe that $Q(z)$, $\frac 1 Q(z)$ are both locally analytic at $z=a_\pm$, see \eqref{defQ}.
Then the following matrix
\[
\mathcal P_+(z) := {\bf A} \bigl(\zeta_+(z)\bigr) Q(z)^{-\frac {\s_3}2} {\rm e}^{\frac 2 3\zeta_+(z)^\frac 32 \s_3}
\]
exhibits discontinuities across $(\Gamma_c\cup\scr L_+\cup \scr L_- \cup \Gamma_m)\cap \DD_+$ with jump matrices given exactly by~\eqref{JT0zeta}, as a consequence of the RHP satisfied by the matrix
${\bf A}(\zeta)$ shown on Figure~\ref{Airy1}. Completely analogous expressions hold for $\mathcal P_-(z)$.

\begin{figure}[t]\centering
\includegraphics[width=0.34\textwidth]{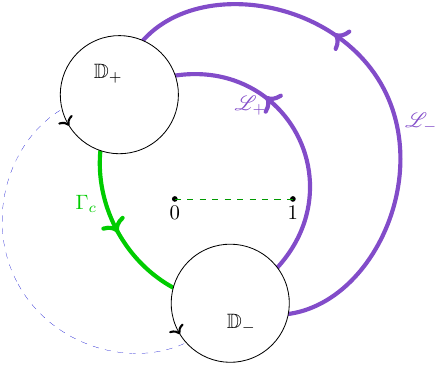}
\caption{The support of the RHP for the error matrix $\scr E(z)$.}
\label{RHPE0}
\end{figure}

The final approximation to the matrix $T_0(z)$ \eqref{defT} is then given by (see the definition of $M_{_Q}$ in \eqref{defMQ})
\[
{\bf G}(z):=
\begin{cases}
M_{_Q}(z), & z\not\in \DD_+\cup \DD_-,
\\
\displaystyle M_{_Q}(z) Q(z)^{\frac {\s_3}2} \begin{bmatrix}
1 & -1\\
{\rm i} & {\rm i} \\
\end{bmatrix}\frac{\ds \zeta_\pm(z)^{-\sigma_3/4}}{4\sqrt{\pi}} \mathcal P_\pm (z),& z\in \DD_\pm.
\end{cases}
\]
The matrix ${\bf G}(z)$ has the same jumps as $T_0(z)$ within $\DD_\pm$ and on $\Gamma_m$. It has additional jump discontinuities across the boundaries $\pa\DD_\pm$ of the form
$
{\bf G} (z_+)= {\bf G} (z_-)\bigl(\1 + \mathcal O\bigl(n^{-1}\bigr) \bigr)$, $z\in \pa\DD_\pm$.
Consequently, the matrix
\be
\scr E(z):= T_0(z) {\bf G}(z)^{-1}\la{errormatrix}
\ee
has only jump discontinuities on $\pa\DD_+\cup \pa \DD_- \cup \scr L_+\cup \scr L_-$ with all jump matrices that can be verified to be of the form $\1 + \mathcal O\bigl(n^{-1}\bigr)$, see Figure~\ref{RHPE0}.

It is then a standard result referred to in the literature as a ``small norm theorem'' \cite{Deiftbook,DKMVZ} that the solution of this last RHP $\scr E(z)$ exists and is uniformly close to the identity matrix.
Reversing the order of the various transformations, this proves that $\forall s\in \C \setminus {\rm EoT}$, the solution~$Y_n(z;s)$ of Riemann--Hilbert Problem~\ref{RHPY1} exists (at least for $n$ large enough).

More precisely, we can state as a result of the above discussion the following.

\begin{Theorem}\la{theoremnozeros}
For any closed set $\mathcal K$ in the complement of the closure of ${\rm EoT}$, there is $N_0\in \N$ such that the solution of Riemann--Hilbert Problem~{\rm\ref{RHPY1}}, and hence no zeros of $\mathcal T_n(s)$ \eqref{taun} lie in~$\mathcal K$ for $n\geq N_0$.
\end{Theorem}
We only give a sketch of the proof, which is a relatively standard argument. We summarize the main steps here for the reader with some experience in the Deift--Zhou type of analysis.

 As discussed previously in this section, the solvability of Riemann--Hilbert Problem~\ref{RHPY1} relies on the solvability of the RHP for the error matrix $\scr E(z)$ \eqref{errormatrix}, whose jump matrices, $J_{\scr E}(z)$, can be bound uniformly, along the contours of the discontinuities, by
\be
\sup_{z\in \pa\DD_+\cup \pa \DD_- \cup \scr L_+\cup \scr L_-}
\le\|\1 - J_{\scr E}(z)\ri\| < \frac C n.
\label{estim}
\ee

The constant $C$ depends on the radii of the local disks $\DD_\pm$, which is only limited by the distance of the branch points $\pm \frac {{\rm i} s}2$ from $z=1$, and it is bounded away from zero uniformly as $s\in \mathcal K$. The other source of potential loss of control over the error term is the distance of the main and complementary arcs $\Gamma_m$, $\Gamma_c$ from $z=1$. Indeed, if either arc contains $z=1$, the asymptotic analysis would require the construction of a special local parametrix and the small norm theorem cannot be used in the same way. However, as long as $s\in \mathcal K$ also the distance of~${\Gamma_m\cup \Gamma_c}$ from $z=1$ is bounded away from zero, and hence we can open the lenses as normal. Along the rims of the lenses, $ \scr L_\pm$, and along $\Gamma_c$, the jump matrices of $\scr E$ tend exponentially to the identity, so that the main source of error is coming from the boundaries of $\DD_\pm$.

Thus, in conclusion, as long as both branch points $\pm \frac {{\rm i}s}2$ and $\Gamma_c\cup \Gamma_m$ remains at finite positive distance from $z=1$ the constant $C$ in \eqref{estim} can be chosen uniformly with respect to $s\in \mathcal K$ so that for $n$ sufficiently large the small norm theorem comes into effect, and uniformly so with respect to $s$.

\subsection[Asymptotic analysis inside EoT]{Asymptotic analysis inside $\boldsymbol{{\rm EoT}}$}
\label{insideEoT}

\begin{figure}[t]\centering
\includegraphics[width=0.45\textwidth]{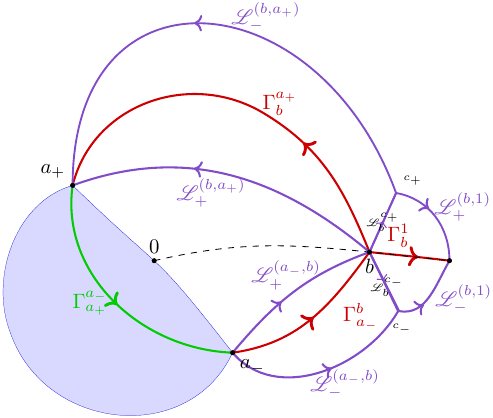}\quad
\includegraphics[width=0.45\textwidth]{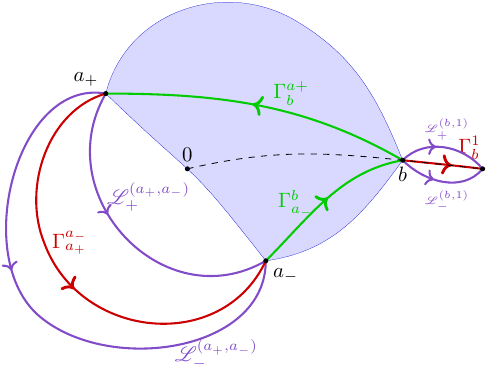}
\caption{The contours $\Gamma_m$ (red) and $\Gamma_c$ (green) for two antipodal values $s_1=-s_2$ of $s$ with $\Re (s_1)>0$ (left) and $\Re(s_2)<0$ (right). Specifically, here $s_1 = 0.48 + 0.48{\rm i} = - s_2$. The dashed line indicates the branch cut of the function $Q(z)$, which runs partially along $\Gamma_b^1$. The components of $\Gamma_c$ (green arcs) can be chosen freely within the shaded region, which indicates the region where $\Re\varphi<0$.}
\label{Contoursinside}
\end{figure}

With reference to Figure~\ref{Contoursinside}, we have already observed that the level sets of $\Re \varphi(z;s)$ depend only on $s^2$ and hence are invariant under the map $s \mapsto -s$. The same is not true for the branch cuts $\Gamma_m$ and the regions of positivity of $\Re \varphi$. The decision of where to place the branch cuts is forced by the ``sign distribution'' (i.e., in which regions we have $\Re \varphi>0$). This is mandated by the behaviour near infinity $\Re \varphi = 2\ln |z| + \mathcal O(1)$ for the outer region, and by the behaviour near the origin $\Re \varphi = \Re \frac s z + \mathcal O(1)$ for the two bounded regions. Before we fork the analysis according to the sign of $\Re (s)$, we make some common preparation.

As shown in Figure~\ref{Contoursinside}, one of the arcs of $\Gamma_m$ consists of an arc joining $z=1,b$, denoted $\Gamma_b^1$ (the orientation of the arc is suggested by the index placements). Thus, we need to preemptively modify the initial Riemann--Hilbert Problem~\ref{RHPY} so that the contour $\gamma$ is ``retracted'' in part along $\Gamma_b^1$. We leave the details to the reader and we report the resulting RHP, which is now the starting point of the following analysis.

\begin{Remark}\label{MRN}
Since the analysis is the complete parallel to that in Section~\ref{outsideEoT}, we recycle the symbols $M$, $W$, $T_0$ for the solutions of corresponding Riemann--Hilbert problems because, although they are different functions, they fulfill the same role. Since moreover we have to repeat the analysis for $\Re(s)>0$ and $\Re(s)<0$, we will recycle the symbols within each separate case.
\end{Remark}
\begin{problem}\label{RHPYg1}
Let $\gamma = \Gamma_b^1 \cup \gamma_b$ consist of the arc $\Gamma_b^1$ and a loop, $\gamma_b$, starting and ending at~${z=b}$ and containing the origin. Let $Q(z)$ in \eqref{defQ} be defined with the branch cut extending from $z=0$ to $z=b$ and then along $\Gamma_b^1$, as shown in Figure~{\rm\ref {Contoursinside}} by the dashed line.
Given $n\in \mathbb N$,
find a~${2\times 2}$ matrix-valued function $Y(z)= Y_n(z)$ such that
\begin{enumerate}\itemsep=0pt
\item[$(1)$] $Y(z)$ and $Y^{-1}(z)$ are holomorphic and bounded in $\C\setminus \gamma$.
\item[$(2)$] The boundary values along $z\in \gamma$ satisfy
\begin{gather}
Y(z_+) = Y(z_-) \begin{bmatrix}
1&Q(z){\rm e}^{\frac {ns}z}\\
0 & 1
\end{bmatrix},\qquad \forall z\in \gamma_b,
\nonumber\\
Y(z_+) = Y(z_-) \begin{bmatrix}
1&\varkappa Q(z_+) {\rm e}^{\frac {ns}z}\\
0 & 1
\end{bmatrix}, \qquad \forall z\in \Gamma_b^1,\qquad
\text{where}\nonumber\\
\label{defkappa}
\varkappa:=
 {\rm e}^{-2{\rm i}\pi { {\rho}}}-1.
\end{gather}
\item[$(3)$] As $z\to \infty$, the matrix $Y_n(z)$ admits an asymptotic expansion of the form~\eqref{asympY}.
\end{enumerate}
\end{problem}
The first transformation is the same as in \eqref{defW} and leads to the RHP for $W$ as follows.
\begin{problem}
\label{RHPWs>0}The matrix $W(z)$ is analytic in $\C \setminus \gamma = \C\setminus (\Gamma_m\cup \Gamma_c)$ and satisfies
\begin{gather*}
W(z_+)= W(z_-) J_W(z), \qquad z\in \gamma,\qquad
W(z) = \1 + \mathcal O\le(\frac 1 z \ri), \qquad z\to \infty,
\end{gather*}
where
\begin{gather*}
J_W(z)
=\begin{bmatrix}
{\rm e}^{-\frac n2\le( \varphi(z_+)-\varphi(z_-)\ri) } & Q(z) {\rm e}^{ \frac n2 \le(\varphi(z_+)+\varphi(z_-)\ri)}\\
0 & {\rm e}^{\frac n2 \le(\varphi(z_+)-\varphi(z_-) \ri)}
\end{bmatrix},\qquad z\in \Gamma\setminus \Gamma_{b}^1,
\\
J_W(z)
=\begin{bmatrix}
{\rm e}^{-\frac n2\le( \varphi(z_+)-\varphi(z_-)\ri) } & \varkappa Q(z_+) {\rm e}^{ \frac n2 \le(\varphi(z_+)+\varphi(z_-)\ri)}\\
0 & {\rm e}^{\frac n2 \le(\varphi(z_+)-\varphi(z_-) \ri)}
\end{bmatrix},\qquad z\in \Gamma_{b}^1.
\end{gather*}
\end{problem}
The second transformation is also similar to \eqref{defT}. However, from this point onwards, the details of the transformation depend on the case $\Re(s)>0$ or $\Re(s)<0$ and thus are given in Sections~\ref{secmodels>0} and~\ref{secmodels<0}, respectively.

We will not carry out completely the error analysis, which would require the construction of appropriate local parametrices near the points $z= b, a_\pm, 1$. These are more or less known in the literature. We mention, for the sake of the reader with experience in the Deift--Zhou method, that\looseness=-1
\begin{enumerate}\itemsep=0pt
\item[(1)] Near the points $z= a_\pm, b$ the local parametrices are constructed from Airy functions \cite{BertolaBothner,DKMVZ}.
\item[(2)] Near the point $z=1$ the local parametrix is constructed in terms of functions \cite{VanlessenStrong}.
\end{enumerate}
We instead focus on the construction of the global (outer) parametrix which solves a {\it model problem}. The pragmatic reason is that this computation will produce a formula for the approximate location of the zeros of $\mathcal T_n(s)$ which can be actually tested numerically, and whose result is evident in Figure~\ref{Figone}.
Some details on the construction of the local parametrices are contained in Section~\ref{paramg1} and Appendix~\ref{SecA3}.

\subsubsection[Second transformation, model problem and its solution: The case $\Re(s)>0$]{Second transformation, model problem and its solution: The case $\boldsymbol{\Re(s)>0}$}
\label{secmodels>0}

 With reference to Figure~\ref{Contoursinside}, left pane, we denote by $\Lambda_{\pm}^{(x,y)}$ the region bounded by $\Gamma_x^y$ and $\scr L_\pm^{(x,y)}$, for $x,y\in \{a_+,a_-,b,1\}$ and we call them {\it lens regions} (as opposed to their boundaries, which we refer to as the lens arcs).

{\bf Definition of $\boldsymbol{T_0}$ for $\boldsymbol{\Re(s)>0}$.}
We refer to the left pane of Figure~\ref{Contoursinside} and then set
\[
T_0(z):=
\begin{cases}
\ds W(z) \begin{bmatrix}
1 & 0 \\
\displaystyle\frac{ {\rm e}^{-n\varphi(z)}}{ Q(z)} & 1
\end{bmatrix},& z\in \Lambda_-^{(a_\pm,b)},
\vspace{1mm}\\
W(z) \begin{bmatrix}
1 & 0 \\
\displaystyle-\frac{ {\rm e}^{-n\varphi(z)}}{ Q(z)} & 1
\end{bmatrix},& z\in \Lambda_+^{(a_\pm,b)}.
\end{cases}
\]
 On the other hand, for the lens $\Lambda_\pm^{(b,1)}$ adjacent to $\Gamma_b^1$, we set instead
\be
\label{defTb1}
T_0(z):=
\begin{cases}
\ds W(z) \begin{bmatrix}
1 & 0 \\
\displaystyle\frac{ {\rm e}^{-n\varphi(z)-2{\rm i}\pi { {\rho}}}}{\varkappa Q(z)} & 1
\end{bmatrix},& z\in \Lambda_-^{(b,1)},
\vspace{1mm}\\
W(z) \begin{bmatrix}
1 & 0 \\
\displaystyle-\frac{ {\rm e}^{-n\varphi(z)}}{\varkappa Q(z)} & 1
\end{bmatrix}, & z\in \Lambda_+^{(b,1)}.
\end{cases}
\ee
The main arcs are as in Figure~\ref{Contoursinside}, left pane, thus
\[
\Gamma_m = \Gamma_b^1 \cup \Gamma_{a_-}^b \cup \Gamma_{b}^{a_+}, \qquad
\Gamma_c = \Gamma_{a_+}^{a_-}.
\]
Recall from Section~\ref{gfuncg1} that we have defined
\[
\varphi(z;s) := \int_{a_-}^z \frac 2 {w^2} \sqrt{w^2 + \frac {s^2}4 + \frac {Aw^2}{w-1}}\d w,
\]
with the constant $A$ determined by the condition of all periods being purely imaginary and the integration path chosen in $\C \setminus \Gamma_m$.
The effective potential satisfies the following boundary value relations:
\begin{gather}
\varphi(z_+)+ \varphi(z_-) = 0, \qquad z\in \Gamma_{a_-}^b,
\la{547}\\
\varphi(z_+)+ \varphi(z_-) = 2\Omega_1, \qquad z\in \Gamma_{b}^1,\\
\varphi(z_+)+ \varphi(z_-)= 2\Omega_2, \qquad z\in \Gamma_{b}^{a_+}, \\
\varphi(z_+)- \varphi(z_-) = 4{\rm i}\pi, \qquad z\in \Gamma_{c}\la{550},
\end{gather}
where the constants $\Omega_1$, $\Omega_2$ are given by
\be
\label{defOmega12}
\Omega_1:= \int_{b}^{a_+} \varphi'(z_+) \d z \in {\rm i}\R_-, \qquad
\Omega_2:= \int_{b}^{1} \varphi'(z_-) \d z \in {\rm i}\R_+.
\ee
From Cauchy's residue theorem, one also has
\[
2\int_{a_-}^{b} \varphi'(z_-)\d z +2\int_{b}^{1} \varphi'(z_-)\d z +2\int_{b}^{a_+} \varphi'(z_-)\d z =4{\rm i}\pi,
\]
with all the integrals in ${\rm i}\R_+$.

A direct computation, using also the properties \eqref{547}--\eqref{550} shows that the matrix $T_0(z)$ satisfies the following RHP.

\begin{problem}
\label{RHPT0g1}
The matrix $T_0(z)$ satisfies the conditions
\[
T_0(z_+)= T_0(z_-) J_{T_0}(z), \qquad z\in \gamma,\qquad
T_0(z) = \1 + \mathcal O\le(\frac 1 z \ri), \qquad z\to \infty,
\]
where
\be
\label{JT0s>0}
J_{T_0}(z) =
\begin{cases}
J_{_W}(z), & z\in \Gamma_c = \Gamma_{a_+}^{a_-},\\
 \begin{bmatrix}
1 & 0 \\
\displaystyle \frac{ {\rm e}^{-n\varphi(z)} }{ Q(z)} & 1
\end{bmatrix},
& z\in \scr L_+^{(b,a_+)}\cup \scr L_-^{(b,a_+)}\cup \scr L_-^{(a_-,b)}\cup \scr L_+^{(a_-,b)},
\vspace{1mm}\\
 \begin{bmatrix}
1 & 0 \\
\displaystyle \frac{ {\rm e}^{-n\varphi(z)} }{\varkappa Q(z)} & 1
\end{bmatrix},
& z\in \scr L_+^{(b,1)},
\vspace{1mm}\\
 \begin{bmatrix}
1 & 0 \\
\displaystyle \frac{ {\rm e}^{-n\varphi(z)} }{ Q(z)} \le(1 +\frac 1 \varkappa\ri) & 1
\end{bmatrix},
& z\in \scr L_b^{c_+},
\vspace{1mm}\\
 \begin{bmatrix}
1 & 0 \\
\displaystyle \frac{ {\rm e}^{-n\varphi(z)} }{ Q(z)} \left(\frac {{\rm e}^{-2{\rm i}\pi { {\rho}}} } \varkappa -1\right) & 1
\end{bmatrix},
& z\in \scr L_b^{c_-},
\vspace{1mm}\\
\begin{bmatrix}
1 & 0 \\
\displaystyle \frac{ {\rm e}^{-n\varphi(z)-2{\rm i}\pi { {\rho}}} }{\varkappa Q(z)} & 1
\end{bmatrix},
& z\in \scr L_-^{(b,1)},
\vspace{1mm}\\
\begin{bmatrix}
 0 & \varkappa Q(z_+) {\rm e}^{n\Omega_1}\\
\displaystyle -\frac {{\rm e}^{-n\Omega_1}} {\varkappa Q(z_+)} & 0
\end{bmatrix},
 & z\in \Gamma_b^{1},
\vspace{1mm}\\
\begin{bmatrix}
 0 & Q(z) {\rm e}^{n\Omega_2}\\
\displaystyle -\frac {{\rm e}^{-n\Omega_2}} { Q(z)} & 0
\end{bmatrix},
 & z\in \Gamma_b^{a_+},
\vspace{1mm}\\
\begin{bmatrix}
 0 & Q(z) \\
\displaystyle -\frac {1} { Q(z)} & 0
\end{bmatrix},
 & z\in \Gamma_{a_-}^{b}.
\end{cases}
\ee
\end{problem}

{\bf Model problem.}
The model problem that we have to solve is the following one.

\begin{problem}
\label{RHPMs>0}
Find a matrix-valued function $M_{_Q}$, analytic in $\C\setminus \Gamma_m$ such that
\begin{gather}
M_{_Q}(z_+)= M_{_Q}(z_-) \begin{bmatrix}
0 & Q(z) \\
\displaystyle\frac{-1}{Q(z)} & 0
\end{bmatrix},\qquad z\in \Gamma_{a_-}^{b},
\nn\\
M_{_Q}(z_+)= M_{_Q}(z_-) \begin{bmatrix}
0 & Q(z) {\rm e}^{n\Omega_2}\\
\displaystyle \frac{-{\rm e}^{-n\Omega_2}}{Q(z)} & 0
\end{bmatrix},\qquad z\in \Gamma_b^{a_+},
\nn\\
M_{_Q}(z_+)= M_{_Q}(z_-) \begin{bmatrix}
0 & \varkappa Q(z_+) {\rm e}^{n\Omega_1}\\
\displaystyle \frac{-{\rm e}^{-n\Omega_1}}{\varkappa Q(z_+)} & 0
\end{bmatrix}, \qquad z\in \Gamma_b^{1},
\label{jumpMQ}
\end{gather}
with $\varkappa$ as in \eqref{defkappa} and $Q$ as in \eqref{defQ}.
Furthermore, the following local behaviours hold:
\begin{gather*}
M_{_Q}(z) = \1 + \mathcal O\bigl(z^{-1}\bigr), \qquad |z|\to \infty,\\
M_{_Q}(z) = \mathcal O\le(\frac 1 {(z-q)^\frac 1 4}\ri), \qquad q= 1,b,a_\pm, \qquad z\to q.
\end{gather*}
\end{problem}
To further normalize the problem, we need to construct a Szeg\H{o} function, along the line of what we already have done in the case of the outside of ${\rm EoT}$.

{\bf The Szeg\H{o} function.}
Let $R(z)$ denote the radical function
\begin{align}
R(z) &=\sqrt{(z-1)(z-b)(z-a_+)(z-a_-)} \nonumber\\
&= \sqrt{
z^4 + (A-2)z^3 + \le(\frac {s^2}4 - A+1\ri) z^2 - \frac {s^2}2 z + \frac {s^2}4},\label{defR}
\end{align}
with the branch cuts chosen along the three arcs of $\Gamma_m = \Gamma_b^1\cup \Gamma_{b}^{a_+} \cup \Gamma_{a_-}^b$, and with the overall determination such that $R(z) \simeq z^2$ as $z\to\infty$.
Consider the following expression:
\begin{align}
S(z) ={}& R(z) \left(
\int_b^1\frac{\ln Q(w_+) + \ln \varkappa-\nu}{R(w_+)(w-z)} \frac{\d w}{2{\rm i}\pi}
+
\int_{a_-}^b\frac{\ln Q(w)}{R(w_+)(w-z)} \frac{\d w}{2{\rm i}\pi}\right.\nonumber\\
&\left.
+
\int_b^{a_+} \frac{\ln Q(w) }{R(w_+)(w-z)} \frac{\d w}{2{\rm i}\pi}
 \right).\label{defSzegos>0}
\end{align}
The constant $\nu$ is chosen so that the function $S(z)$ is bounded as $|z|\to\infty$.
We can actually simplify \eqref{defSzegos>0} quite a bit. Consider, for example, the integral
\[
I_1:= \int_b^1\frac{\ln Q(w_+) + \ln \varkappa-\nu}{R(w_+)(w-z)} \frac{\d w}{2{\rm i}\pi}.
\]
Since, along $\Gamma_b^1$ we have $Q(w_+) = Q(w_-) {\rm e}^{2{\rm i}\pi { {\rho}}}$ and $R(w_+)=-R(w_-)$, we can convert $I_1$ into
\[
I_1 = \frac 1 2 \int_{b}^1 \frac {\ln Q(w_+) \d w}{R(w_+)(w-z)2{\rm i}\pi} + \frac 1 2 \int_{1}^b \frac {\ln Q(w_-) \d w}{R(w_-)(w-z)2{\rm i}\pi} + \int_{b}^1 \frac{{\rm i}\pi { {\rho}} + \ln \varkappa -\nu \d w}{R(w_+)(w-z)2{\rm i}\pi}.
\]
Along the same lines, denoting the other two integrals in \eqref{defSzegos>0} by $I_2$, $I_3$, we have
\begin{gather*}
I_2 = \frac 1 2\int_{a_-}^b \frac {\ln Q(w) \d w}{R(w_+)(w-z)2{\rm i}\pi} + \frac 1 2\int^{a_-}_b \frac {\ln Q(w) \d w}{R(w_-)(w-z)2{\rm i}\pi},
\\
I_3 = \frac 1 2\int^{a_+}_b \frac {\ln Q(w) \d w}{R(w_+)(w-z)2{\rm i}\pi} + \frac 1 2\int_{a_+}^b \frac {\ln Q(w) \d w}{R(w_-)(w-z)2{\rm i}\pi}.
\end{gather*}
Adding $I_1+I_2+I_3$, we observe that the integrals involving $\ln Q(w)$ form a closed loop from $b$ that goes around the whole $\Gamma_m$ in the clockwise direction without intersecting the cut of $Q(z)$.
Using \eqref{defQ}, we observe that we have
\[
\ln Q(z_+) = \ln Q(z_-) + \begin{cases}
 2{\rm i}\pi{K},& z\in (-\infty,0),\\
2{\rm i}\pi{\rho}, & z\in (0,1).
\end{cases}
\]
By using Cauchy's theorem, we conclude that
\begin{align}
S(z)
={}&\frac 1 2\ln Q(z)+ R(z) \left( \int_{-\infty}^0 \frac { -{K}\d w}{2R(w)(w-z)}\right. \nonumber\\
&\left.-\frac 1 2 \int_b^0 \frac {{ {\rho}} \d w}{R(w)(w-z)}
+\int_b^1 \frac {({\rm i}\pi { {\rho}} + \ln \varkappa -\nu) \d w}{R(w_+)(w-z)2{\rm i}\pi}
\right).
\label{defSzegos>0fin}
\end{align}
Observe that the branch cut of $\ln Q(z)$ on $(-\infty,0)$ is compensated by the branch cut of the first integral in the above expression, so that $S(z)$ is actually analytic across the ray $(-\infty,0)$.
From~\eqref{defSzegos>0fin}, it is then simple to ascertain that to make $S(z)$ bounded at infinity we need to choose the value of $\nu$ as follows (see \eqref{defkappa} for $\varkappa$):
\be
\label{defnus>0}
\nu = {\rm i}\pi { {\rho}}\le(1- \frac {\int_b^0 \frac {\d w}{R(w)}}{\int_{b}^1 \frac {\d w}{R(w_+)}} \ri) + \ln \varkappa.
\ee
The expression \eqref{defSzegos>0} makes it clear, through the use of the Sokhostki--Plemelji formula, that the following properties of the Szeg\H{o} function hold.

\begin{Proposition}[Szeg\H{o} function for $\Re(s)>0$]
\label{propSzegos>0}
The function $S(z)$ defined in \eqref{defSzegos>0} or equivalently \eqref{defSzegos>0fin} is analytic and locally bounded on $\C\cup \{\infty\}\setminus \Gamma_m$, and with boundary conditions
\[
S(z_+)+ S(z_-) =
\begin{cases}
\ln \le(\varkappa Q(z_+)\ri) - \nu, & z\in \Gamma_b^1,\\
\ln Q(z), & z\in \Gamma_{a_-}^b \cup \Gamma_b^{a_+}.
\end{cases}
\]
Furthermore, $S(z)$ is bounded near $z=b,a_\pm$ and near $z=1$ it has the behaviour {\rm\cite[\emph{Chapter}~5]{Gakhov}}
\[
S(z) = -\frac { {\rho}} 2\ln(z-1) + \mathcal O(1).
\]
\end{Proposition}

\begin{figure}[t]\centering
\includegraphics[width=0.38\textwidth] {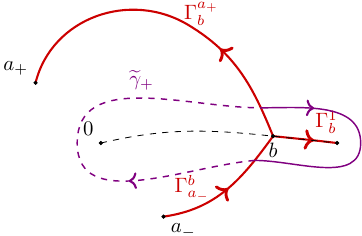}
\qquad
\includegraphics[width=0.38\textwidth] {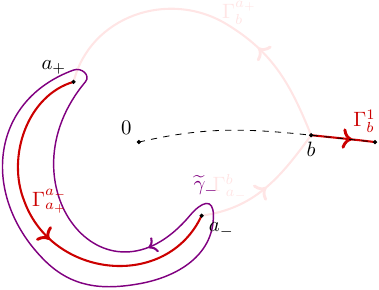}

\vspace{-1mm}

\caption{The alternative computation of the Szeg\H{o} function and the contour $\wt \gamma$. For $\Re (s)>0$ in the left pane, for $\Re(s)<0$ in the right pane. The dashed part of the contour $\wt \gamma$ corresponds to the second sheet of the radical function $R(z)$ (the one where $R(z) \simeq -z^2$ at infinity). The two contours are homologically equivalent.}
\label{figSzego}
\end{figure}

{\bf Alternative description/derivation of the Szeg\H{o} function.}
Proceeding from the desired properties in Proposition \eqref{propSzegos>0} and differentiating, we obtain
\[
S'(z_+)+S'(z_-) = \frac {Q'(z)}{Q(z)}, \qquad z\in \Gamma_m.
\]
Then one can seek a solution of the form
\be
\label{S'}
S'(z)=\frac 1 2 \frac {Q'(z)}{Q(z)} + \eta(z),
\ee
where $\eta(z)$ must be a function analytic off $\Gamma_m$ with $\eta(z_+)+ \eta(z_-)= 0$ on $\Gamma_m$ and such as to cancel out the singularities of $\frac {Q'}Q$ in \eqref{S'} outside of $\Gamma_m$.
Since $\frac {Q'}Q = \frac { {\rho}}{z-1} + \frac{ {K} - {\rho}} z$, we see that
\[
\eta(z) = \frac { {\rho} - {K}} 2 \frac{R(0)}{z R(z)} - \frac C{R(z)},
\]
which cancels the pole at $z=0$ in the expression \eqref{S'}. Here $R$ is given by \eqref{defR}.
The constant~$C$ is determined by the condition that $S(z_+)+S(z_-)= \ln Q(z)$ on the arcs $\Gamma_{b}^{a_\pm}$, which implies that integral of $\eta$ along the contour $\wt \gamma$ on the Riemann surface of $W^2=R(z)^2$ (see Figure~\ref{figSzego}, left pane) vanishes. Specifically, it is given by
\be
C= \frac {{( {\rho} - {K})} R(0)}2\frac{\oint_{\wt \gamma_+} \frac {\d z}{z W}}{\oint_{\wt \gamma} \frac {\d z}{W}}.
\la{673}
\ee
Using the explicit expression of $R$ in \eqref{defR}, we note that $R(0) = -\frac{s}2$ and can simplify \eqref{673} to
\[
C=C_+= \frac { ( {K} - {\rho})}{4} \frac {s \oint_{\wt \gamma_+} \frac {\d z}{z W} }{\int_b^1 \frac {\d z}{ R(z_+)}},
\]
where we have used the fact that {\samepage
\[
\oint_{\wt \gamma} \frac {\d z}{R(z)} = - \int_{b}^1 \frac {\d z}{R(z_+)}
\] (see Figure~\ref{figSzego}, left pane).}

When $\Re(s)<0$ instead, we have to ask that the antiderivative of \eqref{S'} is continuous on~$\C \setminus \Gamma_m$ and then the contour $\wt \gamma$ is different.
In this case then the constant $C$ is given by
\be
C=C_-= \frac {{ ( {\rho} - {K})} R(0)}2\frac{\oint_{\wt \gamma_-} \frac {\d z}{z W}}{\oint_{\wt \gamma_-} \frac {\d z}{W}}
=\frac {{ ( {K} - {\rho})} s }4\frac{\oint_{a_+}^{a_-} \frac {\d z}{z R(z_+)}}{\oint_{a_+}^{a_-} \frac {\d z}{R(z_+)}}.\label{Cs<0}
\ee
We observe that, in fact, the contours $\wt\gamma_\pm$ are homologically equivalent on the Riemann surface punctured at the two points $z=0$ on both sheets. However, the definition of the domain of~$R(z)$ has been chosen with different branch cuts. Keeping this in mind, we can use for $C$ the expression~\eqref{Cs<0} in both cases with the understanding that in the case $\Re (s)>0$ the indication of the boundary value $R(z_+)$ is irrelevant.

In either case, for future reference, we observe that (recalling the definition of $R$ \eqref{defR})
\[
S'(0) = -\frac { {\rho}} 2 \frac {R'(0)}{R(0)} - \frac {C_\pm}{R(0)}
=\frac { {\rho}} 2 +
\frac {{ ( {K} - {\rho})} }2\frac{\oint_{a_+}^{a_-} \frac {\d z}{z R(z_+)}}{\oint_{a_+}^{a_-} \frac {\d z}{R(z_+)}}.
\]

With the Szeg\H{o} function at hand, we can normalize the Riemann--Hilbert Problem~\ref{RHPMs>0}.
We define a new model problem
\bea
M(z):= {\rm e}^{S(\infty) \s_3} M_{_{Q}}(z) {\rm e}^{-S(z) \s_3}.
\label{MtoMQ}
\eea
The effect of this normalization is to turn the jump matrices \eqref{jumpMQ} into constant ones as detailed in the following problem.

\begin{problem}\label{RHPM}
Find a matrix-valued function $M(z)$ analytic in $\C \setminus \Gamma_m$ such that
\begin{gather}
M(z_+)= M(z_-) \begin{bmatrix}
0 & 1 \\
-1 & 0
\end{bmatrix}, \qquad z\in \Gamma_{a_-}^{b},
\nn
\\
M(z_+)= M(z_-) \begin{bmatrix}
0 & {\rm e}^{n\Omega_2}\\
{-{\rm e}^{-n\Omega_2}} & 0
\end{bmatrix},\qquad z\in \Gamma_b^{a_+},
\nn
\\
M(z_+)= M(z_-) \begin{bmatrix}
0 & {\rm e}^{n\Omega_1+\nu}\\
-{\rm e}^{-n\Omega_1-\nu}& 0
\end{bmatrix},\qquad z\in \Gamma_b^{1}
\label{jumpM}
\end{gather}
and satisfying
\begin{gather}
M(z) = \1 + \mathcal O\bigl(z^{-1}\bigr),\qquad |z|\to \infty,\label{boundinfty}\\
M(z) = \mathcal O\le(\frac 1 {(z-q)^\frac 1 4}\ri), \qquad q= 1,b,a_\pm, \quad z\to q.
\label{boundq}
\end{gather}
\end{problem}
The solution of the Riemann--Hilbert Problem~\ref{RHPM} is relatively standard and it requires the use of the Jacobi theta function.

{\bf Solution of the Riemann--Hilbert Problem~\ref{RHPM}.}
With the same definition of $R$ in~\eqref{defR}, we also define
\be
h(z):= \frac 1 {\le((z-1)(z-b)(z-a_+)(z-a_-)\ri)^\frac 1 4},
\label{defh}
\ee
with the same branch cuts as $R$ along $\Gamma_m$ and the determination chosen so that $h(z) \simeq \frac 1 z $ as~${z\to \infty}$.
A careful analysis of the phases shows that
\begin{gather}
\label{hjumps}
h(z_+) ={\rm i} h(z_-),\qquad z\in \Gamma_{a_-}^b,\qquad
h(z_+) =-{\rm i} h(z_-), \qquad z\in \Gamma_b^1\cup \Gamma_b^{a_+}.
\end{gather}
We also need the \emph{Abel map}. Let
\begin{gather}
\label{defomega12}
\omega_1:= \int_b^1 \frac {\d z}{R(z_+)}, \qquad \omega_2:= \int_b^{a_+} \frac {\d z}{R(z_-)},\\
\label{defAbel}
\Abel(z):= \int_{a_-}^z \frac {\d w}{2\omega_1 R(w)}, \qquad \tau:= \frac {\omega_2}{\omega_1},
\end{gather}
where the contour of integration runs in the simply connected domain $\C\setminus \Gamma $.
A direct inspection reveals the following.

\begin{Lemma}[properties of the Abel map]\label{lemmaAbel}
The following relations hold:
\[
\Abel(z_+)=-\Abel(z_-) +
\begin{cases}
\phantom{-}0,& z\in \Gamma_{a_-}^b,\\
-1,& z\in \Gamma_b^{a_+},\\
-\tau, & z\in \Gamma_b^1.
\end{cases}
\]
\end{Lemma}
Let $\vartheta(u;\tau)$ be the Riemann (Jacobi) theta function (also denoted $\theta_4$ in \href{https://dlmf.nist.gov/20.2.E4}{DLMF 20.2.4} but with a different normalization for $u$)
\[
\vartheta(u;\tau) := \sum_{n\in \Z} {\rm e}^{{\rm i}n^2\pi \tau + 2{\rm i}\pi n u}.
\]
The function $\vartheta$ vanishes at $u = \frac {\tau + 1}2 + k + \ell \tau$ for all $k, \ell\in \Z$ and satisfies the quasi-periodicity properties
\be
\vartheta(u+k+ \ell \tau;\tau) = {\rm e}^{-2{\rm i}\pi \ell u - {\rm i}\pi \ell^2 \tau}\vartheta(u;\tau).
\label{thetaperiods}
\ee
Consider the following two row-vectors:
\begin{gather}
\bs \phi (z;\infty)= [\phi_1(z;\infty), \phi_2(z;\infty)],\qquad
\phi_1(z;\infty)= \frac { \vartheta\le(\Abel(z)-\Abel(\infty) - \frac {\tau+1}2 +
G\ri)h(z)}{\vartheta\le(\Abel(z)-\Abel(\infty) - \frac{\tau+1}2\ri)}{\rm e}^{{\rm i}\pi K \Abel(z)},
\nonumber\\
\phi_2(z;\infty)= \frac {-{\rm i} \vartheta\le(-\Abel(z)-\Abel(\infty) - \frac {\tau+1}2 +
G\ri)h(z)}{\vartheta\le(-\Abel(z)-\Abel(\infty) - \frac{\tau+1}2\ri)}{\rm e}^{-{\rm i}\pi K \Abel(z)}
\label{rowphi}
\end{gather}
and
\begin{gather}
\bs\psi(z;\infty) = [\psi_1(z;\infty), \psi_2(z;\infty)],\nn\\
\psi_1(z;\infty) = \frac {-{\rm i} \vartheta\le(\Abel(z)+\Abel(\infty) - \frac {\tau+1}2 +
G\ri)h(z)}{\vartheta\le(\Abel(z)+\Abel(\infty) - \frac{\tau+1}2\ri)}{\rm e}^{{\rm i}\pi K \Abel(z)},
\nn\\
\psi_2(z;\infty) = \frac { -\vartheta\le(-\Abel(z)+\Abel(\infty) - \frac {\tau+1}2 +
G\ri)h(z)}{\vartheta\le(-\Abel(z)+\Abel(\infty) - \frac{\tau+1}2\ri)}{\rm e}^{-{\rm i}\pi K \Abel(z)}.
\label{rowpsi}
\end{gather}
Furthermore, all entries are bounded by \eqref{boundq} by the very definition of $h(z)$ \eqref{defh} and the fact that $\vartheta$ is an entire function with zeros only at the half-period.

Let us investigate the behaviour of $\bs \phi$, $\bs \psi$ near $z=\infty$.
Observe that $\phi_2$ and $\psi_1$ tend to zero as~${z\to\infty}$ \big(due to the factor $h(z) = \mathcal O\bigl(z^{-1}\bigr)$\big), while $\phi_1$, $\psi_2$ have to be computed using l'Hopital's rule given that the $\vartheta$ in the denominator also tends to zero.

Then a direct computation shows that
\[
\lim_{z\to\infty} \begin{bmatrix}
\bs\phi\\
\bs \psi\end{bmatrix} =
\begin{bmatrix} \displaystyle \frac {\vartheta\le(G- \frac {\tau + 1}2\ri)}{\omega_1 \vartheta'\le(-\frac {\tau + 1}2\ri)}{\rm e}^{{\rm i}\pi K\Abel(\infty)} & 0
\\0 & \displaystyle \frac {\vartheta\le(G- \frac {\tau + 1}2\ri)}{\omega_1 \vartheta'\le(-\frac {\tau + 1}2\ri)}{\rm e}^{-{\rm i}\pi K\Abel(\infty)}
\end{bmatrix}.
\]

Thus, as long as the common factor $\vartheta\le(G- \frac {\tau + 1}2\ri)\neq 0$ we can define (the dependence on the parameters $G$, $K$ is understood in the right side)
\be
\label{defwhM}
\wh M(z;G,K,\infty):= \frac {\omega_1 \vartheta'\le(-\frac{\tau+1}2\ri) {\rm e}^{-{\rm i}\pi K \Abel(\infty)\s_3} }{\vartheta\le(G- \frac {\tau + 1}2\ri)}
\begin{bmatrix}
\bs \phi(z;\infty)\\
\bs \psi(z;\infty)
\end{bmatrix}.
\ee
\begin{Remark}\label{remarknormalizationpoint}
We have emphasized in the notation that $\infty$ as a point plays a role in the expression. We will use later the same formula, but replacing $\infty$ with the point $z=0$.
It is important to point out that, irrespectively of what point we replace instead of $\infty$, the matrix satisfies the same boundary relations \eqref{jumpwhM} below, and also the same behaviour near the branch-points~${z=1,b,a_\pm}$. The only difference is that if we replace $\infty$ by a point $z_0$, the matrix~$\smash{\wh M(z;G,K,z_0)}$ will then vanish at infinity and have a simple pole at $z=z_0$ with singular part proportional to the identity matrix
\[
\lim_{z\to\infty} \wh M(z;G,K,z_0)={\bs 0}, \qquad \wh M(z;G,K,z_0) =\frac{ \omega_1h(0) }{\Abel'(z_0)}\frac \1{z-z_0} + \mathcal O(1), \qquad z\to z_0.
\]
 We are going to use this observation later.
\end{Remark}
It is a direct verification using the definition of the Abel map \eqref{defAbel}, the properties of the theta function \eqref{thetaperiods}, and the jump relation of $h$ \eqref{hjumps}, that each row $\bs \phi$, $\bs \psi$ satisfies the three boundary value relations similar to those in \eqref{jumpM}
\begin{gather}
\wh M(z_+)= \wh M(z_-) \begin{bmatrix}
0 & 1 \\
-1 & 0
\end{bmatrix}, \qquad z\in \Gamma_{a_-}^{b},
\nn
\\
\wh M(z_+) = \wh M(z_-) \begin{bmatrix}
0 & -{\rm e}^{{\rm i}\pi K}\\
{\rm e}^{-{\rm i}\pi K} & 0
\end{bmatrix}, \qquad z\in \Gamma_b^{a_+},
\nn
\\
\wh M(z_+) = \wh M(z_-) \begin{bmatrix}
0 & -{\rm e}^{-2{\rm i}\pi(G - \frac {K\tau}2) }\\
{\rm e}^{2{\rm i}\pi(G - \frac {K\tau}2) }& 0
\end{bmatrix}, \qquad z\in \Gamma_b^{1}.
\label{jumpwhM}
\end{gather}

By matching these to the boundary relations \eqref{jumpM}, we deduce the following.

\begin{Proposition}\label{propM}
The solution of the Riemann--Hilbert Problem~{\rm\ref{RHPM}} is given by $M(z) = \wh M(z;G,K,\infty)$ in~\eqref{defwhM} with the values of the constants $G$, $K$ given by
\be
\label{defGs>0}
G= \frac {1}{2{\rm i}\pi} \le(-n\Omega_1 - \nu + n \Omega_2\tau\ri) + \frac {\tau+1}2,\qquad
K = \frac {n\Omega_2}{{\rm i}\pi} + 1.
\ee
\end{Proposition}

\begin{proof}
Comparing the boundary relations \eqref{jumpwhM} and \eqref{jumpM}, we obtain the system
\[
\ds G - \frac {K \tau}2 = - \frac{n\Omega_1+\nu}{2{\rm i}\pi} + \frac 1 2,
\qquad
\ds K = \frac{n\Omega_2}{{\rm i}\pi} +1,
\]
from which the relations follow.
\end{proof}

The solvability of the Riemann--Hilbert Problem~\ref{RHPM} and hence of the Riemann--Hilbert Problem~\ref{RHPMs>0} depends entirely on the non-vanishing of the expression
\be
\vartheta\le(\frac {n\Omega_1 + \nu - n \Omega_2\tau }{2{\rm i}\pi} \ri)
\label{674}
\ee
with $\Omega_1$, $\Omega_2$ defined by \eqref{defOmega12} and $\nu$ by \eqref{defnus>0}
Considering that the zeros of $\vartheta$ are for $u=\frac {\tau +1}2 + \ell + k\tau$, with $k,\ell\in \Z$, we obtain the quantization conditions for the periods of $\varphi$
\begin{gather}
(H_1,H_2)\in \Z^2 \nn,\\
H_1:= \frac{n\Omega_1}{2{\rm i}\pi} + \Re\le(\frac {\nu} {2{\rm i}\pi} \ri) - \frac {\Re \tau}{\Im\tau} \Im\le(\frac {\nu}{2{\rm i}\pi}\ri) + \frac 1 2,\nn
\\
H_2:= \frac{n\Omega_2}{2{\rm i}\pi} - \frac {\Im\le(\frac{\nu}{2{\rm i}\pi}\ri)}{\Im \tau} + \frac 1 2.\label{quantcond}
\end{gather}
Observe that both $H_1$, $H_2$ are real functions of $s$ since $\Omega_1,\Omega_2\in {\rm i}\R$. The modular parameter $\tau$ is also, in a very implicit way, a function of $s$ since the branch-points of the radical $R$ \eqref{defR} are determined by the Boutroux conditions \eqref{Boutroux_cond}. Ditto for $\nu$, which depends on $s$ implicitly via the formula \eqref{defnus>0}.

\begin{Remark}[explanation of Figure~\ref{Figone}.]
\label{remexp2}
The two quantization conditions \eqref{quantcond} can be interpreted as describing a mesh of level sets of the two functions $H_1$, $H_2$ (both functions of the complex parameter $s$). In the various panels in Figure~\ref{Figone} these are precisely forming the mesh of curves that populate the interior of ${\rm EoT}$. The intersection points of this mesh are the points where the quantization conditions \eqref{quantcond} hold, and hence where the solution of the model problem Riemann--Hilbert Problem~\ref{RHPMs>0} ceases to exist. They also approximate very precisely the zeros of the rational solutions (in fact better than expected), with only obvious deviations near the boundaries of the two halves of ${\rm EoT}$, since there the elliptic curve of the radical $R$ \eqref{defR} degenerates (i.e., two branch-points come together).
\end{Remark}

\subsubsection[Second transformation, model problem and its solution: The case $\Re (s)<0$]{Second transformation, model problem and its solution: The case $\boldsymbol{\Re (s)<0}$}
\label{secmodels<0}

 With reference to Figure~\ref{Contoursinside}, right pane, we denote by \smash{$\Lambda_{\pm}^{(x,y)}$} the region bounded by $\Gamma_x^y$ and \smash{$\scr L_\pm^{(x,y)}$}, for $x,y\in \{a_+,a_-,b,1\}$ and we call them {\it lens regions} (as opposed to their boundaries, which we refer to as the lens arcs).

{\bf Definition of $\boldsymbol{T_0}$ for $\boldsymbol{\Re(s)<0}$.}
We refer to the right pane of Figure~\ref{Contoursinside} and then set
\begin{gather*}
T_0(z):=
\begin{cases}
\ds W(z) \begin{bmatrix}
1 & 0 \\
\displaystyle \frac{ {\rm e}^{-n\varphi(z)}}{ Q(z)} & 1
\end{bmatrix},& z\in \Lambda_-^{(a_+,a_-)},
\vspace{1mm}\\
W(z) \begin{bmatrix}
1 & 0 \\
\displaystyle-\frac{ {\rm e}^{-n\varphi(z)}}{ Q(z)} & 1
\end{bmatrix}, & z\in \Lambda_+^{(a_+,a_-)},
\end{cases}
\end{gather*}
while, for the lens \smash{$\Lambda_\pm^{(b,1)}$} adjacent to $\Gamma_b^1$ we have the same as in \eqref{defTb1}.
 The construction is similar to the previous case, but with the notable difference that now the main arcs $\Gamma_m$ consists of two disjoint arcs (see Figure~\ref{Contoursinside}, right pane)
\[
\Gamma_m = \Gamma_{b}^1 \cup \Gamma_{a_+}^{a_-}, \qquad
 \Gamma_c = \Gamma_b^{a_+} \cup \Gamma_{a_-}^b.
 \]
 We define the effective potential as before
 \[
 \varphi(z;s) := \int_{a_-}^z \frac 2 {w^2} \sqrt{w^2 + \frac {s^2}4 + \frac {Aw^2}{w-1}}\d w,
 \]
 where now, however, the branch cuts of the radical in the integrand consist of the two arcs $\Gamma_{b}^1 $, $ \Gamma_{a_+}^{a_-}$ of Figure~\ref{Contoursinside}, right pane, and $\varphi$ has additionally a branch cut $[1,\infty)$ and another one running along $\Gamma_b^{a_+}$.
 Now the effective potential satisfies the following boundary value relations:\looseness=1
 \begin{gather*}
\varphi(z_+)+\varphi(z_-) = 0, \qquad z\in \Gamma_{a_+}^{a_-},
\\
\varphi(z_+)+\varphi(z_-) = 2\Omega_1, \qquad z\in \Gamma_{b}^{1},
\nn
\\
\varphi(z_+)-\varphi(z_-) = 0, \qquad z\in \Gamma_{a_-}^{b},
\\
 \varphi(z_+)-\varphi(z_-) = 2\Omega_2, \qquad z\in \Gamma^{a_+}_{b},
\\
 \varphi(z_+)-\varphi(z_-) = 4{\rm i}\pi, \qquad z\in [1,\infty),
 \end{gather*}
 where now
 \begin{gather}
 \label{defOmega12bis}
 \Omega_1:= \int_{a_-}^b \varphi'(w)\d w,\qquad \Omega_2 = \int_{a_-}^{a_+} \varphi'(w_+)\d w.
 \end{gather}

A direct computation shows that the matrix $T_0(z)$ satisfies the following Riemann--Hilbert problem.

\begin{problem}\label{RHPT0g1s<0}
The matrix $T_0(z)$ satisfies the conditions
\[
T_0(z_+)= T_0(z_-) J_{T_0}(z), \qquad z\in \gamma,\qquad
T_0(z) = \1 + \mathcal O\le(\frac 1 z \ri), \qquad z\to \infty,
\]
where $($we set for brevity $Q_\pm:= Q(z_\pm)$, $\varphi_\pm := \varphi(z_\pm)$ below$)$
\[
J_{T_0}(z) =
\begin{cases}
J_{_W}(z) =
 \begin{bmatrix}
{\rm e}^{-n\Omega_2} & Q{\rm e}^{\frac n2 (\varphi_++\varphi_-)} \\
0 & {\rm e}^{n\Omega_2}
\end{bmatrix},
 & z\in \Gamma_{b}^{a_+},\\[10pt]
 J_{_W}(z) =
 \begin{bmatrix}
1 & Q{\rm e}^{n \varphi } \\
0 &1
\end{bmatrix},
 & z\in \Gamma^{b}_{a_-},\\[10pt]
 \begin{bmatrix}
1 & 0 \\
\displaystyle\frac{ {\rm e}^{-n\varphi} }{ Q} & 1
\end{bmatrix},
& z\in \scr L_+^{(a_+,a_-)}\cup \scr L_-^{(a_+,a_-)},
\\[10pt]
 \begin{bmatrix}
1 & 0 \\
\displaystyle \frac{ {\rm e}^{-n\varphi} }{\varkappa Q} & 1
\end{bmatrix},
& z\in \scr L_+^{(b,1)},
\\[10pt]
\begin{bmatrix}
1 & 0 \\
\displaystyle \frac{ {\rm e}^{-n\varphi-2{\rm i}\pi { {\rho}}} }{\varkappa Q} & 1
\end{bmatrix},
& z\in \scr L_-^{(b,1)},
\\[10pt]
\begin{bmatrix}
 0 & \varkappa Q_+ {\rm e}^{n\Omega_1}\\
\displaystyle -\frac {{\rm e}^{-n\Omega_1}} {\varkappa Q_+} & 0
\end{bmatrix},
 & z\in \Gamma_b^1,
\\[10pt]
\begin{bmatrix}
 0 & Q \\
 -\frac {1} { Q} & 0
\end{bmatrix},
 & z\in \Gamma_{a_+}^{a_-}.
\end{cases}\]
\end{problem}

\begin{problem}\label{RHPMQs<0}
Find a matrix-valued function $M_{_Q}$, analytic in $\C\setminus \Gamma_m$ such that
\begin{gather*}
M_{_Q}(z_+) = M_{_Q}(z_-), \qquad z\in \Gamma_{a_-}^b\cup [1,\infty),\nn\\
M_{_Q}(z_+)= M_{_Q}(z_-) \begin{bmatrix}
0 & Q(z) \\
\displaystyle \frac{-1}{Q(z)} & 0
\end{bmatrix}, \qquad z\in \Gamma_{a_+}^{a_-},
\\
M_{_Q}(z_+)= M_{_Q}(z_-) \begin{bmatrix}
 {\rm e}^{-n\Omega_2} & 0\\
 0 & {\rm e}^{n\Omega_2}
\end{bmatrix},\qquad z\in \Gamma_b^{a_+},
\\
M_{_Q}(z_+)= M_{_Q}(z_-) \begin{bmatrix}
0 & \varkappa Q(z_+) {\rm e}^{n\Omega_1}\\
\displaystyle \frac{-{\rm e}^{-n\Omega_1}}{\varkappa Q(z_+)} & 0
\end{bmatrix}, \qquad z\in \Gamma_b^{1},
\end{gather*}
with $\varkappa$ as in \eqref{defkappa} and $Q$ as in \eqref{defQ}.
Furthermore, the following local behaviours hold:
\begin{gather*}
M_{_Q}(z) = \1 + \mathcal O\bigl(z^{-1}\bigr), \qquad |z|\to \infty,\\
M_{_Q}(z)= \mathcal O\le(\frac 1 {(z-q)^\frac 1 4}\ri), \qquad q= 1,b,a_\pm, \qquad z\to q.
\end{gather*}
\end{problem}

Like before, to further normalize the problem we need to construct an appropriate Szeg\H{o} function.

{\bf The Szeg\H{o} function.}
Let $R(z)$ denote the radical function \eqref{defR}, but now with the branch cuts of the radical chosen along $\Gamma_m = \Gamma_b^1\cup \Gamma_{a_+}^{a_-}$ (right pane of Figure~\ref{Contoursinside}).
Consider the following expression:
\be
\label{defSzegos<0}
S(z) = R(z) \le(
\int_b^1\frac{\ln Q(w_+) + \ln \varkappa-\nu}{R(w_+)(w-z)} \frac{\d w}{2{\rm i}\pi}
+
\int_{a_+}^{a_-}\frac{\ln Q(w)}{R(w_+)(w-z)} \frac{\d w}{2{\rm i}\pi}
 \ri).
\ee
Following analogous considerations as in the case $\Re (s)>0$ (see Section~\ref{secmodels>0}), we finally obtain the same expression \eqref{defSzegos>0fin},
where the condition that $S(z)$ is bounded at infinity imposes the same constraint on $\nu$ as in \eqref{defnus>0}.
Similarly to Proposition~\ref{propSzegos>0}, we have now the following.

\begin{Proposition}[Szeg\H{o} function for $\Re(s)<0$]\label{propSzegos<0}
The function $S(z)$ defined in \eqref{defSzegos<0} or equivalently \eqref{defSzegos>0fin} is analytic and bounded on $\C\cup \{\infty\}\setminus \Gamma_m$, and with boundary conditions
\[
S(z_+)+ S(z_-) =
\begin{cases}
\ln \le(\varkappa Q(z_+)\ri) - \nu,& z\in \Gamma_b^1,\\
\ln Q(z), & z\in \Gamma_{a_+}^{a_-}.
\end{cases}
\]
Furthermore, $S(z)$ is bounded near $z=b,a_\pm$ and near $z=1$ it has the behaviour
\[
S(z) = -\frac { {\rho}} 2\ln(z-1) + \mathcal O(1).
\]
\end{Proposition}
With this new Szeg\H{o} function, we normalize Riemann--Hilbert Problem~\ref{RHPMQs<0} and define the new model problem
\[
M(z):= {\rm e}^{S(\infty) \s_3} M_{_{Q}}(z) {\rm e}^{-S(z) \s_3}.
\]
The new matrix $M$ solves now the following problem.
\begin{problem}
\label{RHPMs<0}
Find a matrix-valued function $M(z)$ analytic in $\C \setminus \Gamma_m$ such that
\begin{gather}
M(z_+)= M(z_-) \begin{bmatrix}
0 & 1 \\
-1 & 0
\end{bmatrix}, \qquad z\in \Gamma_{a_+}^{a_-},
\nn
\\
M(z_+)= M(z_-) \begin{bmatrix}
{\rm e}^{-n\Omega_2}\\
0 & {\rm e}^{n\Omega_2}
\end{bmatrix},\qquad z\in \Gamma_b^{a_+},
\nn
\\
M(z_+)= M(z_-) \begin{bmatrix}
0 & {\rm e}^{n\Omega_1+\nu}\\
-{\rm e}^{-n\Omega_1-\nu}& 0
\end{bmatrix}, \qquad z\in \Gamma_b^{1}
\label{jumpMs<0}
\end{gather}
and satisfying the same \eqref{boundinfty}, \eqref{boundq}.
\end{problem}

{\bf Solution of Riemann--Hilbert Problem~\ref{RHPMs<0}.}
With the same definition of $R$ in \eqref{defR}, we also set the same definition of $h$ as in \eqref{defh}
where however now the domain consists of~$\C$ minus the branch cuts~${\Gamma_b^1\cup \Gamma_{b}^{a_+}\cup \Gamma_{a_+}^{a_-}}$, and the determination chosen so that $h(z) \simeq \frac 1 z $ as $z\to \infty$.
A careful analysis of the phases of $h$ shows that
\begin{gather*}
h(z_+) = -{\rm i} h(z_-),\qquad z\in \Gamma_{b}^1,\qquad
h(z_+) =-{\rm i} h(z_-),\qquad z\in \Gamma_{a_+}^{a_-},\\
h(z_+) =-h(z_-),\qquad z\in \Gamma_{b}^{a_+}.
\end{gather*}
The Abel map is defined as in \eqref{defAbel} but without the boundary value in the definition of $\omega_2$ in~\eqref{defomega12} since now the contour $\Gamma_{b}^{a_+}$ is not a branch cut of the radical $R$.
This time the Abel map satisfies somewhat different relations, a consequence of the different choice of branch cuts.
\begin{Lemma}[properties of the Abel map]
\label{lemmaAbels<0}
The following relations hold:
\begin{gather*}
\Abel(z_+)=-\Abel(z_-) +
\begin{cases}
0,& z\in \Gamma_{a_+}^{a_-},\\
-\tau, & z\in \Gamma_b^1,
\end{cases}
\qquad
\Abel(z_+) = \Abel(z_-)-1, \qquad z\in \Gamma_b^{a_+}.
\end{gather*}
\end{Lemma}
Consider now the following two row-vectors:
\begin{gather*}
\bs \phi (z)= [\phi_1(z), \phi_2(z)],\qquad
\phi_1(z) = \frac { \vartheta\le(\Abel(z)-\Abel(\infty) - \frac {\tau+1}2 +
G\ri)h(z)}{\vartheta\le(\Abel(z)-\Abel(\infty) - \frac{\tau+1}2\ri)}{\rm e}^{{\rm i}\pi K \Abel(z)},
\nn\\
\phi_2(z) = \frac {{\rm i} \vartheta\le(-\Abel(z)-\Abel(\infty) - \frac {\tau+1}2 +
G\ri)h(z)}{\vartheta\le(-\Abel(z)-\Abel(\infty) - \frac{\tau+1}2\ri)}{\rm e}^{-{\rm i}\pi K \Abel(z)}
\end{gather*}
and
\begin{gather*}
\bs\psi(z) = [\psi_1(z), \psi_2(z)],\qquad
\psi_1(z) = \frac { \vartheta\le(\Abel(z)+\Abel(\infty) - \frac {\tau+1}2 +
G\ri)h(z)}{\vartheta\le(\Abel(z)+\Abel(\infty) - \frac{\tau+1}2\ri)}{\rm e}^{{\rm i}\pi K \Abel(z)},
\\
\psi_2(z) = \frac {{\rm i} \vartheta\le(-\Abel(z)+\Abel(\infty) - \frac {\tau+1}2 +
G\ri)h(z)}{\vartheta\le(-\Abel(z)+\Abel(\infty) - \frac{\tau+1}2\ri)}{\rm e}^{-{\rm i}\pi K \Abel(z)}.
\end{gather*}
These are essentially the same formul\ae\ as \eqref{rowphi}, \eqref{rowpsi} except for a minor modification of the normalization.
Accordingly, we define \smash{$\wh M(z;G,K,\infty)$} exactly as in \eqref{defwhM}.
The boundary relations for this matrix \smash{$\wh M$} are now
\begin{gather*}
\wh M(z_+)= \wh M(z_-) \begin{bmatrix}
0 & 1 \\
-1 & 0
\end{bmatrix},\qquad z\in \Gamma_{a_+}^{a_-},
\\
\wh M(z_+)= \wh M(z_-)
\begin{bmatrix}
-{\rm e}^{-{\rm i}\pi K} & 0 \\
0 & -{\rm e}^{{\rm i}\pi K}
\end{bmatrix},
\qquad z\in \Gamma_b^{a_+},
\\
\wh M(z_+)= \wh M(z_-) \begin{bmatrix}
0 & {\rm e}^{-2{\rm i}\pi(G - \frac {K\tau}2) }\\
-{\rm e}^{2{\rm i}\pi(G - \frac {K\tau}2) }& 0
\end{bmatrix},\qquad z\in \Gamma_b^{1}.
\end{gather*}

By matching these to the boundary relations \eqref{jumpMs<0}, we deduce the following.
\begin{Proposition}\label{propM<0}
The solution of the Riemann--Hilbert Problem~{\rm\ref{RHPMs<0}} is given by $\wh M(z;G,K)$ in \eqref{defwhM} with the values of the constants $G$, $K$ given by
\be
\label{defGs<0}
G= \frac 1{2{\rm i}\pi} \le(n\Omega_1 + \nu + n \Omega_2\tau\ri) + \frac {\tau}2,\qquad
K = \frac {n\Omega_2}{{\rm i}\pi} + 1.
\ee
\end{Proposition}
The proof is entirely similar to Proposition~\ref{propM}.
The same observation holds that the solvability of the Riemann--Hilbert Problem~\ref{RHPMs<0} and hence of the Riemann--Hilbert Problem~\ref{RHPMs<0} depends entirely on the non-vanishing of the expression
\be
\label{meshs<0}
\vartheta\le(\frac {n\Omega_1 + \nu - n \Omega_2\tau }{2{\rm i}\pi} + \frac 12 \ri)
\ee
with $\Omega_1$, $\Omega_2$ defined by \eqref{defOmega12bis} and $\nu$ by \eqref{defnus>0}. Note the slight difference from \eqref{674}.
The quantization conditions for the periods of $\varphi$ now read
\begin{gather}
\wt H_1:= \frac{n\Omega_1}{2{\rm i}\pi} + \Re\le(\frac {\nu} {2{\rm i}\pi} \ri) - \frac {\Re \tau}{\Im\tau} \Im\le(\frac {\nu}{2{\rm i}\pi}\ri) \in \Z,\nonumber
\\
\wt H_2:= \frac{n\Omega_2}{2{\rm i}\pi} + \frac {\Im\le(\frac{\nu}{2{\rm i}\pi}\ri)}{\Im \tau} + \frac 1 2 \in \Z.\label{quantcond<0}
\end{gather}
The mesh of level sets $ \wt H_1\in \Z \ni \wt H_2$ are the gridlines shown in Figure~\ref{Figone} in the portion $\Re (s)<0$ of each pane.
\subsubsection{Local parametrices}
\label{paramg1}
The construction of the local parametrices is similar to that of Section~\ref{Errorg0}. The only parametrix that is less commonly encountered (but see \cite{BertolaAdvances}) is the one that needs to be defined at the point~$b$ in the case $\Re (s)>0$, when three main arcs meet at the same point (see Figure~\ref{Contoursinside}, left pane).
Let~$\DD_x$ denote a small disk centered at $x\in\{a_+,a_-, b,1\}$ and not containing any of the other branch-points $a_+$, $a_-$, $b$, $1$. Inside each of these disks one has to define an appropriate local solution of the Riemann--Hilbert Problem~\ref{RHPT0g1} or the Riemann--Hilbert Problem~\ref{RHPT0g1s<0} (according to the sign of $\Re(s)>0$ or $\Re(s)<0$, respectively).
Such a local solution (a.k.a.\ ``local parametrix'') is constructed in terms of Airy functions or Bessel functions. More specifically,
\begin{enumerate}\itemsep=0pt
\item[(1)] Inside $\mathbb D_{a_+}$, $\mathbb D_{a_-}$ the local parametrix is constructed in a manner analogous to what we illustrated in Section~\ref{Errorg0}.
\item[(2)] Inside $\mathbb D_{b}$ the local parametrix is constructed also in terms of Airy functions, arranged however in a different formula along the lines explained in Appendix~\ref{SecA3}.
\item[(3)] Inside $\mathbb D_1$ the local parametrix is constructed in terms of Bessel functions. This can be found in (see \cite[Section~3, p.~155--156]{VanlessenStrong}), or (see \cite[Section~4.2]{BertoKatTov}), and we will not enter into details since it is mainly irrelevant to the formulas that are of interest in this paper.
\end{enumerate}

 \subsubsection{Summary: The approximating mesh of location of zeros}
 \label{summary}
 The jump matrices of the error matrix $\scr E$ are of the form
 \[
 J_{\scr E}(z) =M_{_Q}(z) \le(\1 + \mathcal O\bigl(n^{-1}\bigr)\ri) M_{_Q}^{-1}(z). \label{JE}
 \]
 While in the region outside ${\rm EoT}$ the model solution $M_{_Q}$ does not depend on $n$ (see \eqref{modelMg0}) and hence $J_{\scr E} = \1 + \mathcal O\bigl(n^{-1}\bigr)$ since the matrix $M_{_Q}$ is $n$-independent. On the other hand, when~${s\in {\rm EoT}}$, the solution $M$ is instead given by Proposition~\ref{propM} and consequently $M_{_Q}$ \eqref{MtoMQ}) depends on $n$.

 The validity of the estimate $J_{\scr E} = \1 + \mathcal O\bigl(n^{-1}\bigr)$ depends on the matrices $M$, $M^{-1}$ remaining bounded, uniformly in $n$, with respect to $z$ on the contour.
 Given that $\det (M)\equiv 1$, the source of the potential unboundedness are the terms in the denominators in \eqref{defwhM}. Of these, the only term that may potentially vanish is the term $\vartheta\le(G- \frac {\tau+1}2\ri)$, which becomes \eqref{674} or \eqref{meshs<0} (depending on the sign of $\Re(s)$). Therefore, the correct error estimate is
 \[
 J_{\scr E}(z) =\1 + \mathcal O\le(\frac 1{n \vartheta\le(G- \frac {\tau+1}2\ri)^2} \ri),
 \]
 with $G$ given by either \eqref{defGs>0} or \eqref{defGs<0} (depending on the sign of $\Re (s)$).

Thus the sufficient condition of asymptotic solvability of the Riemann--Hilbert Problem~\ref{RHPY1} is that
\[
 \vartheta\le(\frac {n \Omega_1(s) - n\Omega_2(s)\tau(s) +\nu(s)}{2{\rm i}\pi} + \frac 1 2\mathcal X_{\{\Re s>0\}}(s)\ri)= o\le(\frac 1{\sqrt n}\ri),
\]
with $\mathcal X_{\{\Re s>0\}}(s)$ denoting the indicator function of the right half-plane.
This condition fails inside the union of small disks of radii \smash{$\mathcal O\bigl(n^{-\frac 3 2}\bigr)$} in the $s$-plane centered at the points where~${\vartheta =0}$, namely, the intersection of the mesh of lines indicated by \eqref{quantcond} (for $\Re (s)>0$) or \eqref{quantcond<0} (for~${\Re (s)<0}$).

The actual numerics shows a seemingly much faster convergence of the mesh to the actual position of the zeros (see Figure~\ref{Figone}), which appears to be one of those serendipitous situations where the approximation works ``better than expected''.

\begin{figure}[t]	\centering
\includegraphics[width=0.48\textwidth]{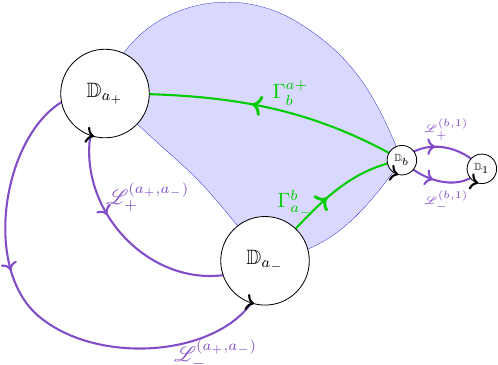} \quad
\includegraphics[width=0.48\textwidth]{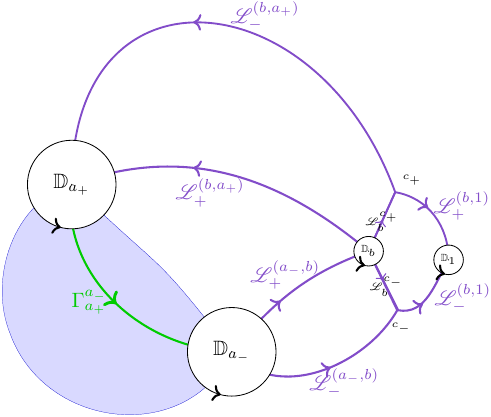}
\caption{The RHPs for the error terms in the two cases $\Re (s)>0$ (left) and $\Re(s)<0$. Compare with Figure~\ref{Contoursinside}.}
\end{figure}

 \section{Approximation of the Hamiltonian}
 The Hamiltonian is obtained from the Taylor expansion near $z=0$ of the solution of Riemann--Hilbert Problem~\ref{RHPY1} outside of ${\rm EoT}$, and Riemann--Hilbert Problem~\ref{RHPYg1} inside ${\rm EoT}$, according to Proposition \ref{tauPV}.

 In either case, the chain of transformations ($Y\to W\to T_0 \sim M_{_Q}$) and the subsequent approximation imply that in a neighbourhood of $z=0$ we have
 \[
 Y_n(z) = {\rm e}^{-n\frac \ell 2\s_3} W(z){\rm e}^{n(g(z) + \frac \ell 2)\s_3}=
 {\rm e}^{-n\frac \ell 2\s_3} \scr E (z) M_{_Q} (z){\rm e}^{n(g(z) + \frac \ell 2)\s_3},
 \]
 where $\scr E(z)$, as discussed in Section~\ref{summary}, is a matrix that is \smash{$\1+ \mathcal O\bigl(\frac 1{n \vartheta(G- \frac {\tau+1}2)^2} \bigr)$} near the origin (as long as $s$ has finite distance from the boundary of ${\rm EoT}$ and the imaginary axis within it).

We thus have
\[
Y_n^{-1} Y_n' |_{z=0} =\bigl({\rm e}^{-ng\s_3} \bigl(M_{_Q}^{-1} M_{_Q}' + M^{-1}_{_Q} \scr E^{-1} \scr E' M_{_Q} \bigr) {\rm e}^{ng\s_3}+ n g'(z) \s_3 \bigr)\big|_{z=0}.
\]
Given the expression of $M$ in either Proposition~\ref{modelMg0} (for $s\not\in {\rm EoT}$), Proposition~\ref{propM} for $s\in {\rm EoT}\cap\{\Re (s)> 0\}$, or Proposition~\ref{propM<0} $s\in {\rm EoT}\cap\{\Re (s)< 0\}$, we have
 \[
Y_n^{-1} Y_n' |_{z=0} =\bigl( {\rm e}^{(-ng-S)\s_3} \bigl(M^{-1} M' + M^{-1} \scr E^{-1} \scr E' M\bigr){\rm e}^{(S+ ng)\s_3} +\bigl(S'+ n g'(z)\bigr) \s_3 \bigr)|_{z=0},
\]
where clearly the expressions for $g$, $S$ also depend on which region we are considering.

We are interested in the $(1,1)$ entry and hence the conjugation in the first term by the diagonal matrix ${\rm e}^{(S-ng)\s_3}$ is immaterial. Furthermore, the term containing $\scr E$ yields a sub-leading contribution which we ignore for the purpose of this computation.
In principle, we should fork the computation according to the three regions: (i)~outside ${\rm EoT}$, (ii)~inside ${\rm EoT}$ with $\Re (s)>0$, (iii)~inside ${\rm EoT}$ with $\Re(s)<0$. However, the computation inside ${\rm EoT}$ would be rather formal because we have not specified the correct orders of approximations when $s$ is in a neighbourhood of one of the zeros.
Additionally, the explicit approximation is of no particular interest to us and thus for simplicity we decided to forego it entirely in this paper.

{\bf Hamiltonian outside $\boldsymbol{{\rm EoT}}$.}
 Using the explicit expression of $M(z)$ in Proposition~\ref{modelMg0}, we have
 $ M^{-1}M' = \frac {{\rm i}\s_2}s $
 and hence this term does not contribute being off-diagonal.
 Using~\eqref{S(z)} and the fact that $R(0)=-\frac s 2$, \smash{$ R(1) = - \sqrt{1+\smash{\frac{s^2}4}\vphantom{\big|}}$} with the root's determination such that $R(1;s)\simeq -\frac s 2$ as $|s|\to\infty$, we find
\smash{$
 S'(0) =
 \frac { {K} }s + \frac {2{\rho}}{s^2 + s \sqrt{s^2+4}}$}.
 Using then the explicit expression of $g(z)$ derived from \eqref{phig0}
 \[
 g'(z) = \frac 1 2 \varphi'(z;s) + \frac{s}{2z^2} = \frac{\sqrt{z^2+\frac {s^2}4}}{z^2} + \frac s {2z^2}.
 \]
 Since $g'(z)$ is regular at $z=0$, we need to take the root's determination near $z=0$ that tends to $-\frac s 2$ and hence
 \[
 g'(z) =-s \frac{\sqrt{1+\frac {4z^2}{s^2} }}{2z^2} + \frac s {2z^2} = -\frac 1 s + \mathcal O(z) \Rightarrow g'(0;s) = -\frac 1 s.
 \]
 We thus conclude the following.

 \begin{Proposition} \label{H_Vapproxout}
 For $s$ in closed subsets outside ${\rm EoT}$, we have the uniform approximation
\[
 H_V(s) = \frac{n-{K}}s
 - \frac {2{\rho}}{s^2 + s \sqrt{s^2+4}}
 +\frac { {\rho}} 2 + \mathcal O\bigl(n^{-1}\bigr).
\]
 \end{Proposition}

\section{Conclusion}
In this paper, we have considered the particular scaling where we send $n\to\infty$ and rescale only the independent variable $t = ns$ \eqref{42}. However, as evidenced also in Figure~\ref{Figone}, if $ {\rho}$ or $ {K}$ or both are large relative to $n$, the approximation needs to be modified.
 If we let $ {\rho} = n \varrho$, $ {K} = n \varkappa$, then we would need to construct a different $g$-function where $\theta$ in~\eqref{42} is replaced by
\[
\theta(z;s) = n\le(\frac s z+ \varrho \ln \le(1-\frac 1 z \ri) + \varkappa \ln z\ri).
\]
Consequently, the construction of the $g$-function, even under the one-cut assumption would significantly change. In practice, this means that the set ${\rm EoT}$ would have a different shape that depends on $\varrho$, $\varkappa$. While conceptually there is no major difference, we found that there are practical and significant obstacles in obtaining an effective description of the $g$-function under these assumptions. A separate analysis is needed but it is deferred to a future publication.

A separate, long term question is whether the other families of rational solutions described in Theorem~\ref{thmKitaev} can be similarly framed in terms of semiclassical orthogonal polynomials. Irrespectively, the isomonodromic approach discussed in the Introduction should be available and hence the corresponding asymptotic analysis should be accessible. These are also issue that we defer to future investigations.

\vspace{-1mm}

\appendix
\section{Airy parametrices}\label{app1}
\begin{figure}[t]\centering
\begin{tikzpicture}[scale=1.3]

\draw [line width=1,postaction={decorate,decoration={markings,mark=at position 0.51 with {\arrow[line width=2pt]{>}}}}] (0,0) to node[pos=0.67, above]{$\begin{bmatrix}1 & 1\\0 & 1 \end{bmatrix}$} (2,0);

\draw [line width=1,postaction={decorate,decoration={markings,mark=at position 0.51 with {\arrow[line width=2pt]{<}}}}] (0,0) to node[pos=0.67, above right]{$\begin{bmatrix}1 & 0\\1 & 1 \end{bmatrix}$} (120:2);
\draw [line width=1,postaction={decorate,decoration={markings,mark=at position 0.51 with {\arrow[line width=2pt]{<}}}}] (0,0) to node[pos=0.67, below right]{$\begin{bmatrix}1 & 0\\1 & 1 \end{bmatrix}$} (-120:2);

\draw [line width=1,postaction={decorate,decoration={markings,mark=at position 0.51 with {\arrow[line width=2pt]{<}}}}] (0,0) to node[pos=0.67, above]{$\begin{bmatrix} 0 & 1\\-1 & 0\end{bmatrix}$} (-2,0);
\end{tikzpicture}

\caption{The jump behaviour of the Airy parametrix.}
\label{Airy1}

\end{figure}
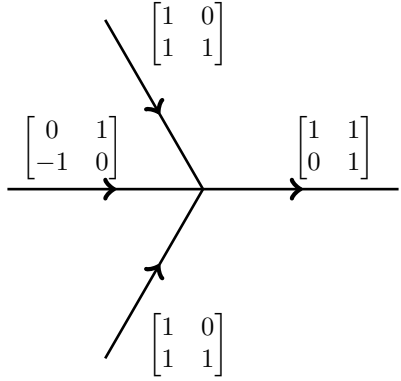

The complete construction of the approximation to the RHP of the main body of the paper requires the definition of a local solution to the final RHP which is known in the literature as an ``Airy parametrix''. While this is quite standard, it may be useful for the reader to find here its complete and self contained definition. The origin of these definitions can be traced back to~\cite{DKMVZ} but here we refer to \cite[Appendix~A]{BertolaBothner}.
We define\vspace{-1mm}
\[
	A_0(\zeta) = \begin{bmatrix}
	\displaystyle\frac{\d}{\d\zeta}\textnormal{Ai}(\zeta) &\displaystyle {\rm e}^{{\rm i}\frac{\pi}{3}}\frac{\d}{\d\zeta}\textnormal{Ai}\bigl({\rm e}^{-{\rm i}\frac{2\pi}{3}}\zeta\bigr) \\
	\textnormal{Ai}(\zeta) & \displaystyle {\rm e}^{{\rm i}\frac{\pi}{3}}\textnormal{Ai}\bigl({\rm e}^{-{\rm i}\frac{2\pi}{3}}\zeta\bigr)\\
	\end{bmatrix},\qquad\zeta\in\mathbb{C},\vspace{-1mm}
\]
where $\textnormal{Ai}(\zeta)$ is the Airy function, namely, a particular solution to Airy's equation
$f(\zeta)''=\zeta f(\zeta)$
satisfying the following asymptotic behaviour as $\zeta\rightarrow\infty$ in the sector $-\pi<\operatorname{arg} \zeta<\pi$:\vspace{-1mm}
\begin{equation*}
	\textnormal{Ai}(\zeta) = \frac{\zeta^{-1/4}}{2\sqrt{\pi}}{\rm e}^{-\frac{2}{3}\zeta^{3/2}}\left(1-\frac{5}{48}\zeta^{-3/2}+\frac{385}{4608}\zeta^{-6/2}+\mathcal O\bigl(\zeta^{-9/2}\bigr)\right).\vspace{-1mm}
\end{equation*}
We finally define the following piecewise-analytic matrix-valued function:{\samepage
\begin{gather}
		{\bf A} (\zeta) =
 \begin{cases}
 A_0(\zeta), & \operatorname{arg} \zeta\in\bigl(0,\frac{2\pi}{3}\bigr), \\
 A_0(\zeta)\begin{bmatrix}
 1 & 0 \\
 -1 & 1 \\
 \end{bmatrix}, & \operatorname{arg}\zeta\in\bigl(\frac{2\pi}{3},\pi\bigr), \\[10pt]
 A_0(\zeta)\begin{bmatrix}
 1 & -1\\
 0 & 1\\
 \end{bmatrix}
 & \operatorname{arg}\zeta\in\bigl(-\frac{2\pi}{3},0\bigr) , \\[10pt]
A_0(\zeta)\begin{bmatrix}
0 & -1 \\
 1 & 1\\
 \end{bmatrix}, & \operatorname{arg}\zeta\in\bigl(-\pi,-\frac{2\pi}{3}\bigr),
 \end{cases}\label{AiryParm}
\end{gather}
which solves the RHP with jumps for $\operatorname{arg} \zeta=-\pi,-\frac{2\pi}{3},0,\frac{2\pi}{3}$ as depicted in Figure~\ref{Airy1}.}

The definition \eqref{AiryParm} is crafted in such a way that the following asymptotic expansion holds in any direction:
\[
{\bf A} (\zeta)=
\frac{\zeta^{\sigma_3/4}}{2\sqrt{\pi}}\begin{bmatrix}
-1 & {\rm i}\\
1 & {\rm i}\\
\end{bmatrix}\left\{\1+\frac{1}{48\zeta^{3/2}}\begin{bmatrix}
1 & 6{\rm i}\\
6{\rm i} & -1\\
\end{bmatrix}+\mathcal O\bigl(\zeta^{-6/2}\bigr)\right\}{\rm e}^{-\frac{2}{3}\zeta^{3/2}\sigma_3}.
\]

\subsection{The Airy parametrix for a three main-arc intersection}
\label{SecA3}
The basic building block discussed in the previous section can be used to construct a more complicated local parametrix that needs to be used in a neighbourhood of a $3$-arc intersection, like the case of the point $z=b$ for the asymptotic in ${\rm EoT}$ for $\Re(s)>0$, Section~\ref{secmodels>0}.

We thus first state and solve a model RHP and then show how it proves of use to the case at hand.

\begin{figure}[t]\centering
\includegraphics[width=0.495\textwidth]{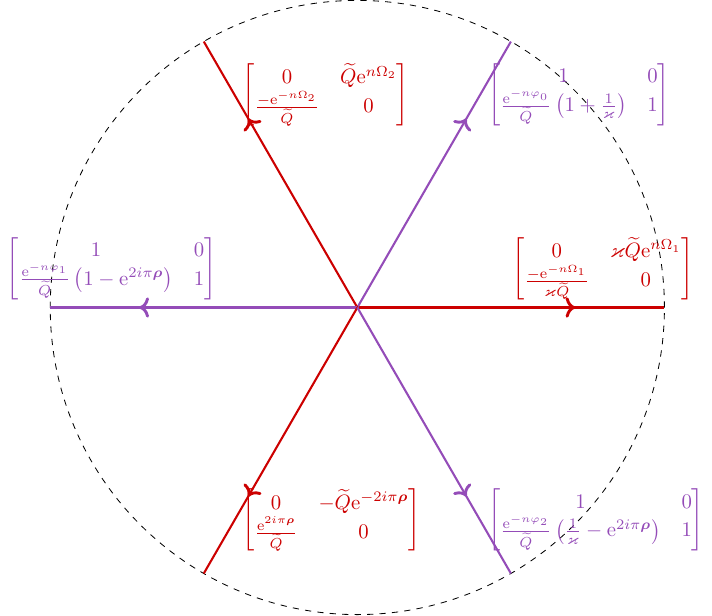}
\includegraphics[width=0.495\textwidth]{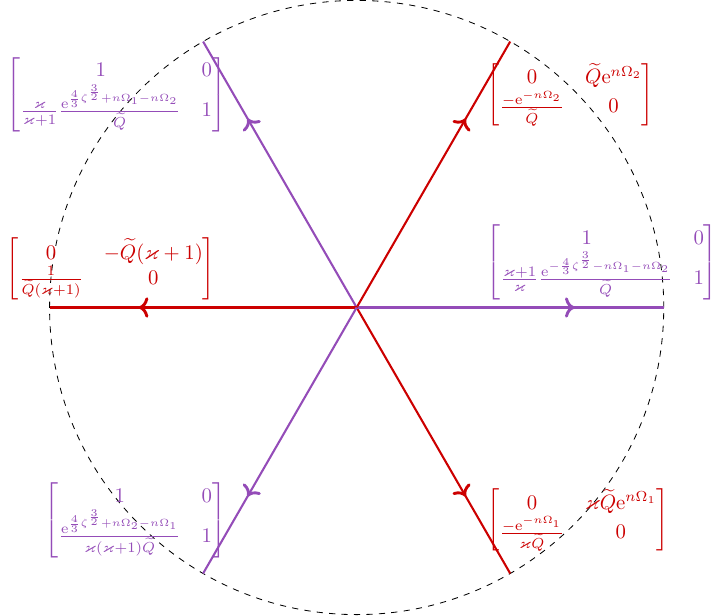}

\caption{The local Riemann--Hilbert Problem~\ref{RHPT0g1} in $\DD_b$ (left pane) and its transformation to the $\zeta$ plane.}\label{FigA3}
\end{figure}

Consider the jump matrices of the Riemann--Hilbert Problem~\ref{RHPT0g1} on the arcs that originate or terminate at~${z=b}$ (see \eqref{JT0s>0}). They are summarized in the left pane of Figure~\ref{FigA3}. There, we have denoted by~$\varphi_0$,~$\varphi_1$,~$\varphi_2$ the restrictions of $\varphi$ to the sectors in $\DD_b$. These sectors are defined as follows: $\scr S_0$ bounded by the arcs $\Gamma_b^1$, $\Gamma_b^{a_+}$,
$\scr S_1$ bounded by the arcs $\Gamma_b^{a_+}$, $\Gamma_{a_-}^b$ and
$\scr S_2$ bounded by the arcs~\smash{$\Gamma_{a_-}^b$},~\smash{$\Gamma_b^1$}, respectively (see Figure~\ref{Contoursinside}, left pane).
They are related to each other as follows:
\begin{gather*}
 \varphi_0(z) + \varphi_1(z) = 2\Omega_2, \qquad z\in \Gamma_{b}^{a_+},\\
 \varphi_0(z) + \varphi_2(z) = 2\Omega_1, \qquad z\in \Gamma_{b}^{1},\\
 \varphi_1(z) + \varphi_2(z) = 0, \qquad z\in \Gamma_{a_-}^{b}.
\end{gather*}
Observing that $\varphi_0(b) = -\Omega_1- \Omega_2 $, we can define the local conformal map $\zeta(z)$ by the expression
\[
\frac 43 \zeta^\frac 32 = n\varphi_0(z) + n(\Omega_1 + \Omega_2), \qquad z\in \scr S_0,
\]
where the radical of $\zeta$ is intended in the sense of principal determination (with the branch cut extending along $\zeta\in \R_-$).
If we perform the analytic extension of this definition to the sectors~$\scr S_1$, $\scr S_2$, we obtain the following relations:
\begin{gather*}
n\varphi_1(z) = -n\varphi_0(z) + 2\Omega_2 =- \frac 43 \zeta^\frac 32 - n\Omega_1 + n\Omega_2,\qquad z\in \scr S_1,\\
n\varphi_2(z) = -n\varphi_0(z) + 2\Omega_1 =- \frac 43 \zeta^\frac 32 - n\Omega_2 + n\Omega_1,\qquad z\in \scr S_2.
\end{gather*}
With this definition of $\zeta$, the arc $\Gamma_{a_-}^b\cap \DD_b$ is mapped to the negative $\zeta$ axis.

Moreover, we have denoted by \smash{$\wt Q(z)$} the analytic extension of $Q(z)$ in the full disk $\DD_b$ from the upper part, so that
\[
\wt Q(z) =
\begin{cases}
Q(z), & \{z\mid\Im \zeta(z)> 0\}\cap \DD_b,\\
{\rm e}^{2{\rm i}\pi { {\rho}} }Q(z), & \{z\mid \Im \zeta(z)<0 \}\cap \DD_b.
\end{cases}
\]
With these notation in place and recalling the relationship between $\varkappa$ and ${\rm e}^{2{\rm i}\pi { {\rho}}}$ \eqref{defkappa}, the jump matrices are as indicated in Figure~\ref{FigA3}, left pane. Note that we have also re-oriented some of the arcs. Note that the jump matrix on the left ray is the result of the multiplication of the jumps on the two arcs \smash{$\scr L_+^{(a_-,b)}$}, \smash{$\scr L_+^{(b,a_+)}$}.

With these preparations, the solution of the RHP with the jump matrices indicated in the right pane of Figure~\ref{FigA3} is written as follows in terms of the matrix ${\bf A}(z)$ defined in \eqref{AiryParm}:
\[
\wt {\bf P}(\zeta) =
\begin{cases}
\displaystyle{\bf A}(\zeta) {\rm e}^{(\frac 2 3 \zeta^{\frac 32} + \frac{\Omega_1-\Omega_2}2) \s_3} \left(\frac {\varkappa}{(\varkappa+1)\wt Q(z)} \right)^{\frac {\s_3}2}
{\rm e}^{\frac {{\rm i}\pi\s_3}2},& \arg(\zeta) \in \bigl( \frac {\pi} 3, \pi \bigr),\\
\displaystyle
{\bf A}(\zeta) {\rm e}^{ (\frac 2 3 \zeta^{\frac 32} + \frac{\Omega_2-\Omega_1}2)\s_3} \left(\frac {1}{\varkappa (\varkappa+1)\wt Q(z)} \right)^{\frac {\s_3}2}{\rm e}^{-\frac {{\rm i}\pi\s_3}2}, & \arg(\zeta) \in \bigl( -\pi, -\frac \pi 3 \bigr),\\
\displaystyle
{\bf A}(\zeta) {\rm e}^{ (\frac 2 3 \zeta^{\frac 32} + \frac{\Omega_2+\Omega_1}2)\s_3}
\begin{bmatrix}
0 & -\wt Q(z)\\
\displaystyle\frac 1{\wt Q(z)} & 0
\end{bmatrix}
 \left(\frac {\varkappa+1}{\varkappa \wt Q(z)} \right)^{\frac {\s_3}2}{\rm e}^{\frac {-{\rm i}\pi\s_3}2}, & \arg(\zeta) \in \bigl( -\frac\pi 3, \frac \pi 3 \bigr).
\end{cases}
\]

\subsection*{Acknowledgements}
The second author completed the work during his tenure as Royal Society Wolfson Visiting
Fellow (RSWVF/R2/242024) at the School of Mathematics in Bristol University. The work was supported in part by the Natural Sciences and Engineering Research Council of Canada (NSERC) grant RGPIN-2023-04747. Both authors thank the anonymous referees for the detailed reports and bibliographic improvements.

\LastPageEnding

\end{document}